\documentclass[a4paper,fleqn,usenatbib]{mnras}

\usepackage[T1]{fontenc}
\usepackage{ae,aecompl}

\usepackage{graphicx}	
\usepackage{amsmath}	
\usepackage{amssymb}	
\usepackage{ulem}       
\usepackage{cancel}     
\usepackage{float}
\usepackage{hyperref}

\title[Jets from SANE Super-Eddington Accretion Disks]{Jets from SANE Super-Eddington Accretion Disks: Morphology, Spectra, and Their Potential as  Targets for ngEHT}

\author[Brandon Curd et al.]{
Brandon Curd$^{1,2}$\thanks{E-mail: brandon.curd@cfa.harvard.edu},
Razieh Emami$^{1}$,
Richard Anantua$^{3,1,2}$,
Daniel Palumbo$^{1,2}$,
\newauthor Sheperd Doeleman$^{1,2}$, 
Ramesh Narayan$^{1,2}$
\\
$^{1}$ Harvard-Smithsonian Center for Astrophysics, 60 Garden Street, Cambridge, MA 02138, USA
\\
$^{2}$ Black Hole Initiative at Harvard University, 20 Garden Street, Cambridge, MA 02138, USA
\\
$^3$ Department of Physics $\&$ Astronomy, The University of Texas at San Antonio, One UTSA Circle, San Antonio, TX 78249, USA
\\
}

\date{Accepted XXX. Received YYY; in original form ZZZ}

\pubyear{2022}

\begin{document}
\label{firstpage}
\pagerange{\pageref{firstpage}--\pageref{lastpage}}
\maketitle

\begin{abstract}
We present general relativistic radiation magnetohydrodynamics (GRRMHD) simulations of super-Eddington accretion flows around supermassive black holes (SMBHs) which may apply to tidal disruption events (TDEs). We perform long duration ($t\geq81,200\, GM/c^3$) simulations which achieve mass accretion rates $\gtrsim 11$ times the Eddington rate and produce thermal synchrotron spectra and images of their jets. The jet reaches a maximum velocity of $v/c \approx 0.5-0.9$, but the density weighted outflow velocity is $\sim0.2-0.35c$. Gas flowing beyond the funnel wall expands conically and drives a strong shock at the jet head while variable mass ejection along the jet axis results in internal shocks and dissipation. For a $T_i/T_e=1$ model, the radio/submillimeter spectra peak at $>100$ GHz and the luminosity increases with BH spin, exceeding $\sim 10^{41} \, \rm{erg\, s^{-1}}$ in the brightest models. The emission is extremely sensitive to $T_i/T_e$ as some models show an order of magnitude decrease in the peak frequency and up to four orders of magnitude decline in their radio/submillimeter luminosity as $T_i/T_e$ approaches 20. Assuming a maximum VLBI baseline distance of $10 \ {\rm{G}}\lambda$, 230 GHz images of $T_i/T_e=1$ models shows that the jet head may be bright enough for its motion to be captured with the EHT (ngEHT) at $D\lesssim110$ (180) Mpc at the $5\sigma$ significance level. Resolving emission from internal shocks requires $D\lesssim45$ Mpc for both the EHT or ngEHT. The 5 GHz emission in each model is dimmer ($\lesssim10^{36} \ {\rm{erg\, s^{-1}}}$) than upper limits placed on TDEs with no radio emission which suggests jets similar to our models may have gone undetected in previous observations. Our models suggest that the ngEHT may be utilized for $>230$ GHz radio/submillimeter followup of future TDEs.
\end{abstract}




\section{Introduction}

The central black holes (BHs) of galaxies throughout the universe gain some fraction of their mass from the tidal disruption of stars in the near vicinity (see \citealt{2021MNRAS.500.3944P} for a recent study). Random scatterings of stars orbiting the BH occasionally places them on a chance orbit that will cross the tidal radius ($R_t$), beyond which the star's self gravity is weaker than the tidal forces acting on it and it is subsequently disrupted into an elongated stream of gas \citep{1975Natur.254..295H,1988Natur.333..523R,1989IAUS..136..543P,1989ApJ...346L..13E}. The bright transient which occurs as a result of the disruption is known as a tidal disruption event (TDE). The most well studied TDE is that of a near solar mass star being disrupted by a 
supermassive BH (SMBH). In these events, the transient light curve roughly follows a $L\propto (t/t_{\rm{fb}})^{-5/3}$ decay. Here, the fallback time
\begin{equation} \label{eq:tfb}
t_{\rm{fb}} = 3.5\times10^{6} \, {\rm{s}} \, \left(\dfrac{M_{\rm{BH}}}{10^6 M_\odot}\right)^{1/2} \left(\dfrac{M_*}{M_\odot}\right)^{-1}\left(\dfrac{R_*}{R_\odot}\right)^{3/2},
\end{equation}
where $M_*$ is the mass of the disrupted star and $R_*$ is its radius, is the characteristic decay time for the TDE \citep{2013MNRAS.435.1809S}. This behaviour originates from the gas dynamics during the disruption, which causes the star to be disrupted into an elongated stream of gas with a binding energy distribution which dictates that mass flows back towards pericenter at the `fallback rate' $\dot{M}_{\rm{fb}}\propto (t/t_{\rm{fb}})^{-5/3}$ \citep{2013MNRAS.435.1809S}. The earliest theoretical works on the subject proposed that the radiation emitted by TDEs would originate from a small scale, geometrically thick accretion disk located near $\sim R_t$ due to circularization of the stream with gas temperatures leading to X-ray/UV emission (e.g. see \citealt{1988Natur.333..523R}).

Thanks to various surveys in the optical/UV and X-ray, the number of known TDEs has grown substantially since their initial discovery via \textit{ROSAT} (see \citealt{2015JHEAp...7..148K,2021ARA&A..59...21G} for a review). We now understand that TDE emission is quite complicated as not all TDEs have bright X-ray emission. Also, the optical/UV emission follows the same power-law behaviour as expected for the X-ray. This suggests that the optical/UV component originates from reprocessed X-rays emitted from the hot accretion disk. This scenario is supported observationally as several Bowen-TDEs, which show Bowen fluorescence (requiring reprocessing of higher energy photons), have been detected \citep{2019ApJ...887..218L,2021ApJ...908....4V}. 

Radio studies in the $5-8.4$ GHz range have only detected emission from a handful of TDEs \citep{2020SSRv..216...81A}. Three of these are likely so-called ``jetted'' TDEs, which launched powerful relativistic jets with bulk lorentz factor $\Gamma > 10$ and had a peak radio luminosity of $L_{\rm{peak, radio}}\sim10^{40}-10^{42} \, \rm{erg\, s^{-1}}$. The other events, which emitted significantly less energy in the radio with $L_{\rm{peak, radio}}\sim10^{37}-10^{39} \, \rm{erg\, s^{-1}}$, have been modeled using sub-relativistic outflows, internal jet shocks, and off-axis jets. An important difference in each of these mechanisms is the site of acceleration for electrons which eventually produce the observed synchrotron emission. For example, in the external shock picture the electrons are accelerated in the shock between the outflow and the circumnuclear medium (CNM) while in the internal acceleration picture the electrons are accelerated inside of the jet itself.

It is worth noting that radio emission in ``non-jetted'' TDEs (TDEs with no highly relativistic jet) typically appears $>30$ days after the peak emission. This may simply be due to the initially turbulent evolution of the disk, which may suppress the formation of a funnel region, through which the gas may be accelerated into a jet \citep{2021MNRAS.507.3207C}. However, \citet{2020MNRAS.495.1374B} present models which appear to have formed a funnel. It is also possible that our understanding of the disk formation process itself is incomplete. For instance, the onset of disk formation may not correlate to the time at which optical/UV/X-ray emission emerges.

TDEs are of great interest for studies of accretion physics and stellar populations close to the BH. For TDEs of Sun-like stars around BHs with mass $M_{\rm{BH}}\lesssim 10^{7.5}\,M_\odot$, the peak fallback rate will exceed the Eddington rate $\dot{M}_{\rm{Edd}}$ \citep{2013MNRAS.435.1809S}, which opens up the possibility that the accretion rate is actually super-Eddington. For example, a solar mass star disrupted around a $10^6 \, M_\odot$ BH will have a peak fallback rate that is $\sim100$ times the Eddington rate, and the fallback rate should remain above Eddington for a few fallback times.  Hydrodynamic simulations of the early stages of disk formation in a TDE suggest that as much as 20\% of the returning stream may actually cross the horizon through an accretion flow; however, the current library of published simulations has yet to cover a substantial range of the TDE parameter space and the effects of magnetism have largely been ignored in long term simulations \citep{2009ApJ...697L..77R,2014ApJ...783...23G,2015ApJ...804...85S,2016MNRAS.461.3760H,2016MNRAS.455.2253B,2019arXiv191010154L,2020MNRAS.495.1374B,2021MNRAS.504.4885B,2021MNRAS.507.3207C,2022MNRAS.510.1627A}.

The exact nature of the accretion disk structure is still uncertain, but previous works have applied global GRRMHD accretion disk simulations to model the accretion flow \citep{2018ApJ...859L..20D,2019MNRAS.483..565C}. The emission properties of TDEs are possibly described by thick accretion disks \citep{2018ApJ...859L..20D,2019MNRAS.483..565C}. In particular, magnetically arrested disks (or MADs, \citealt{2003PASJ...55L..69N}) around spinning BHs produce powerful jets and have emission properties remarkably similar to that of \textit{Swift} J1644+57 \citep{2019MNRAS.483..565C}. For non-jetted TDEs, the `standard and normal evolution' (or SANE) disks may be appropriate to describe X-ray TDEs, but the optical/UV component was uncertain in \citet{2019MNRAS.483..565C} since it originated from the outer radius of the torus, which was initialized using an equilibrium torus model rather than from following the entire disk formation process in a TDE. SANE models may also apply to jetted TDEs, as a funnel region in super-Eddington outflows can result in highly energetic outflows capable of explaining even jetted events \citep{2015MNRAS.453.3213S,2020MNRAS.499.3158C}. A numerical study of the radio/submillimeter emission from outflows driven by SANE super-Eddington accretion flows and its comparison with the radio emission in TDEs is thus well motivated.

In this work, we model the radio/submillimeter emission (via thermal synchrotron) from the outflows of super-Eddington accretion flows using GRRMHD simulations similar to those presented in \citet{2019MNRAS.483..565C}. However, in this work we use a grid with more resolution in the jet and run the simulation substantially longer such that the outflow reaches radii similar to the emission scales for known radio TDEs \citep{2020SSRv..216...81A}. We perform simulations of SANE accretion disks around BHs of mass $M_{\rm{BH}}=(5,10)\times 10^6\,M_\odot$ and a dimensionless BH spin of $a_*=(0,0.5,0.9)$ and measure the jet power and morphology across the parameter space. We perform general relativistic ray tracing (GRRT) of each model to produce spectra and images of the jet emission. We examine the viability of detecting and resolving the jet at 230 GHz. Based on our results, we suggest that the Event Horizon Telescope (EHT) or the next generation Event Horizon Telescope (ngEHT) could be suited for radio/submillimeter follow-up of TDEs. A direct probe of jet launching from super-Eddington accretion disks and radio/submillimeter emission in TDEs are both of significant interest and are expected to increase our understanding of BH accretion and the environment of BHs.

The paper is structured as follows. In Section \ref{sec:TDErates}, we motivate radio/submillimeter observations of nearby TDEs by computing the expected number of TDEs within $D<60$ Mpc. In Section \ref{sec:nummethods}, we outline the numerical methods used in this work. In Section \ref{sec:results}, we describe the accretion flow and outflow properties. In Section \ref{sec:imaging}, we analyse GRRT radio/submillimeter spectra and images of each model and determine the viability of resolving the jets in each simulation with the EHT/ngEHT. We compare our results with current TDE radio observations in Section \ref{sec:discussion} and conclude in Section \ref{sec:conclusions}.

\section{TDE Rates} \label{sec:TDErates}

\begin{figure*}
    \centering{}
	\includegraphics[width=\textwidth]{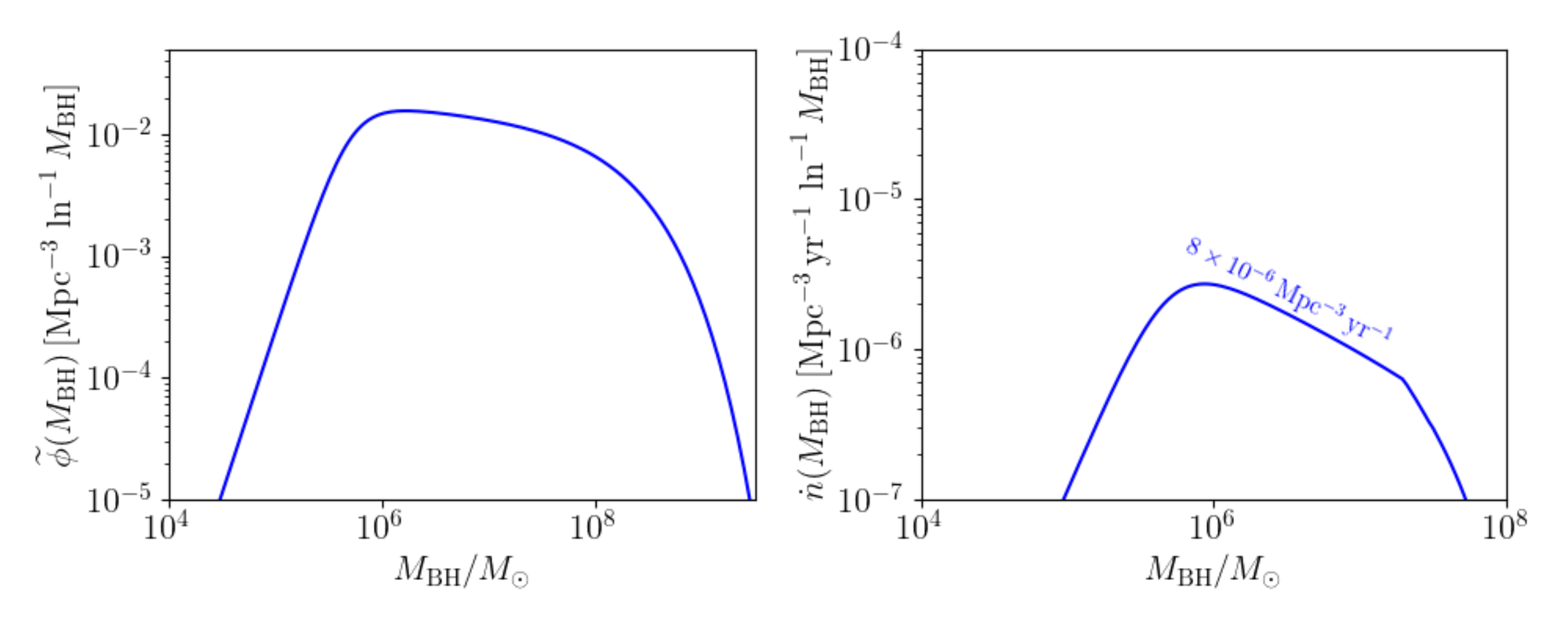}
    \caption{Here we show the BHMF as calculated in the text (left panel). We then show the volumetric TDE rate (right panel). The text above the curve shows the TDE rate when the curve is integrated over BH mass.}
    \label{fig:BHMF}
\end{figure*}

\begin{figure}
    \centering{}
	\includegraphics[width=\columnwidth]{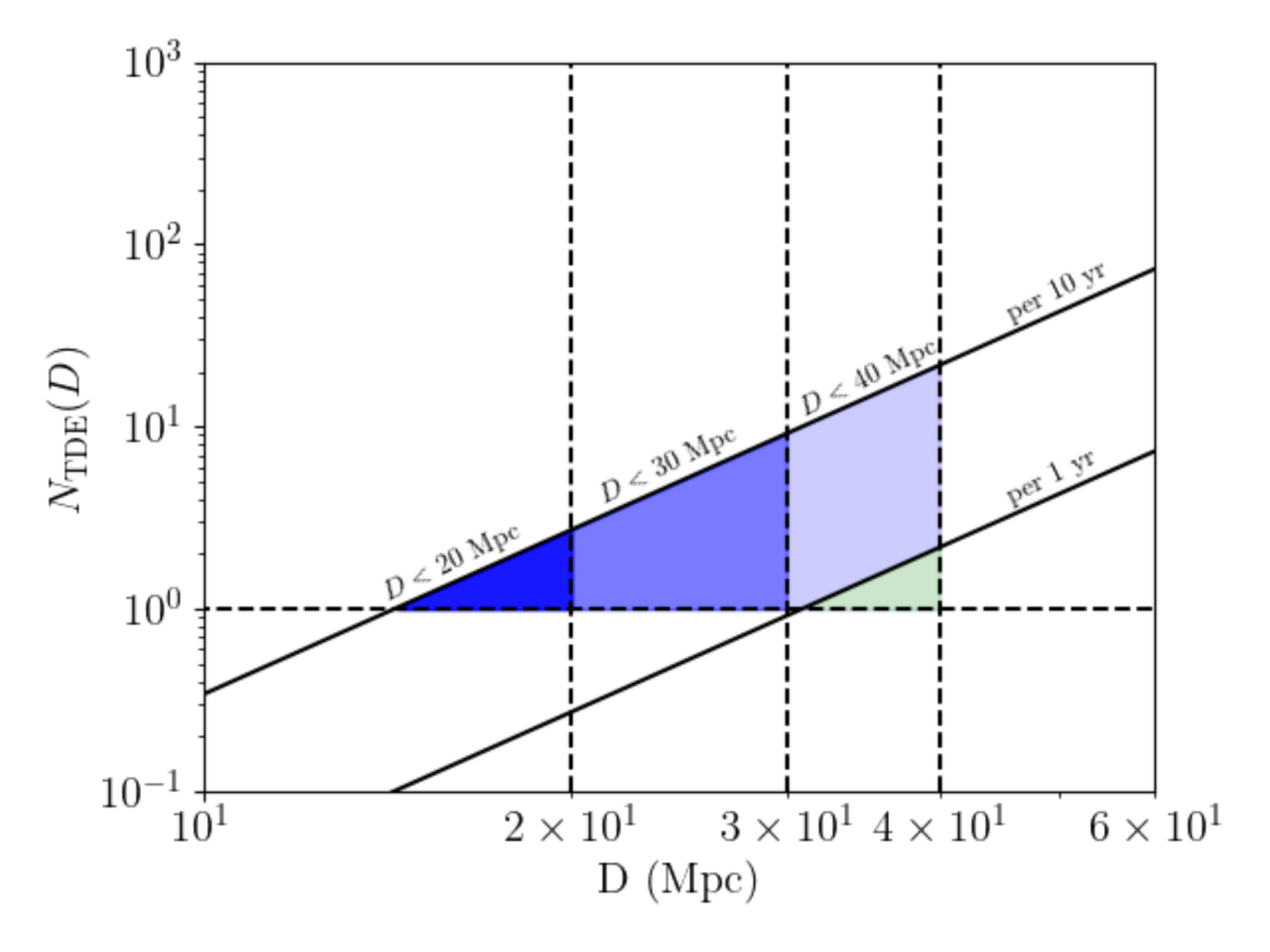}
    \caption{Here we estimate the number of total TDEs expected given a limiting observable distance. The shaded regions denote the range of parameter space where the expected number of TDEs in $1$ year (green) and $10$ years (blue) exceeds $1$.}
    \label{fig:rates}
\end{figure}

In this section, we motivate follow-up observations of TDEs in the radio/submillimeter band by quantifying the number of TDEs in a given volume of the nearby universe assuming conservative TDE rates. We utilize methods described in \citet{2016MNRAS.455..859S} and the interested reader should see their work for a detailed discussion of TDE rates. The number of possible TDEs per year grows rapidly at higher redshift since the number of BHs in an integrated volume of space varies as $N_{\rm{BH}}(D)\propto D^3$. Similarly, assuming estimates of the TDE rates are applicable broadly within the local universe (see \citealt{2016MNRAS.455..859S,2020MNRAS.497.2276P}), we can estimate the number of TDEs per year within a finite distance so long as we know  the mass dependent TDE rate of a given BH ($\Gamma(M_{\rm{BH}})$) as well as the black hole mass function (BHMF, $\widetilde{\phi}(M_{\rm{BH}}))$. The BHMF varies at larger cosmic distance, but here we consider only the local universe ($z<0.1$), where the BHMF is essentially redshift independent.

We make use of the BHMF adopted by \citet{2016MNRAS.455..859S} to estimate the volumetric TDE rate. See their work for an in depth definition. To estimate the BHMF, we first define the number density based on the Schechter function \citep{1976ApJ...203..297S} using the R band luminosity:
\begin{equation}
    \widetilde{\phi}(L_R)dL_R = \widetilde{\phi}_* \left(\dfrac{L_R}{L_*}\right)^{-1.1}\exp (-L_R/L_*)dL_R,
\end{equation}
where $\widetilde{\phi}_*=4.9\times10^{-3}h_7^3 {\rm{\,Mpc^{-3}}}$ and $L_*=2.9\times10^{10}h_7^{-2}L_\odot$. Here $h_7=1$ is the normalized Hubble constant. Combining the above Schechter function with the Faber-Jackson law, $\sigma\approx 150 {\rm{\,km\, s^{-1}}} (L_R/10^{10}L_\odot)^{1/4}$, with the \citet{2013ApJ...764..184M} calibration of the $M_{\rm{BH}}-\sigma$ relation, we arrive at the BHMF (left panel in Figure \ref{fig:BHMF}):
\begin{multline}
  \widetilde{\phi}(M_{\rm{BH}})d\ln M_{\rm{BH}} = 2.56\widetilde{\phi}_* f_{\rm{occ}} \left(\dfrac{M_{\rm{BH}}}{10^8 M_\odot}\right)^{-1.07}\\
  \times \exp \left(-0.647\left(\dfrac{M_{\rm{BH}}}{10^8 M_\odot}\right)\right)d\ln M_{\rm{BH}}.
\end{multline}
Here $f_{\rm{occ}}$ is the occupation fraction, which accounts for the expectation that lower mass galaxies may not host a SMBH at their center. Following \citet{2016MNRAS.455..859S}, we define the occupation fraction as:
\begin{equation}
f_{\rm{occ}} = 
     \begin{cases}
     0.5 \tanh \bigg{[} \ln \left(\dfrac{M_{\rm{bulge}}}{M_c} \right) \times  \nonumber \\
       ~~~~~~ 2.5^{8.9 - \log_{10}(M_c / M_\odot)}\bigg{]} + 0.5, 
       \quad  M_{\rm{bulge}} < 10^{10}\, M_\odot \\
       1. \quad \quad \quad \quad \quad \quad \quad \quad \quad \quad ~~~~~~~~~~~~M_{\rm{bulge}} > 10^{10}\, M_\odot
     \end{cases}
\end{equation}
We relate $M_{\rm{bulge}}$ to the BH mass with the $M_{\rm{BH}}-M_{\rm{bulge}}$ relation from \citet{2013ApJ...764..184M}. As was noted by \citet{2016MNRAS.455..859S}, the value of $M_c = 10^{8.5}\, M_\odot$ is the most likely case, so we fix $M_c = 10^{8.5}\, M_\odot$ in our analysis.

We also account for the impact of different mass stars on the TDE rate by incorporating the Kroupa initial mass functions (IMF):
\begin{equation}
\chi_{\rm{Kro}} = \dfrac{dN_*}{dM_*} \propto 
     \begin{cases}
       (M_*/M_\odot)^{-1.3}, \quad & 0.08 < (M_*/M_\odot) < 0.5  \\
       (M_*/M_\odot)^{-2.3} , \quad  & 0.5 < (M_*/M_\odot) < 1. \\
     \end{cases}
\end{equation}
Note that we have followed \citet{2016MNRAS.455..859S} and only assume stars in the range of $0.08-1\, M_\odot$ are accounted for. This is reasonable particularly if we are considering older galaxies where high mass stars would have already evolved and died. We normalize the IMF such that $\int\chi_{\rm{Kro}} dM_* = 1$.

For the TDE rate, $\Gamma_{\rm{TDE}}(M_{\rm{BH}})$, we use the upper limit estimate from \citet{2016MNRAS.455..859S}:
\begin{equation}
\label{gamma-s16}
  \Gamma_{\rm{S16}} = 2.9\times10^{-5}{\rm{\, yr^{-1}}}\left(\dfrac{M_{\rm{BH}}}{10^8\, M_\odot}\right)^{-0.404}.
\end{equation}
This estimate comes from core/cusp galaxies using a sample of roughly 200 objects. We note that the TDE rate can be increased in galaxies with denser stellar cores \cite[see][for more details]{2020MNRAS.497.2276P}. However, in this work, we restrict ourselves to conservative estimates.

We infer the volumetric TDE rate as a function of BH mass accounting for both the distribution of stars and the fact that some stars will get swallowed and do not contribute to the TDE rate. Therefore the volumetric TDE rate as a function of BH mass would be: 
\begin{equation}
    \dot{n}(M_{\rm{BH}}) = \int_{M_{*, {\rm{min}}}}^{M_{*, {\rm{max}}}} \Gamma(M_{\rm{BH}},M_*) \widetilde{\phi}(M_{\rm{BH}})\chi_{\rm{Kro}}dM_*.
\end{equation}
where $\Gamma(M_{\rm{BH}},M_*)$ is the TDE rate which is given by Equation (\ref{gamma-s16}). We account for cases where the star is swallowed by the BH by setting $\Gamma(M_{\rm{BH}}, M_*)=0$ if $R_t/r_g < r_H$ for the star at the given BH mass. We show the volumetric TDE rate in the right panel in Figure \ref{fig:BHMF} for $M_c=10^{8.5}M_\odot$ and $\Gamma_{\rm{S16}}$. The text above the curve shows the integral $\dot{n}=\int \dot{n}(M_{\rm{BH}}) d \ln M_{\rm{BH}}$ which is the total volumetric TDE rate. Lower cutoff masses for the occupation fraction can significantly increase the TDE rate, but the greatest enhancement comes from assuming that denser stellar cores are present.

To estimate the number of TDEs occurring within a closed volume in a finite amount of time, we use the total volumetric TDE rate, which is constant in time, to obtain:
\begin{equation}
    N_{\rm{TDE}} = \dfrac{4}{3}\pi D^3 \dot{n} \Delta t, 
\end{equation}
where $D$ is the distance in Mpc and $\Delta t$ is the total time in years. Figure \ref{fig:rates} shows the TDE rates given a number of years (1 year and 10 years of observation).

The vertical lines in Figure \ref{fig:rates} show the effect on the number of observable TDEs for a maximal observing distance. Even a modest maximum observing distance of $D<20$ Mpc would provide 2-3 TDEs per decade while extending this range to $D<40$ Mpc would lead to nearly 2 TDEs per year. Assuming these flares are captured by other survey telescopes such as the Large Synoptic Survey Telescope (LSST, \citealt{2019ApJ...873..111I}), which is expected to lead to many more TDE detections in the coming years \citep{2020ApJ...890...73B}, radio/submillimeter follow-up of several TDEs may be possible. We note that we have assumed full sky coverage in our analysis, which is unrealistic, but our calculations demonstrate how the TDE rate varies with distance in order to quantify how beneficial the detector sensitivity may be. In the following sections, we for the first time attempt to quantify the observability of the radio/submillimeter emission from outflows launched by a super-Eddington accretion disks with mass accretion rates similar to the peak accretion rate in TDEs. 

\section{Numerical Methods} \label{sec:nummethods}

The simulations presented in this work were performed using the general relativistic radiation magnetohydrodynamical (GRRMHD) code \textsc{KORAL}
\citep{2013MNRAS.429.3533S,2014MNRAS.439..503S,2014MNRAS.441.3177M,2017MNRAS.466..705S} which solves the conservation equations in a fixed, arbitrary spacetime using finite-difference method. We solve the following conservation equations:
\begin{align}
  (\rho u^\mu)_{;\mu} &= 0, \label{eq:eq5} \\
  (T^\mu_\nu)_{;\mu} &= G_\nu, \label{eq:eq6} \\
  (R^\mu_\nu)_{;\mu} &= -G_\nu, \label{eq:eq7} 
\end{align}
where $\rho$ is the gas density in the comoving fluid frame, $u^\mu$ are the components of the gas four-velocity as measured in the ``lab frame'', $T^\mu_\nu$ is the MHD stress-energy tensor in the ``lab frame'':
\begin{equation} \label{eq:eq9}
  T^\mu_\nu = (\rho + u_g+ p_g + b^2)u^\mu u_\nu + (p_g + \dfrac{1}{2}b^2)\delta^\mu_\nu - b^\mu b_\nu,
\end{equation}
$R^\mu_\nu$ is the stress-energy tensor of radiation, and $G_\nu$ is the radiative four-force which describes the interaction between gas and radiation \citep{2014MNRAS.439..503S}. Here $u_g$ and $p_g=(\gamma_g - 1)u_g$ are the internal energy and pressure of the gas in the comoving frame (linked by adiabatic index $\gamma_g$) and $b^\mu$ is the magnetic field four-vector which is evolved following the ideal MHD induction equation \citep{2003ApJ...589..444G}. In the \textsc{KORAL} simulations, we assume a single temperature plasma where the ion temperature ($T_i$) and the electron temperature ($T_e$) are identical. This description is adequate in the optically thick regions where collisions are common, but may not be accurate in the extended jet where the gas density is substantially lower \citep{2022arXiv220103526L}. We also assume that the electrons follow a thermal distribution, which is another caveat in the initial \textsc{KORAL} simulations since internal shocks in the jet are expected to accelerate electrons into a non-thermal distribution. Magnetic reconnection events in the disk/wind may also produce regions where a non-thermal electron population exists, but we do not account for the possibility of these effects.

The radiative stress-energy tensor is obtained from the evolved radiative primitives, i.e. the radiative rest-frame energy density and its four velocity following the M1 closure scheme modified by the addition of radiative viscosity \citep{2013MNRAS.429.3533S,2015MNRAS.447...49S}.

The opposite signs of $G_\nu$ in the conservation equations for gas and radiation stress-energy (Eqs. \ref{eq:eq6} and \ref{eq:eq7}) reflect the fact that the gas-radiation interaction is conservative, i.e. energy and momentum are transferred between gas and radiation, \cite[see][for more details]{2017MNRAS.466..705S}. We include the effects of absorption and emission via the electron scattering opacity ($\kappa_{\rm{es}}$), free-free absorption opacity ($\kappa_{\rm{a}}$), and bound-free absorption opacity through the Sutherland Dopita model \citep{1993ApJS...88..253S} and assume a solar metal abundance for the gas. We also include the effects of thermal synchrotron and Comptonization \citep{2015MNRAS.454.2372S,2017MNRAS.466..705S}. 

The simulations presented in this work are conducted in 2D $r-\theta$ coordinates and we implement the mean-field dynamo model described in \citet{2015MNRAS.447...49S} to sustain the magnetic field throughout the simulation.

\subsection{Basic Dynamics} \label{sec:dynamics}

A star which has been captured by a SMBH will be disrupted when it can no longer be held together by its self-gravity. This occurs at radii less than the tidal radius: 
\begin{equation} \label{eq:eq1}
  R_t/r_g = 47 \left(\dfrac{M_{\rm{BH}}}{10^6 M_\odot}\right)^{-2/3} \left(\dfrac{M_*}{M_\odot}\right)^{-1/3}\left(\dfrac{R_*}{R_\odot}\right).
\end{equation}
It is common to describe the disruption in terms of the impact parameter, $\beta$, which is defined as the ratio between the tidal radius and pericenter separation such that $\beta \equiv R_t/R_p$. 

The pericenter separation at which a full disruption of the star is sensitive to the stellar composition since the compactness of the star effects how easily it is disrupted. For Zero Age Main Sequence (ZAMS) stars, those described by a $\gamma=5/3$ polytrope are fully disrupted if $\beta \gtrsim 0.9$ while stars described by a $\gamma=4/3$ polytrope must come within $\beta \gtrsim 2$ \citep{2013ApJ...767...25G,2017A&A...600A.124M}. \citet{2019ApJ...882L..26G} demonstrated that the pericenter separation required for evolved stars is even smaller (sometimes greater than $\beta=3$ based on their findings) as the core is no longer hydrogen dominated due to its compactness. For our purposes, we assume a ZAMS star with a $\gamma=5/3$ polytrope was disrupted for simplicity.

If hydrodynamical forces are neglected, then the change in the specific binding energy of the fluid in the star as a result of the tidal interaction can greatly exceed the internal binding energy of the star \citep{1988Natur.333..523R}. As a result, a spread in binding energy is imparted on the stellar material. \citet{2013MNRAS.435.1809S} find that the spread in orbital energy $\Delta\epsilon$ is insensitive to $\beta$ since the energy is essentially frozen in at the tidal radius. This spread is then given by:
\begin{equation} \label{eq:eq4}
  \Delta\epsilon \approx 4.3\times10^{-4} \left(\dfrac{M_{\rm{BH}}}{10^6 \, M_\odot}\right)^{1/3}\left(\dfrac{M_*}{M_\odot}\right)^{2/3}\left(\dfrac{R_*}{R_\odot}\right)^{-1}c^2.
\end{equation} 
The orbital binding energy of the most/least bound material is given by $\epsilon_{\rm{mb}} = \epsilon_* - \Delta\epsilon/2$ and $\epsilon_{\rm{lb}} = \epsilon_* + \Delta\epsilon/2$. Here $\epsilon_*$ is the initial orbital binding energy of the star. For parabolic orbits, which have $\epsilon_*=0$, the spread in binding energy leads to half of the mass remaining bound and the other half being ejected. In this work, we assume the star was disrupted on a parabolic orbit since the majority of TDEs will be of such stars.

The spread in binding energy is one of the most crucial parameters that defines the TDE evolution. In particular, the fallback time (Equation \ref{eq:tfb}) and the peak mass fallback rate:
\begin{equation} \label{eq:mdotfbpeak}
  \dfrac{\dot{M}_{\rm{fb, peak}}}{\dot{M}_{\rm{Edd}}} \approx  133 \left(\dfrac{M_{\rm{BH}}}{10^6 M_\odot}\right)^{-3/2} \left(\dfrac{M_*}{M_\odot}\right)^{2}\left(\dfrac{R_*}{R_\odot}\right)^{-3/2},
\end{equation}
are direct consequences of the spread in binding energy.

\subsection{Definitions}
In this section, we discuss the units adopted throughout the text and provide brief descriptions of quantities used to study the \textsc{KORAL} simulation data. 

Throughout this work, we use gravitational units to describe physical parameters. For distance we use the gravitational radius $r_g\equiv GM_{\rm{BH}}/c^2$ and for time we use the gravitational time $t_g\equiv GM_{\rm{BH}}/c^3$, where $M_{\rm{BH}}$ is the mass of the BH. Often, we set $G=c=1$, so the above relations would be equivalent to $r_g=t_g=M_{\rm BH}$. \footnote{For a BH mass of $10^6\,M_\odot$, the gravitational radius and time in CGS units are $r_g = 1.48\times 10^{11}$ cm and $t_g = 4.94$ s, respectively.} Occasionally, we restore $G$ and $c$ when we feel it helps to keep track of physical units.

We adopt the following definition for the Eddington mass accretion rate:
\begin{equation} \label{eq:mdotEdd}
  \dot{M}_{\rm{Edd}} = \dfrac{L_{\rm{Edd}}}{\eta_{\rm NT} c^2},
\end{equation}
where $L_{\rm{Edd}} = 1.25\times 10^{38}\, (M_{\rm{BH}}/M_\odot)\, {\rm erg\,s^{-1}}$ is the Eddington luminosity, $\eta_{\rm{NT}}$ is the radiative efficiency of a thin disk around a BH with spin parameter $a_*$ (which is often referred to as the Novikov-Thorne efficiency):
\begin{equation} \label{eq:etaNT}
  \eta_{\rm{NT}} = 1 - \sqrt{1 - \dfrac{2}{3 r_{\rm{ISCO}}}},
\end{equation}
and $r_{\rm{ISCO}}=3+Z_2 - \sqrt{(3-Z_1)(3+Z_1+2Z_2)}$ is the radius of the Innermost Stable Circular Orbit (ISCO, \citealt{1973blho.conf..343N}) in the Kerr metric, where $Z_1 = 1 + (1-a_*^2)^{1/3}\left((1+a_*)^{1/3}+(1-a_*)^{1/3}\right)$ and $Z_2 = \sqrt{3a_*^2 + Z_1^2}$. For $a_* =$ 0, 0.5, and 0.9, the efficiency is $\eta_{\rm{NT}}=$ 5.72\%, 8.21\%, and 15.58\%.

We compute the net mass inflow rate as:
\begin{equation} \label{eq:mdotin}
  \dot{M}(r) = -\int_0^\pi \int_0^{2\pi} \sqrt{-g}\rho \,u^r d\phi d\theta.
\end{equation}
We treat the accretion of mass onto the BH as this integral taken at the horizon $r_H$.

We estimate the electron scattering photosphere location for an observer at infinity along the direction $(\theta,\phi)$ by integrating the optical depth radially inward from the outer boundary of the grid. Far from the BH, the curvature of spacetime is negligible, so we simply integrate at constant $(\theta,\phi)$ in the ``lab frame'':
\begin{equation} \label{eq:taues}
  \tau_{\rm{es}}(r) = \int_r^{R_{\rm{max}}} \dfrac{\rho \kappa_{\rm{es}}}{c} \left(u^t - u^r\right)\sqrt{g_{rr}}dr',
\end{equation}
where $\kappa_{\rm{es}} = 0.2(1+X)\kappa_{\rm{KN}}\,{\rm cm^2}$ is the electron scattering opacity, $X$ is the Hydrogen mass-fraction which is assumed to be the Solar abundance $X_\odot=0.7381$, $\kappa_{\rm{KN}}$  is the Klein-Nishina correction factor for thermal electrons \citep{2017MNRAS.466..705S}, and $R_{\rm{max}}$ is the radius corresponding to the outer boundary of the grid. For the gas and radiation temperatures in the simulations presented here, the Klein-Nishina correction is negligible and the electron scattering opacity is essentially the Thomson opacity. In this work, we choose the location of the photosphere as the $\tau_{\rm{es}}=1$ surface.

We define the accretion flow as three distinct regions (disk, wind, and jet) based on the total energy via the Bernoulli parameter. We also make use of the electron scattering opacity, $\tau_{\rm{es}}$, to determine whether the region of the fluid under consideration is optically thin or thick. In optically thick regions, i.e. ($\tau_{\rm{es}} \geq 1$), the radiation is advected with the flow and can contribute to acceleration of the gas so we treat it as contributing to the Bernoulli parameter. Meanwhile, in optically thin regions, i.e. ($\tau_{\rm{es}} < 1$), we assume that only the MHD components are relevant to the total gas energy. We modify the Bernoulli definition used in \citet{2019MNRAS.483..565C} to include optically thin regions where the interaction between gas and radiation can be neglected via:
\begin{equation} \label{eq:Bernoulli}
    \rm{Be} = 
    \begin{cases}
      -\dfrac{(T^t_{\ \, t} + R^t_{\ \, t} + \rho u^t)}{\rho u^t}, \ \, \rm{(optically \, thick)}\\
      -\dfrac{(T^t_{\ \, t} + \rho u^t)}{\rho u^t}, \ \, \rm{(optically \, thin)}\\
    \end{cases}
\end{equation}
The `disk' is made up of bound gas with $\rm{Be} < 0$. This constitutes both the inner accretion disk and the large mass reservoir of the initial torus. Both the `wind' and `jet' are unbound and are generally radially out flowing. The wind is defined as any fluid with $0 < \rm{Be}\leq 0.05$. The jet is any fluid with $\rm{Be}> 0.05$. This choice of cutoff for wind vs. jet is based on the velocity at infinity ($v_\infty$), with the wind having $v_\infty \lesssim 0.3 c$ and the jet having $v_\infty \gtrsim 0.3 c$. It is worth noting that this choice assumes that gas at small radii will not lose/gain energy as it travels outward, which is not guaranteed to be the case outside of steady state regions. A fraction of the positive Bernoulli gas in the simulation domain could remain bound to the BH such as in \citet{2018ApJ...863..158C}; however, the Bernoulli is still a useful approximate definition to characterize the outflow based on energy. We make use of the Bernoulli parameter to find the opening angle of the jet. I.e. we define $\theta_{\rm{jet}}(r)$ such that:
\begin{equation}
    {\rm{Be}}(r,\theta_{\rm{jet}}(r)) = 0.05.
\end{equation}

The total luminosity (the net energy flux) is computed as:
\begin{equation} \label{eq:ltot}
  L_{\rm{net}}(r) = -\int_{0}^{\pi}\int_0^{2\pi}\sqrt{-g} (T^r_{\ \, t} + R^r_{\ \, t} + \rho u^r) d\phi d\theta,
\end{equation}
where we integrate the radial flux of energy carried by gas plus magnetic field ($T^r_{\ \,t}$) and radiation ($R^r_{\ \,t}$), and subtract out the rest-mass energy ($\rho u^r$, since it does not lead to observational consequences for an observer at infinity). When computed at the BH horizon, Equation \ref{eq:ltot} gives the total energy extracted from the accretion flow so we define $L_{\rm{tot}}=L_{\rm{net}}(r_H)$. We also compute the total energy outflowing in the wind and jet as: 
\begin{equation} \label{eq:lMHD}
  L_{\rm{MHD}}(r) = -\int_{0}^{\pi}\int_0^{2\pi}\sqrt{-g} (T^r_{\ \, t} + \rho u^r) d\phi d\theta.
\end{equation}
We compute energy outflow in the wind at $r=1000\, r_g$ by using Equation \ref{eq:lMHD}, but we only sum energy outflows where the gas is unbound, outflowing, and non-relativistic (or $0 < \rm{Be} < 0.05$ and $u^r > 0$). A similar choice is made to compute the total energy outflowing in the jet at $r=1000\,r_g$ but we use the criterion that energy is only summed where $\rm{Be} \geq 0.05$ and $u^r > 0$. Note that at $1000\, r_g$, the wind and jet are in an optically thin region so the Bernoulli is computed without the radiation energy density included.

The radiative bolometric luminosity is given by:
\begin{equation} \label{eq:eq16}
  L_{\rm{bol}}(r) = -\int_{0}^{\pi}\int_0^{2\pi}\sqrt{-g} R^r_{\ \, t} d\phi d\theta,
\end{equation}
which gives the flux of radiation energy through a surface at a given radius. In this work, we measure the flux through a sphere at $r=5000\,r_g$, which lies beyond the outer radius of the initial torus, and define $L_{5000}=L_{\rm{bol}}(5000\,r_g)$. We assume rays crossing the surface reach a distant observer. 

We define the total, wind, jet, and radiation efficiencies as:
\begin{equation}
    \eta_{\rm{tot}} = \dfrac{1}{\dot{M}} L_{\rm{net}}(r=r_H),
\end{equation}
\begin{equation}
    \eta_{\rm{wind}} = \dfrac{1}{\dot{M}} L_{\rm{MHD}}(r=1000\,r_g, 0.05 > {\rm{Be}} > 0, u^r > 0),
\end{equation}
\begin{equation}
    \eta_{\rm{jet}} = \dfrac{1}{\dot{M}} L_{\rm{MHD}}(r=1000\,r_g, {\rm{Be}}\geq0.05, u^r > 0),
\end{equation}
and
\begin{equation}
    \eta_{\rm{rad}} = \dfrac{1}{\dot{M}} L_{\rm{bol}}(r=5000\,r_g),
\end{equation}
In each case, positive values correspond to energy being extracted from the system. 

The magnetic flux is computed as:
\begin{equation} \label{eq:eq20}
  \Phi(r) = \dfrac{1}{2} \int_0^{\pi}\int_{0}^{2\pi}|B^r(r)|dA_{\theta \phi},
\end{equation}
where $B^r$ is the radial component of the magnetic field. We quantify the magnetic field strength at the BH horizon through the normalized magnetic flux parameter \citep{2011MNRAS.418L..79T}:
\begin{equation} \label{eq:eq20}
  \phi_{\rm{BH}} = \dfrac{\Phi(r_H)}{\sqrt{\dot{M}(r_H)}}.
\end{equation} 
For geometrically thick disks the MAD state is achieved once $\phi_{\rm{BH}}\sim 40-50$ \citep[see e.g.][]{2011MNRAS.418L..79T,2012JPhCS.372a2040T}.

To characterize the magnetic field, we define the magnetic pressure ratio: 
\begin{equation}
  \beta_t \equiv \dfrac{(p_{\rm{gas}} + p_{\rm{rad}})}{p_{\rm{mag}}},
\end{equation}
(distinguished from the impact parameter $\beta$) which is used to define the pressure ratio in optically thick regions where the radiation is dynamically important. Note that we also make use of the ratio between the gas and magnetic pressure:
\begin{equation} \label{eq:betag}
  \beta_g \equiv \dfrac{p_{\rm{gas}}}{p_{\rm{mag}}},
\end{equation}
which is used to characterize the optically thin jet.

We quantify the resolution of the fastest growing mode of the magnetorotational instability (MRI, \citealt{1991ApJ...376..214B}) by computing the quantity:
\begin{equation}
  Q_\theta = \dfrac{2\pi}{\Omega dx^\theta}\dfrac{|b^\theta|}{\sqrt{4\pi\rho}}, \label{eq:eq10} 
\end{equation}
where $dx^\theta$ (the grid cell in polar coordinate $\theta$) and $b^\theta$ (the $\theta$ component of the magnetic field) are both evaluated in the orthonormal frame, $\Omega$ is the angular velocity, and $\rho$ is the gas density. Numerical studies of the MRI have shown that values of $Q_\theta$ in excess of at least 10 are needed to resolve the fastest growing mode \citep{2011ApJ...738...84H}. As we are considering 2D simulations in $r-\theta$, we do not consider the $\phi$ MRI quality factor. Throughout each simulation, we find the quality factor $Q_\theta > 10$ at the mid plane for $r<50 \, r_g$.

\begin{table*}
  \centering
  \begin{tabular}{l c c c c c c}
    \hline
    \hline
     & \texttt{m5a0.0-HR} & \texttt{m5a0.5-HR} & \texttt{m5a0.9-HR} & \texttt{m10a0.0-HR} & \texttt{m10a0.5-HR} & \texttt{m10a0.9-HR} \\
    \hline
    $M_{\rm{BH}} \, (M_\odot)$ & $5\times10^6$ & $5\times10^{6}$ & $5\times10^{6}$ & $10^7$ & $10^7$ & $10^7$\\
    $\langle\dot{M}\rangle (\dot{M}_{\rm{Edd}})$ & $\sim 11$ & $\sim 11$ & $\sim 11$ & $\sim 12$ & $\sim 16$ & $\sim25$ \\ 
    $a_*$  & $0$ & $0.5$ & $0.9$ & $0$ & $0.5$ & $0.9$ \\
    $\eta_{\rm{tot}}$  & 4.00\% &  7.19\% & 14.17\% & 4.29\% & 6.42\% & 13.23\% \\
    $\eta_{\rm{wind}}$  & 0.23\% &  0.52\% & 0.88\% & 0.10\% & 0.22\% & 0.72\% \\
    $\eta_{\rm{jet}}$  & 0.38\% &  0.96\% & 2.75\% & 0.24\% & 0.63\% & 1.15\% \\
    $\eta_{\rm{rad}}$  & 3.33\% &  4.64\% & 12.34\% & 3.76\% & 4.39\% & 5.48\% \\
    $R_0 (r_g)$ & $20$ & $20$ & $20$ & $20$ & $20$ & $20$ \\
    Duration ($t_g$) & $83,000$ & $83,000$ & $83,000$ & $83,000$ & $83,000$ & $81,200$ \\
    $N_r \times N_\theta$ & $640\times 256$ & $640\times 256$ & $640\times 256$ & $640\times 256$ & $640\times 256$ & $640\times 256$ \\
    $R_{\rm{in}} \,(r_g) /R_{\rm{out}} \,(r_g)$ & $10/3000$ & $10/3000$ & $10/3000$ & $10/3000$ & $10/3000$ & $10/3000$  \\
    $R_{\rm{min}} \,(r_g) /R_{\rm{max}} \,(r_g)$ & $1.1/10^5$ & $1.1/10^5$ & $1.1/10^5$ & $1.1/10^5$ & $1.1/10^5$ & $1.1/10^5$  \\
    \hline
  \end{tabular}
  \caption{Simulation parameters and properties of the three simulations presented in this work. We specify the BH mass ($M_{\rm{BH}}$), average accretion rate ($\langle\dot{M}\rangle$), spin of the BH ($a_*$), total efficiency computed at $r_H$ ($\eta_{\rm{tot}}$), wind and jet efficiency computed at $1000\,r_g$ ($\eta_{\rm{wind}}$ and $\eta_{\rm{jet}}$), radiative efficiency computed at $5000\, r_g$ ($\eta_{\rm{rad}}$), simulation duration in $t_g$, grid resolution, inner and outer edges of the initial torus, and the inner and outer radial boundaries of the simulation box. Note that the accretion rate ($\langle\dot{M}\rangle$) and efficiencies ($\eta_{\rm{tot}}, \, \eta_{\rm{wind}}, \, \eta_{\rm{jet}}$ and $\eta_{\rm{rad}}$) are computed using time averages over the final $50,000\, t_g$ of each simulation.}
  \label{tab:tab1}
\end{table*}

\begin{figure}
    \centering{}
	\includegraphics[width=\columnwidth]{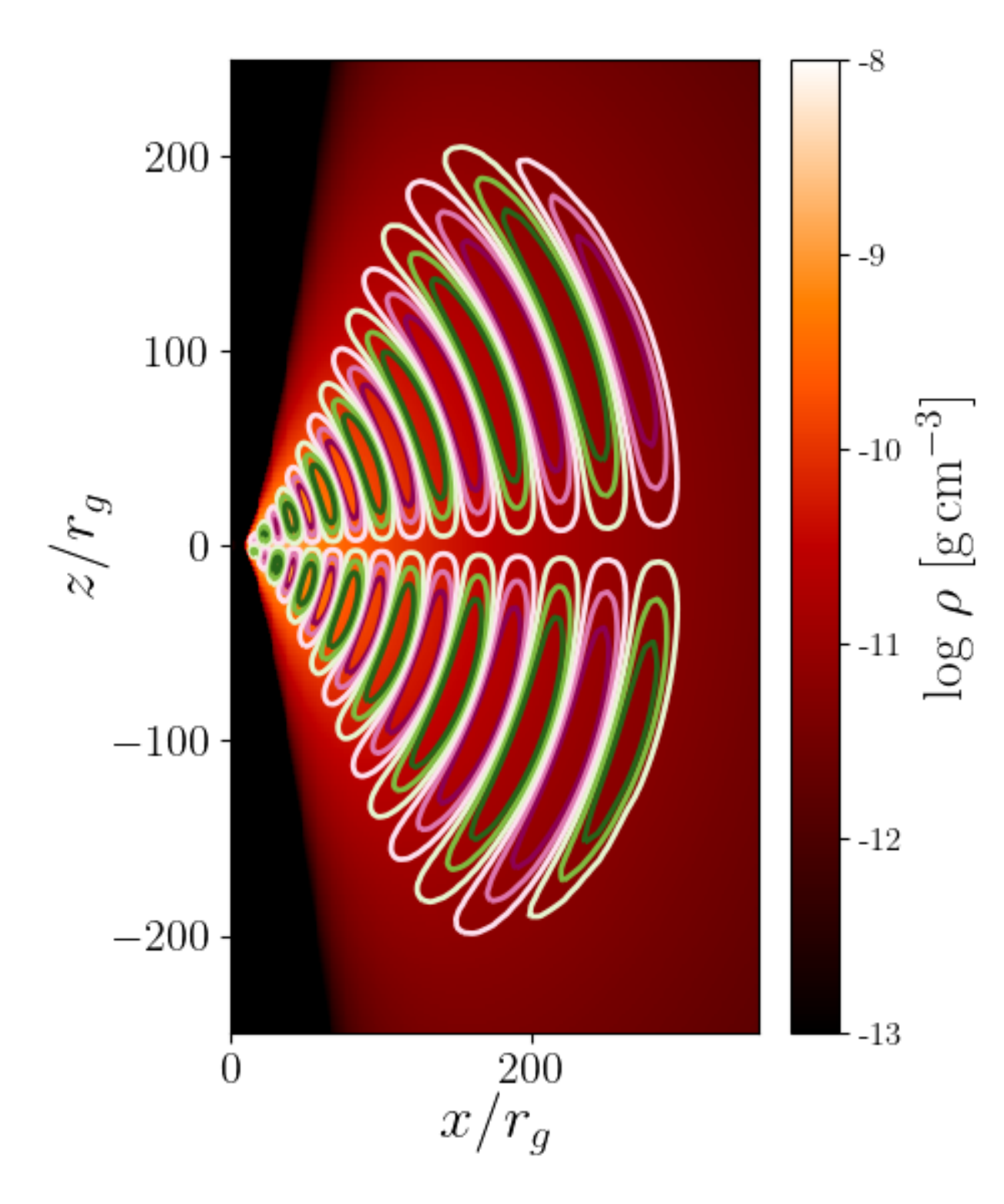}
    \caption{Here we show the initial torus configuration for model \texttt{m5a0.0-HR} in terms of the gas density (colors) and vector potential $A_\phi$ (contours). The green/purple lines denote positive/negative vector potential.}
    \label{fig:inittorus}
\end{figure}

\begin{figure*}
    \centering{}
	\includegraphics[width=\textwidth]{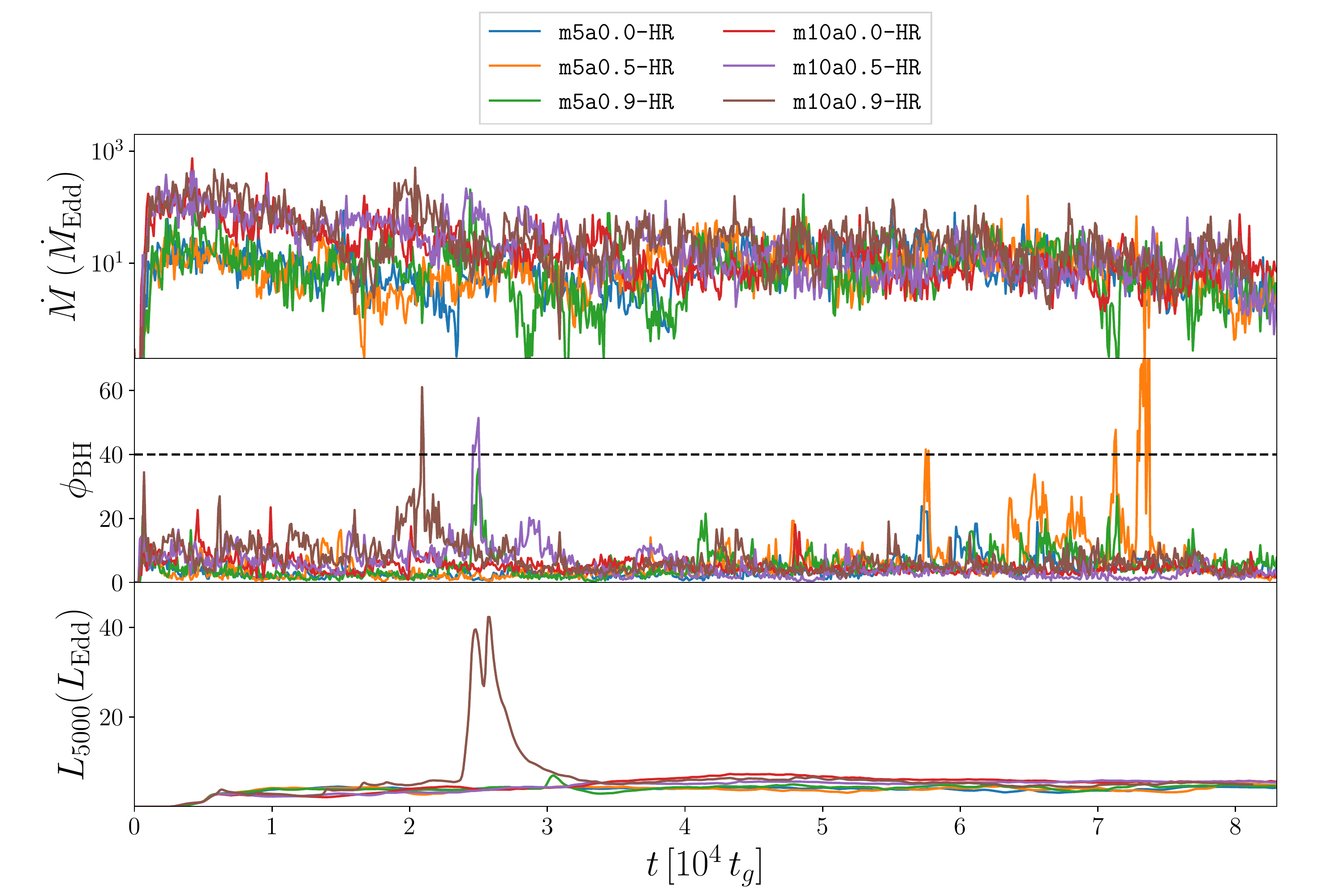}
    \caption{Here we show the mass accretion rate (top), normalized magnetic flux at the BH horizon (middle), and bolometric luminosity (bottom) for each model. Also indicated is $\phi_{\rm{BH}}=40$ (horizontal dashed line), which is approximately the magnetic flux at which a geometrically thick disk will enter the MAD state.}
    \label{fig:supScal}
\end{figure*}

\subsection{Numerical Models/Initial Conditions} \label{sec:initialconditions}

The domain outside of the gas torus is initialized with a low density, hot gas with a density maximum at the BH horizon of:
\begin{equation}
    \rho_{\rm{atm,max}}=4.5\times10^{-8}\left(\dfrac{M_{\rm{BH}}}{M_\odot}\right)^{-1}\ [{\rm{g \, cm^{-3}}}].
\end{equation}
The gas temperature at the BH horizon is set to $T_{\rm{atm,max}}=10^{11}$ K. We assume that the atmosphere follows a profile similar to a spherical accretion flow \citep{1952MNRAS.112..195B} and set the gas density in the atmosphere using a radial profile $\rho_{\rm{atm}}\propto r^{-3/2}$ while the internal energy varies as $u_{g,\rm{,atm}}\propto r^{-5/2}$. The radiation in the atmosphere is initialized with a radiation temperature of $5\times10^4(M_{\rm{BH}}/M_\odot)$ K everywhere, which ultimately defines an atmosphere with negligible radiation energy initially.

To initialize the torus, we assume that the stream rapidly circularized and formed an accretion flow. It is important to note that this model assumes that the circularization process happens on a timescale that is much shorter than the fallback time. We also assume that angular momentum and binding energy are conserved, which allows for simple initial conditions of the initial gas torus to be defined.

Using these inputs, we follow the method described in \citet{2004ApJ...605..307K} to initialize the simulation domain with an equilibrium torus which has its angular momentum vector aligned with the BH spin vector. We set the radius of the density maximum of the torus ($R_0$, see Table \ref{tab:tab1}) to occur at $R_{\rm{circ}}=2R_p$ and the binding energy is initialized using a constant angular momentum torus with $l=\sqrt{R_{\rm{circ}}}$ and $\epsilon = \epsilon_{\rm{mb}}$. This results in a tenuously bound, geometrically thick torus where the inner edge of the torus ($R_{\rm{in}}$) is smaller than $R_0$ and the outer edge ($R_{\rm{out}}$) is at 1000s of $r_g$ (see values in Table \ref{tab:tab1}). In each model, our choice of density maximum results in a late time accretion rate of $\gtrsim11\, \dot{M}_{\rm{Edd}}$.

The magnetic field is initialized in terms of the vector potential ($A_\phi$) using multiple field loops with varying polarity across the mid-plane and in radius (see Figure \ref{fig:inittorus}). This guarantees that the accretion disk remains at a low magnetic flux throughout. Without this initialization, the chance advection of a large poloidal field loop towards the BH horizon can drive the disk towards a MAD state, which we cannot accurately simulate in 2D. The magnetic field is normalized such that the maximum pressure ratio $\beta_{t,\rm{max}}\equiv (p_{\rm{gas,max}}+p_{\rm{rad,max}})/p_{\rm{mag,max}} = 33$. This choice of magnetic pressure is somewhat arbitrary, but is sufficient to resolve MRI in the initial torus. As the simulations in this work are performed in spherical $r-\theta$ coordinates, we employ a mean field dynamo which mimics the dynamo processes that naturally take place in full 3D simulations and prevent the magnetic flux from weakening throughout the simulation.

We use a grid which has more cells both near the poles to resolve the outflow and in the midplane to resolve the disk. The radial grid cells are logarithmically spaced. We set the resolution $N_r \times N_\theta$ such that the cells are roughly 1:1 in most of the simulation domain. We use modified Kerr-Schild coordinates with the inner edge of the domain inside the BH horizon. At the inner radial boundary ($R_{\rm{min}}$), we use an outflow condition while at the outer boundary ($R_{\rm{max}}$), we use a similar boundary condition and in addition prevent the inflow of gas and radiation. Note that our choice of $R_{\rm{min}}$ is such that at least 6 cells in the computational domain lie inside of the horizon. At the polar boundaries, we use a reflective boundary. We employ a periodic boundary condition in azimuth. To maintain numerical stability, we introduce mass in highly magnetized regions of the simulation domain using a floor condition on the magnetization $\sigma \equiv b^2/\rho \leq 100$ throughout each simulation.

We identify the simulations by the BH mass in units of $m_6=M_{\rm{BH}}/10^6 \, M_\odot$ and the dimensionless BH spin $a_*$. For instance, \texttt{m5a0.5-HR} corresponds to a BH mass of $5\times10^6\, M_\odot$ with $a_*=0.5$. See Table \ref{tab:tab1} for descriptions of each model.

\section{Results} \label{sec:results}

\begin{figure}
    \centering{}
	\includegraphics[width=\columnwidth]{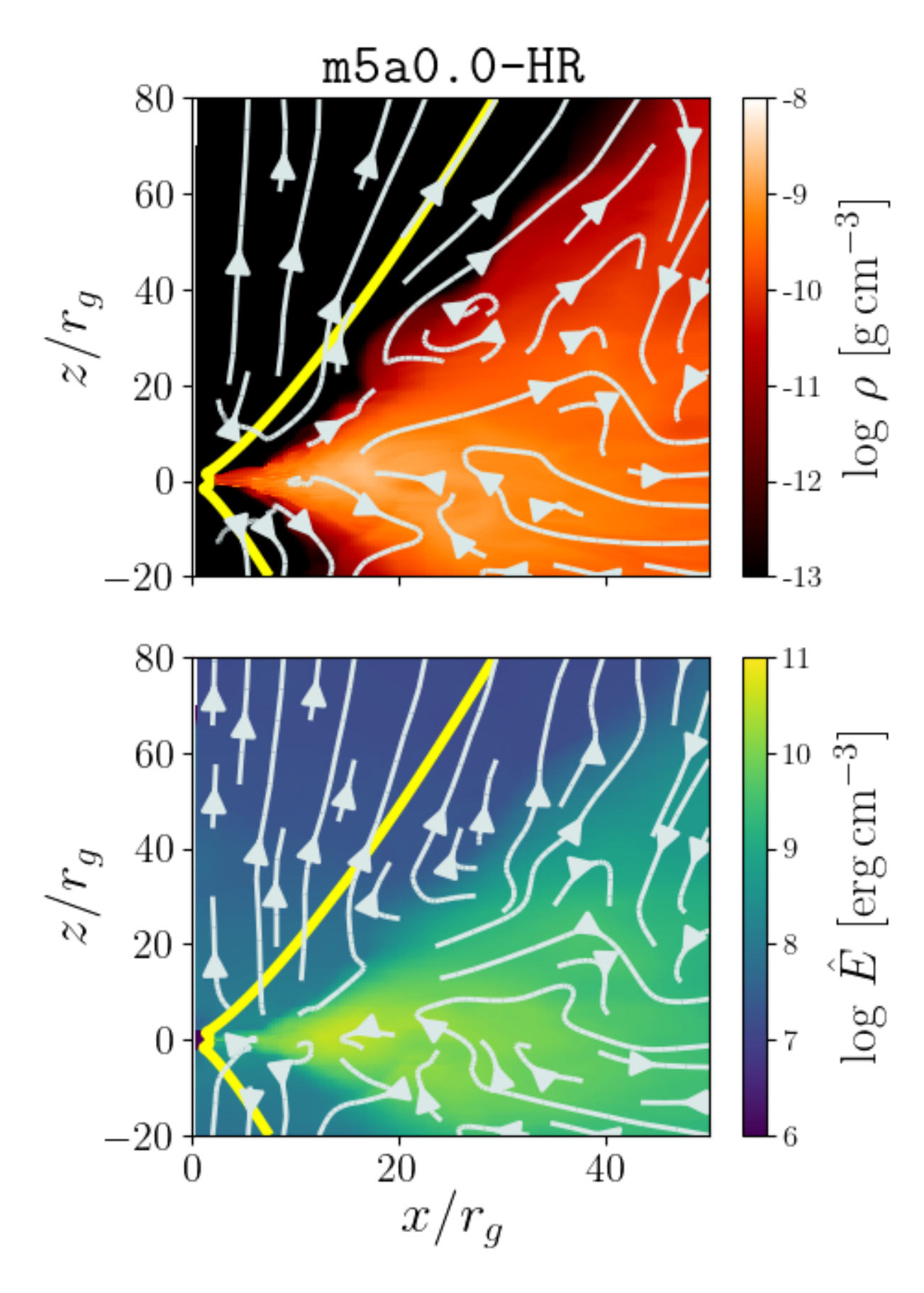}
    \caption{A zoomed in view of the accretion flow and funnel of \texttt{m5a0.0-HR} at $t=83,000\,t_g$. The colors show the gas density (top) and radiation energy density (bottom). The streamlines indicate the fluid velocity (top) and radiative flux (bottom). We also indicate the photosphere ($\tau_{\rm{es}}=1$, yellow line).}
    \label{fig:6s09hi_zoom}
\end{figure}

\begin{figure}
    \centering{}
	\includegraphics[width=\columnwidth]{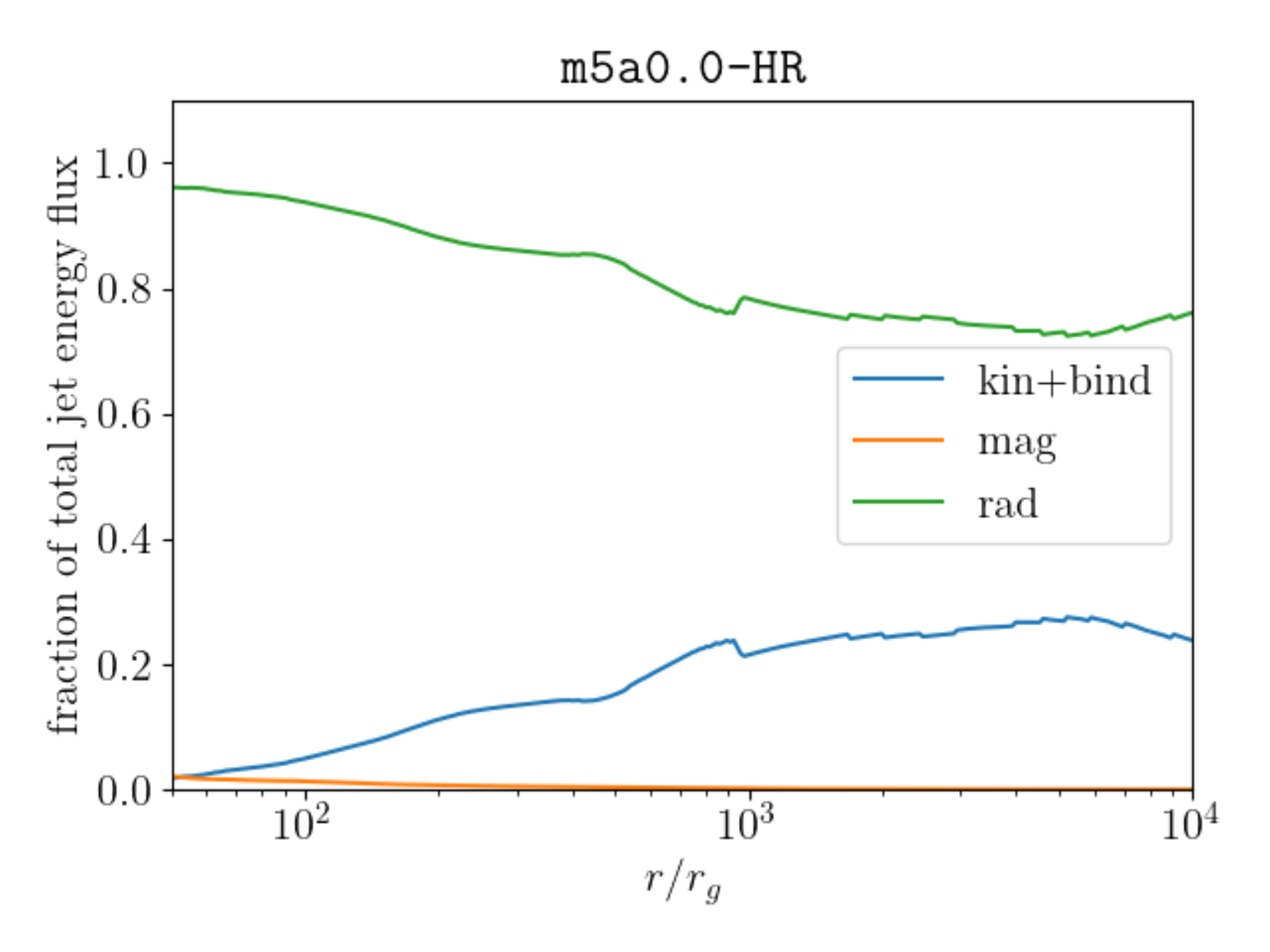}\\ 
	\includegraphics[width=\columnwidth]{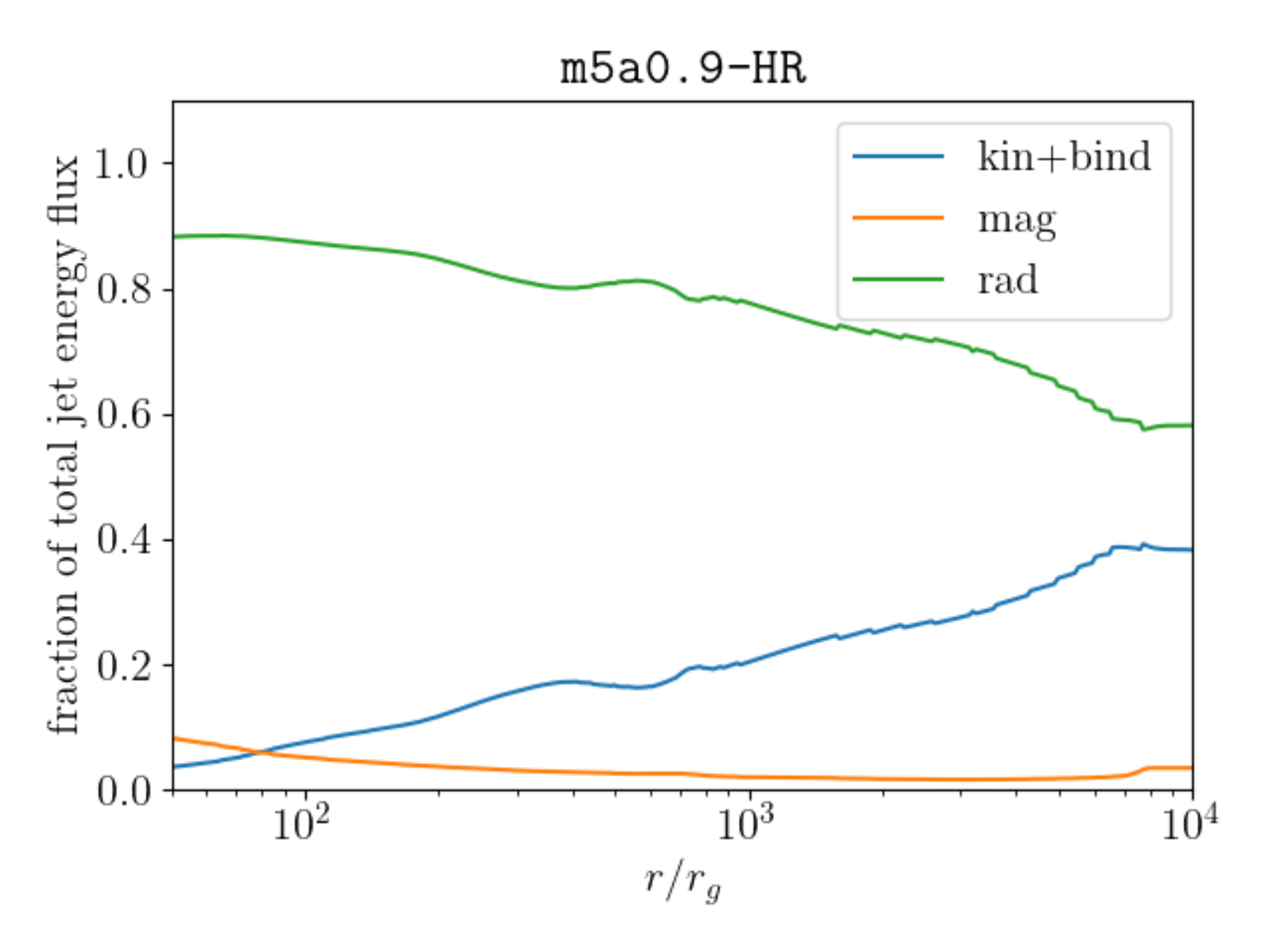}    
	\caption{Here we present the fraction of energy flux in the top jet for model \texttt{m5a0.0-HR} (top) and \texttt{m5a0.9-HR} (bottom) in the form of radiation energy flux ($R^r_{\ \, t}$), kinetic plus gravitational binding energy flux ($\rho u^r u_t + \rho u^r$), and magnetic energy flux ($b^2u^r u_t - b^r b_t$). The data has been time averaged over $t=33,000-83,000\, t_g$. For radii $r\lesssim 5000 r_g$, we observe a continual conversion of radiation energy to kinetic and binding energy. Beyond $5000\, r_g$, the fraction of energy flux in radiation steadily increases. This is likely due to a combination of dissipation of kinetic energy as well as the fact that the radiation is no longer forced to flow through the funnel for $r\gtrsim 3000\, r_g$ and can escape laterally. The magnetic energy flux in the jet also appears to increase with spin, which contributes to accelerating lower density gas along the jet axis.}
    \label{fig:radaccel}
\end{figure}

\begin{figure*}
    \centering{}
	\includegraphics[width=\textwidth]{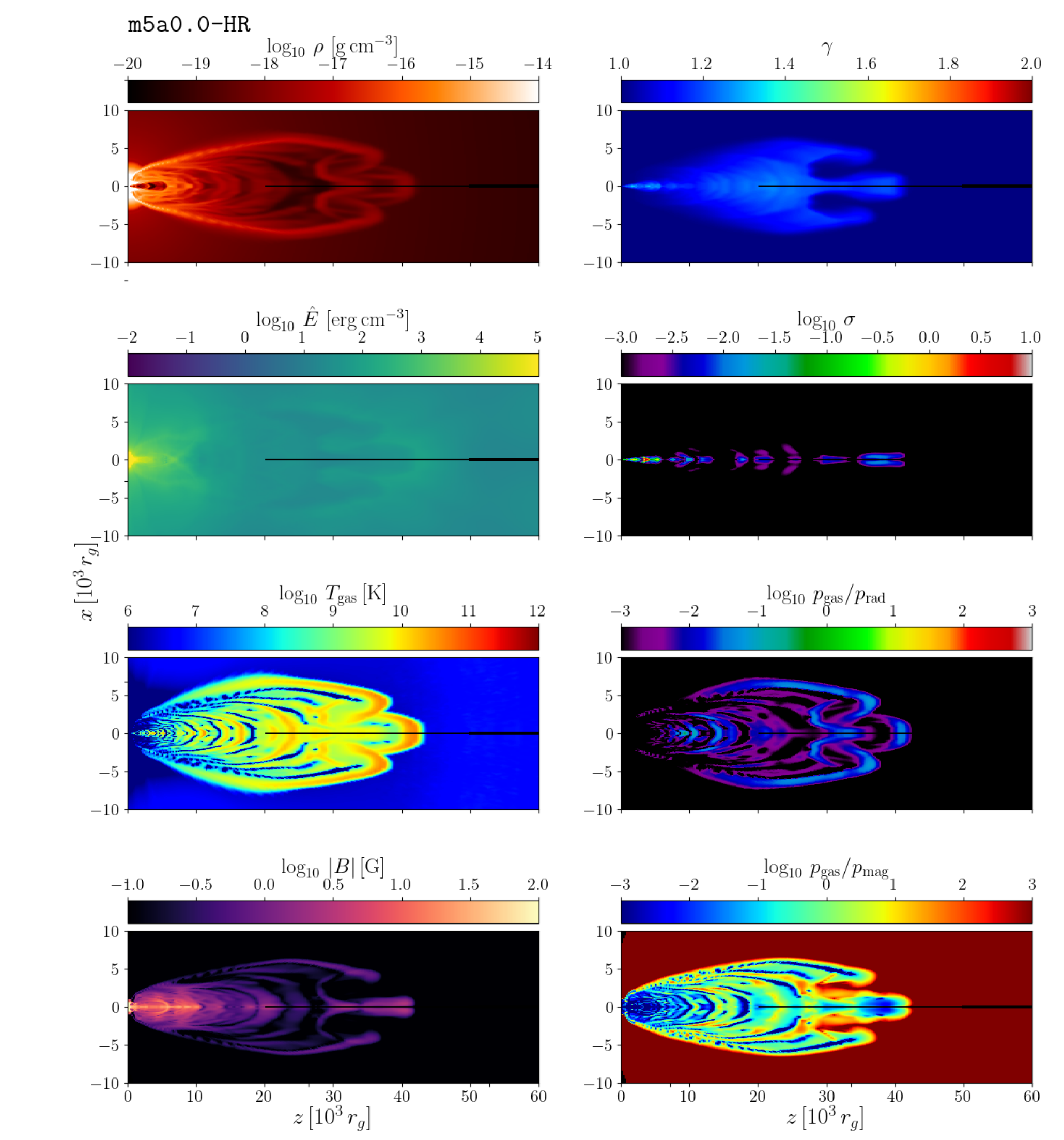}
    \caption{Here we show a zoomed out view of \texttt{m5a0.0-HR} at $t=83,000\,t_g$. The colors show the gas density (top left), radiation energy density (2nd panel, left), gas temperature (3rd panel, left), magnetic field strength (bottom left), gas lorentz factor $\gamma$ (top right), gas magnetization $\sigma$ (2nd panel right), gas pressure to magnetic pressure ratio $p_{\rm{gas}}/p_{\rm{mag}}$ (3rd panel, right), and gas to radiation pressure ratio $p_{\rm{gas}}/p_{\rm{rad}}$ (bottom right).}
    \vspace*{1cm} 
    \label{fig:m5a00HR}
\end{figure*}

\begin{figure*}
    \centering{}
	\includegraphics[width=\textwidth]{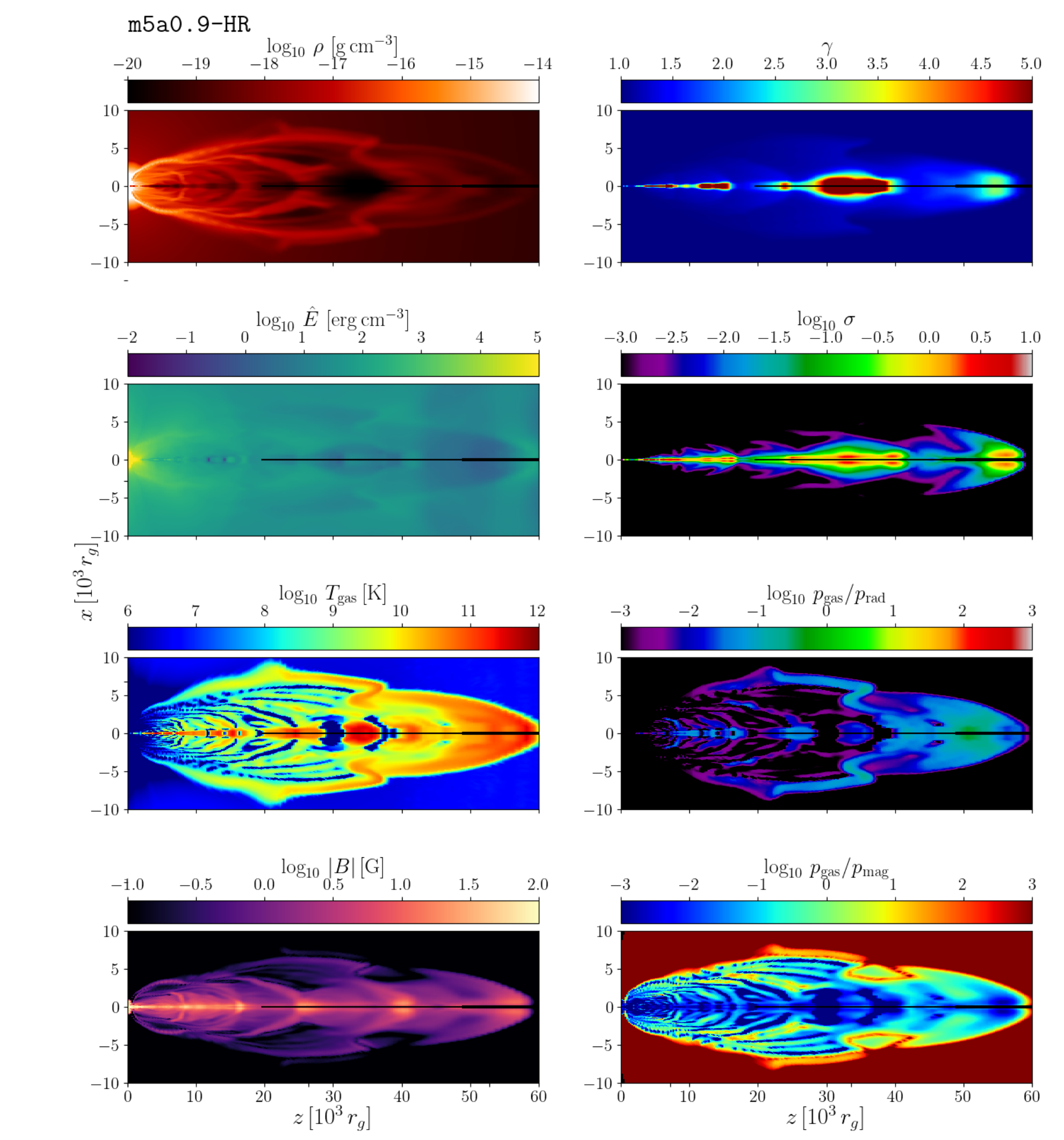}
    \caption{The same as Figure \ref{fig:m5a00HR}, but for model \texttt{m5a0.9-HR} at $t=83,000\,t_g$. Note that we have increased the range of $\lambda$ to highlight the maximum jet velocity. The simple addition of a spinning BH has a noticeable effect in many characteristics of the jet. Namely, the jet propagates farther owing to a high velocity component near the poles with $\gamma > 2$. In addition, the jet is hotter along the jet axis and jet head. Lastly, the magnetic field strength in the jet is larger at larger radii ($r>30,000\, r_g$), which leads to more of the gas becoming strongly magnetized with $\sigma > 1$.}
    \vspace*{1cm} 
    \label{fig:m5a09HR}
\end{figure*}

\begin{figure}
    \centering{}
	\includegraphics[width=\columnwidth]{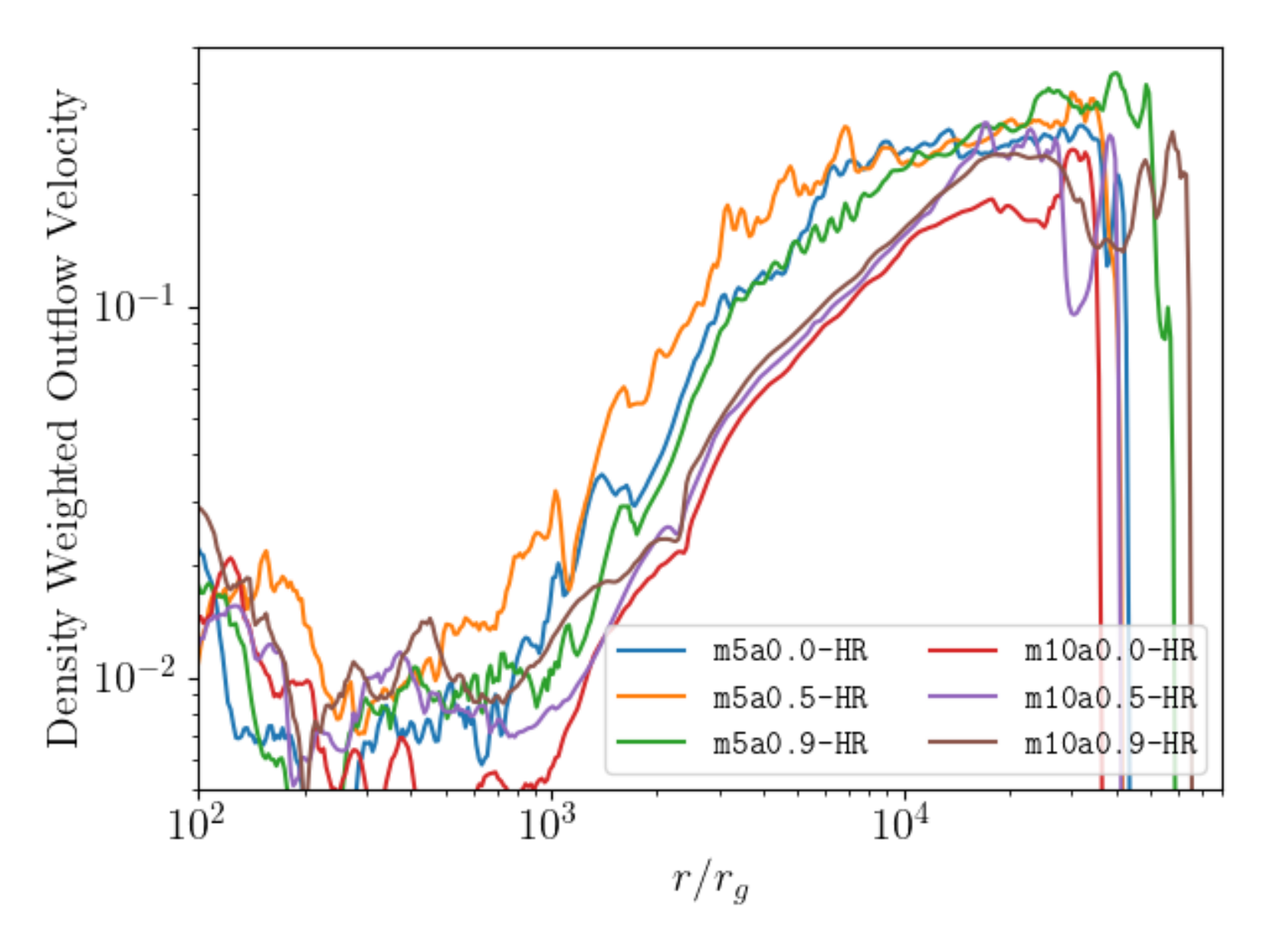}    
	\caption{Here we present radial profiles of the density weighted outflow velocity in the jet for each model at the final snapshot. The behaviour is similar in each model. The density weighted outflow velocity shows acceleration between $10^3 \lesssim r/r_g \lesssim 10^4$. Beyond $10^4\, r_g$, the velocity seems to have reached a plateau of $\sim 0.2-0.35c$.}
    \label{fig:jetvel}
\end{figure}

\subsection{Accretion Flow Properties} \label{sec:properties}

We show the accretion rate, magnetic flux at the BH horizon, and the luminosity of outgoing radiation computed directly from the \textsc{KORAL} data in Figure \ref{fig:supScal}. For each model in Figure \ref{fig:supScal}, the accretion rate does not remain at the initial accretion rate (e.g. $\sim100$ for \texttt{m10a0.0-HR}) and instead goes through phases of high and low accretion, eventually settling into a nearly steady state accretion phase for $t\gtrsim 25,000\, t_g$. The magnetic flux in each simulation is generally $\phi_{\rm{BH}} \lesssim 5$, which is well below the limit for the MAD state ($\phi_{\rm{BH}} \gtrsim 40$). There are, occasional periods where a loop of poloidal magnetic field is advected towards the BH horizon and momentarily drives $\phi_{\rm{BH}}$ near the MAD limit. As indicated by the accretion rate, this does not disrupt the accretion flow (which typically occurs for strongly magnetized accretion flows at the MAD limit in 2D), but the BZ process will extract some spin energy from the BH and momentarily increase the jet power in the case of a spinning BH. These events are extremely short lived and rare, so these deviations are not the driving factor in the energetics and jet evolution. The escaping radiation is slightly super-Eddington at $L_{5000} \approx 3L_{\rm{Edd}}$, though \texttt{m10a0.9-HR} briefly brightened to $\approx 40\,L_{\rm{Edd}}$ between $t\approx25,000-30,000\,t_g$.

Since each of the simulations has similar disk features, we focus on \texttt{m5a0.0-HR} to describe the inner accretion flow properties. Visualizations of the simulation data for \texttt{m5a0.0-HR} are shown in Figures \ref{fig:6s09hi_zoom}. The accretion disk is evidently thick and turbulent, with the turbulence being driven by MRI. The gas accretes onto the black hole primarily along the equatorial plane. The flow is significantly turbulent even at relatively large radii. The turbulent structure of the velocity streamlines is in part the result of material near the black hole gaining energy and being launched back into the disk. For such a low binding energy disk, small perturbations can lead to the material becoming unbound quite easily. Outflows driven by radiation and Poynting flux are evident within $\sim45^\circ$ from the pole. For an in depth description of the accretion flow of similar SANE models, see \citet{2019MNRAS.483..565C}. 

To characterize the acceleration of gas in the jet, we compute the contributions to the energy flux in the $\rm{Be} > 0.05$ region for the top jet. Namely, we compute the kinetic plus gravitational binding energy ($\dot{E}_{\rm{kin+bind}}$), internal ($\dot{E}_{\rm{int}}$), magnetic ($\dot{E}_{\rm{mag}}$), and radiative ($\dot{E}_{\rm{rad}}$) energy fluxes. The energy fluxes are computed as:
\begin{equation}
    \dot{E}_{\rm{kin+bind}}(r) = -\int_0^{\theta_{\rm{jet}}(r)} \int_0^{2\pi} \sqrt{-g} \, (\rho u^r u_t + \rho u^r) \, d\phi d\theta,
\end{equation}
\begin{equation}
    \dot{E}_{\rm{mag}}(r) = -\int_0^{\theta_{\rm{jet}}(r)} \int_0^{2\pi} \sqrt{-g} \, (b^2 u^r u_t - b^r b_t) \, d\phi d\theta,
\end{equation}
and
\begin{equation}
    \dot{E}_{\rm{rad}}(r) = -\int_0^{\theta_{\rm{jet}}(r)} \int_0^{2\pi} \sqrt{-g} \, R^r_{\, \ t} \, d\phi d\theta,
\end{equation}
respectively. We ignore the flux of gas internal energy since the gas is radiation pressure dominated. We illustrate the acceleration of gas by radiation in model \texttt{m5a0.0-HR} in Figure \ref{fig:radaccel}. For radii $50 \lesssim r/r_g \lesssim 5000$, we see a steady conversion of radiative flux to kinetic plus binding energy flux. However, the fraction of energy flux in the form of radiation in the jet begins to increase for $r\gtrsim 5000\, r_g$. We attribute this to two factors. First, there is some conversion of kinetic energy into thermal and radiation energy due to internal shocks within the jet (which we illustrate in Section \ref{sec:dissipation}). In addition, there is also the fact that the radiation is no longer laterally confined by the funnel walls for $r\gtrsim 3000\, r_g$. Once the radiation can escape laterally, the radiation force supplied to gas along the jet axis declines. The fact that the fraction of radiation energy flux in the jet begins to climb for $r\gtrsim 5000\, r_g$ suggests that the optically thick gas is weakly accelerated by radiation at larger radii.
 
We present a snapshot of the jet resulting from the accretion flow in Figure \ref{fig:m5a00HR}. Much of the radiation energy density is contained within the disk near the equatorial plane. Radiation is advected in with the accretion flow and escapes out through the funnel, driving a mildly relativistic outflow ($\gamma \lesssim 1.2$). As described in \citet{2015MNRAS.453.3213S}, this process occurs because optically thick gas flows from the disk into the funnel region and is subsequently accelerated by the radiation streaming through the funnel. The non-spinning BH model \texttt{m5a0.0-HR} reaches similar gas velocities in the jet as the models presented in \citet{2015MNRAS.453.3213S}, which were also of $a_*=0$ BHs. The disk is radiation pressure dominated, but contributions from radiation energy to the total pressure in the jet will be negligible due to the low optical depth. However, the magnetic pressure can become quite large in the jet and sometimes exceeds the gas pressure within the jet. Some regions reach above the magnetization $\sigma$ of unity. The map of the magnetization $\sigma$ shows that it is primarily near the pole where the gas becomes strongly magnetized. 

Despite maintaining $\phi_{\rm{BH}}\lesssim 5$ in each model, the spin $a_* > 0$ models each show increased jet power and relativistic ($\gamma>2$) but low density outflows near the poles. We illustrate this for \texttt{m5a0.9-HR} in Figure \ref{fig:m5a09HR}. We also note that the magnetic field strength is greater along the extended jet and, as a consequence, the jet is more magnetized in general as indicated by the extent of the $\sigma  > 1$ region. The fractional energy flux profiles in model \texttt{m5a0.9-HR} (Figure \ref{fig:radaccel}, bottom panel) illustrate that the simulations show an increase in Poynting flux in the jet as the spin increases. The high velocity component with $\gamma>2$ is likely due to Poynting acceleration, which is more efficient in more magnetized jets \citep{2019MNRAS.490.2200C}. Also of note is the fact that the funnel is optically thin down to $r\approx r_H$ near the poles (Figure \ref{fig:6s09hi_zoom}), which suggests radiative acceleration will be weak near the poles and cannot accelerate the $\gamma >2$ component. In addition, the trend of increasing jet velocity as spin increases was demonstrated in GRMHD SANE models by \citet{2013MNRAS.436.3741P}.

The accretion flow results in a total efficiency $\eta_{\rm{tot}}$ which is slightly less than the NT value in each model (see Table \ref{tab:tab1}). Since the accretion rate is only marginally super-Eddington, radiation can more efficiently escape and the radiative efficiency $\eta_{\rm{rad}}$ is only a few percent smaller than NT as a result. The accretion flow generates a jet and wind with a total energy flux that is a fraction of the accretion power. We find $(\eta_{\rm{wind}} + \eta_{\rm{jet}} ) \approx 0.6-3.6\%$. We also observe a clear trend of increasing energy outflowing in the jet and wind as the spin increases for a given BH mass.

Although the jet power increases and the jet near the polar axis becomes significantly more relativistic as the spin increases, the density weighted outflow velocity is surprisingly similar across BH mass and spin. As we show in Figure \ref{fig:jetvel}, each model only achieves an outflow velocity of $\sim0.2-0.35c$ in a density weighted sense. The acceleration of the gas between $10^3\lesssim r/r_g \lesssim 10^4$ is also clearly illustrated. This is in agreement with the terminal density weighted jet velocity of $\approx0.3c$ reported by \citet{2015MNRAS.453.3213S}, but we have demonstrated that the spin and jet magnetization does not appear to play a role in the velocity of the higher density component of the jet in SANE, super-Eddington accretion disks with $\dot{M}\gtrsim 11\dot{M}_{\rm{Edd}}$.

\begin{figure}
    \centering{}
	\includegraphics[width=\columnwidth]{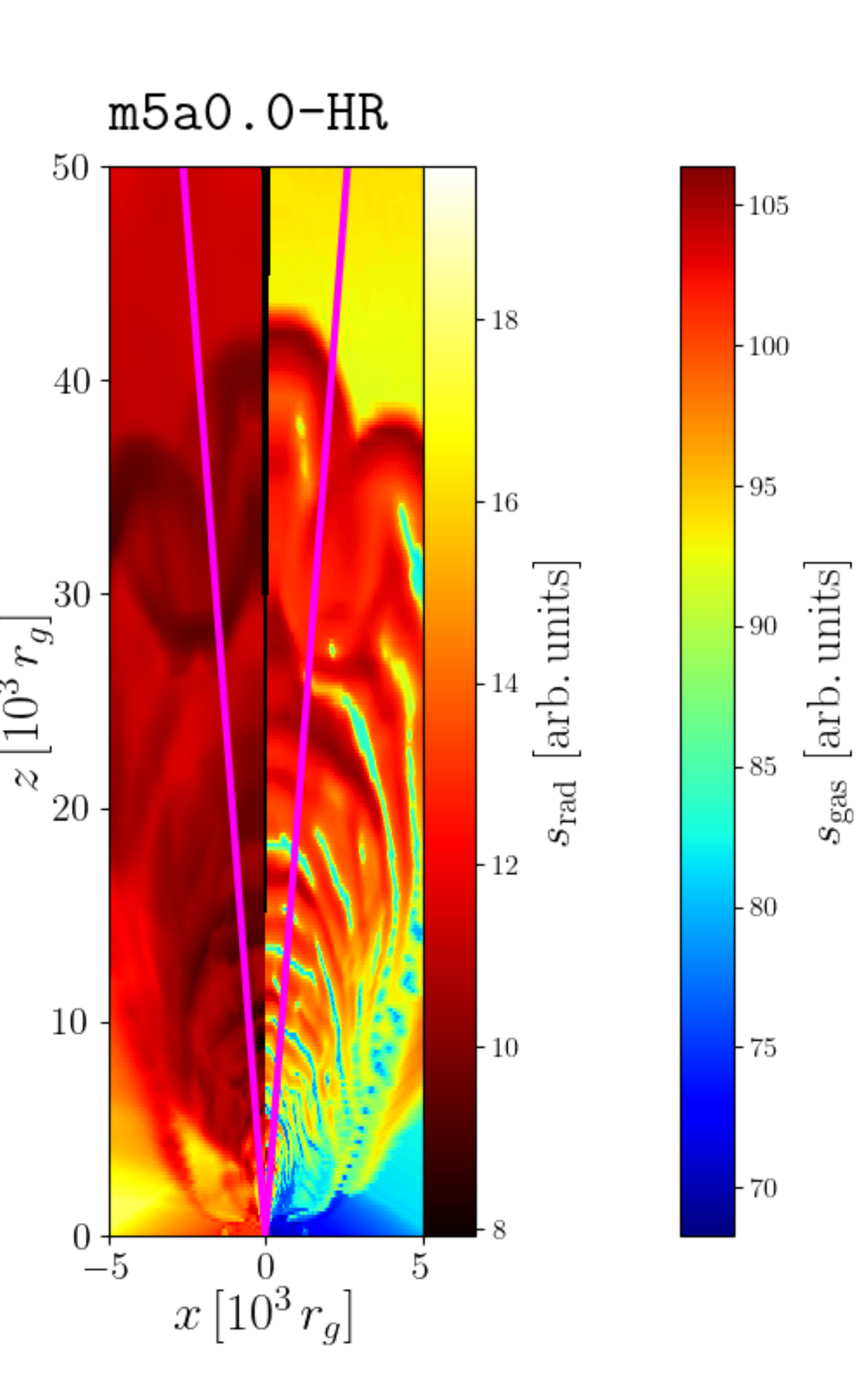}\\
	\includegraphics[width=\columnwidth]{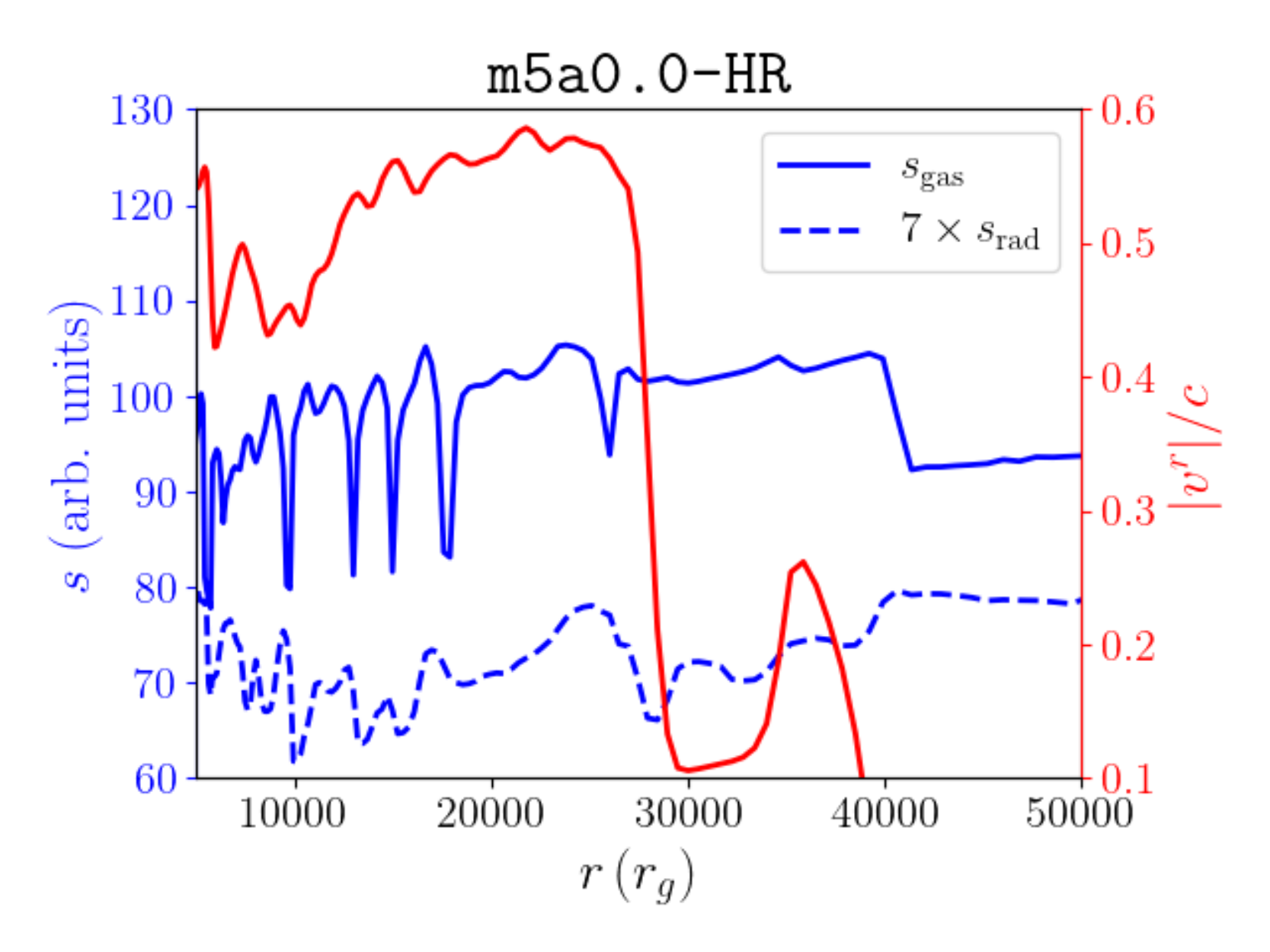}
    \caption{Here we show the gas (right) and radiation (left) entropy in the top panel. The bottom panel shows a trace of the radial velocity and both gas and radiation entropy taken at $\theta=3^\circ$ (indicated as the pink line in the top panel). Dissipation leads to an increasing entropy in both gas and radiation overall. The sharpest jump occurs at the jet head ($z\approx40,000 \, r_g$). There are also several jumps in entropy down the jet axis (beginning at $r\lesssim 30,000\, r_g$ in this snapshot) which are internal shocks driven by fast moving gas shocking with slow moving gas within the jet.}
    \label{fig:entropy}
\end{figure}

\subsection{Dissipation} \label{sec:dissipation}

To search for dissipation in the jet, we examine the gas entropy per unit mass:
\begin{equation}
    s_{\rm{gas}} = \dfrac{1}{\gamma_g - 1}\ln{\left(\dfrac{p_{\rm{gas}}}{\rho^{\gamma_g}}\right)},
\end{equation}
and the log scaled radiation entropy per unit mass:
\begin{equation}
    s_{\rm{rad}} = \log_{10}\left(\dfrac{4aT^3}{3\rho}\right).
\end{equation}

The variable velocity of gas flowing along the jet is expected to lead to shocks, wherein dissipated kinetic energy leads to heating of the gas and thus an increase in internal energy. We consider the radial velocity, gas entropy, and radiation entropy in \texttt{m5a0.5-HR} in Figure \ref{fig:entropy}. The trace of the entropy in both gas and radiation at $\theta\approx3^\circ$ shows quite clearly that (i) a strong shock exists at the jet head, (ii) internal shocks are present within the jet, (iii) the entropy shows similar oscillations to the velocity and kinetic energy flux, and (iv) the entropy generally increases between $z\sim5000-40,000\, r_g$. Taken together, this suggests that the jet head is where dissipation is largest and we are indeed observing dissipation of kinetic energy via internal shocks within the jet at radii smaller than the jet head.

\begin{figure}
    \centering{}
	\includegraphics[width=\columnwidth]{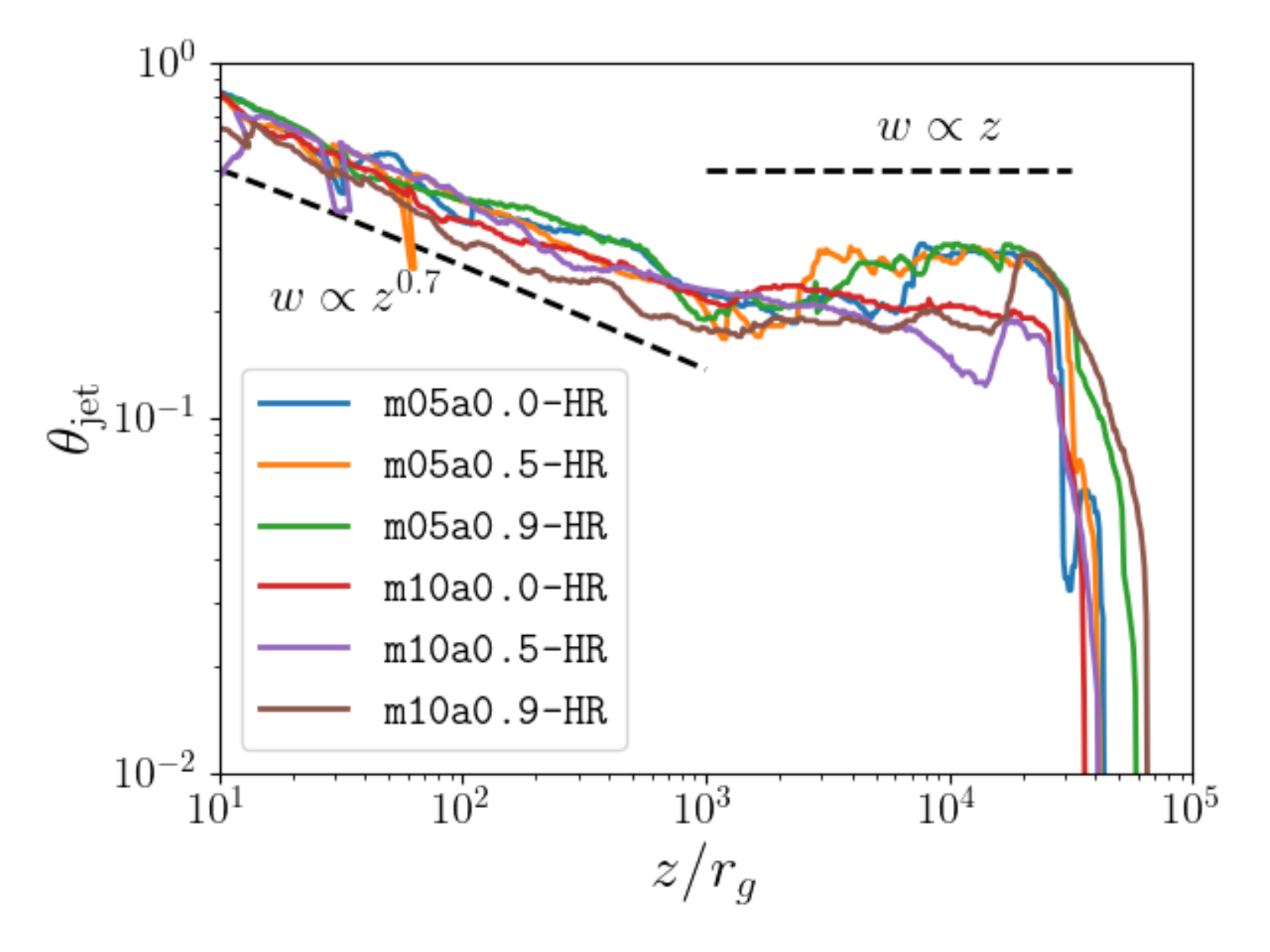}
    \caption{Here we show the half opening angle of the jet (solid lines) for each model. Note that we have symmetrized the data by taking the mean of the half opening angle measurement for the bottom and top jet. We find that the jet expands roughly conically from $z\approx1000\, r_g$ until nearly $z\approx 40,000 \, r_g$ in each model.}
    \label{fig:jetprofiles}
\end{figure}

\subsection{Jet Structure} \label{sec:jetstructure}

We define the jet boundary using the Bernoulli parameter via Equation \ref{eq:Bernoulli} and measure the half opening angle $\theta_{\rm{jet}}$ for both the top and bottom jet as a function of distance along the jet ($z$, see Figure \ref{fig:jetprofiles}). We find nearly identical opening angles for both the top and bottom jet in each model. We find that at distances $r_H/r_g < z/r_g \lesssim 1000$ the jet undergoes very nearly parabolic expansion with a power-law of roughly $\omega \propto z^{0.7}$, where $\omega$ is jet width. The jet appears to maintain a roughly conical structure at jet lengths $1000 \lesssim z/r_g \lesssim 30,000 $ with $\theta_{\rm{jet}}$ maintaining a nearly flat profile. For $z/r_g \gtrsim 40,000$, the jet angle shrinks rapidly at the jet head.

An important caveat that we must point out is that the break between the parabolic and conical region is a byproduct of the choice of initial torus. In each model, the break occurs very close to the radial extent of the initial torus model ($\sim 3000\, r_g$, see Table \ref{tab:tab1}). This point marks the transition from higher density gas, which can provide lateral pressure support, to the substantially lower density atmosphere, which provides negligible pressure support in comparison.

Conically expanding jets were successfully applied to ASASSN-14li in the models of \citet{2018ApJ...856....1P}, who modeled the radio synchrotron emission as a superposition of synchrotron emitting regions in a conically expanding jet. Figure \ref{fig:jetprofiles} demonstrates that SANE super-Eddington disks produce conically expanding jets, at least with our chosen initial torus configuration. The half-opening angles that we find are all larger than the best fit models of \citet{2018ApJ...856....1P}. We note that their model assumed a previously cleared funnel was present while we have assumed a low density atmosphere which does not provide substantial lateral confinement. Future work on TDE jets should explore the effects of the atmosphere and initial torus in greater detail.

\begin{figure}
    \centering{}
	\includegraphics[width=\columnwidth]{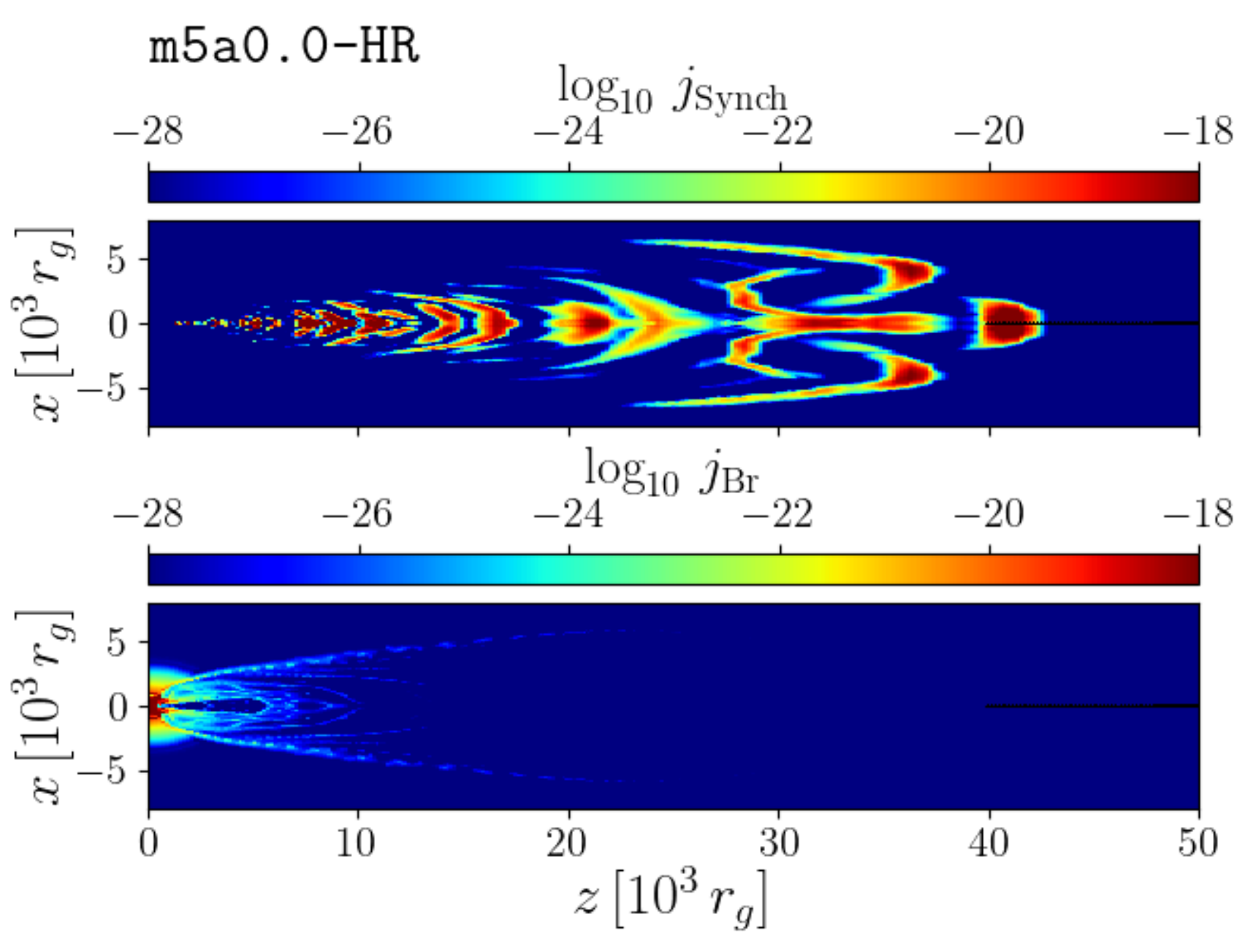}
    \caption{Here we show the synchrotron emissivity (top panel) and the bremsstrahlung emissivity (bottom panel) at 1 GHz for \texttt{m5a0.0-HR}. We note that the spectra as presented in Sections \ref{sec:tdepspec} and \ref{sec:twotempspectra} peak at $>10$ GHz; however, the choice to show the 1 GHz emissivities is to better visualize the spatial difference between bremsstrahlung and synchrotron emission in the jet. The synchrotron emissivity is dominant outside of the optically thick disk. The bremsstrahlung emissivity shown here is representative of the low frequency emissivity as the bremsstrahlung cutoff occurs at $\nu > 10^{16}$ Hz, far beyond the frequencies we consider in this work.}
    \label{fig:emissivity}
\end{figure}

\section{Synchrotron Emission Imaging Analysis} \label{sec:imaging}

In Figure \ref{fig:emissivity}, we compare the emissivity of synchrotron and bremsstrahlung processes in \texttt{m5a0.0-HR}. We compute the emissivities using functions defined in \citet{2017MNRAS.466..705S} since these are the frequency dependent functions currently used by \textsc{KORAL}. Namely, we use the fitting function for the ultrarelativistic synchrotron emissivity in Gaussian-CGS units:
\begin{equation}
  j_{\rm{Synch}}=4.43\times 10^{-30}\nu_M n_e \dfrac{x_M I'(x_M)}{2\theta_e^2} \, \left[\dfrac{{\rm{erg}}}{{\rm{cm^3\, s \, Hz}}}\right],
\end{equation}
where $n_e\equiv \rho/(\mu m_p)$ is the electron number density in a fully ionized gas given a mean molecular weight $\mu$ and proton mass $m_p$, $\theta_e \equiv kT/m_ec^2$ is the dimensionless electron temperature, $x_M = \nu/\nu_M$, $\nu_M = (3/2) eB \theta_e^2/2\pi m_e c$ is the characteristic synchrotron frequency, and $I'(x_M)$ is a fitting function provided by \citet{1996ApJ...465..327M}:
\begin{equation}
    I'(x_M) = \dfrac{4.0505}{x_M^{1/6}}\left(1 + \dfrac{0.40}{x_M^{1/4}} + \dfrac{0.5316}{x_M^{1/2}}\right)\exp(-1.8899x_M^{1/3}).
\end{equation}
The bremsstrahlung emissivity is computed as:
\begin{equation}
    j_{\rm{Br}} = \dfrac{6.8\times 10^{-38}}{4\pi} T^{-1/2} n_e^2 \overline{g} R(T)\exp(-h\nu/kT) \left[\dfrac{{\rm{erg}}}{{\rm{cm^3\, s \, Hz}}}\right],
\end{equation}
where we assume a Gaunt factor $\overline{g}=1.2$ and $R(T)\equiv 1 + 4.4\times10^{-10}(T/ \rm{1\, K})$ is a relativistic correction adopted from \citet{1979rpa..book.....R}. The frequency $\nu_M$ sets the peak of the synchrotron emission in this formulation.

We find that the synchrotron emission dominates the jet while bremsstrahlung dominates in the optically thick disk, thus we expect the jet emission will largely be from synchrotron processes.  The periodic outflows appear as discrete `bubbles' of high synchrotron emission along the jet. Here the gas is hotter and high frequency synchrotron emission ($\nu\gtrsim 10^{11}$) is expected as a result. We note however that these computations were done directly from the \textsc{KORAL} data and do not represent the spatial intensity information as we have not accounted for opacity effects, viewing angle, and resolution, all of which may change which features can be observed. Nevertheless, this analysis demonstrates that the jet emission is not continuous. The previous description is true of each model considered in this work.

We also estimate the Compton-$y$ parameter ($y=\tau_{\rm{es}}k T/m_{\rm{e}}c^2$, \citealt{1979rpa..book.....R}) in the extended jet to test whether Compton effects are important. We find that $y\ll 1$ throughout the bulk of the jet, so Compton effects will also be minimal due to the low opacity. As such, the synchrotron emission alone can provide a reasonable model of the jet emission.

We make use of the GRRT code \texttt{ipole} \citep{2018MNRAS.475...43M} to produce images of the jet that include ray-tracing and radiative transfer effects. Before post-processing the data, we set the gas density to zero, $\rho=0$, in regions where $\sigma > 1$. GRMHD codes inject mass and energy in these regions to keep the simulation stable. As such, the accuracy of the radiation field from these regions is less certain than where $\sigma < 1$. This method, which is the most conservative choice, has been employed in other studies of ray-traced GRMHD simulations \citep{2019MNRAS.486.2873C}. Because \texttt{ipole} does not presently include bremsstrahlung or Compton effects, we also cut out data for $r<5000\, r_g$ prior to ray-tracing.

\begin{figure}
    \centering{}
	\includegraphics[width=\columnwidth]{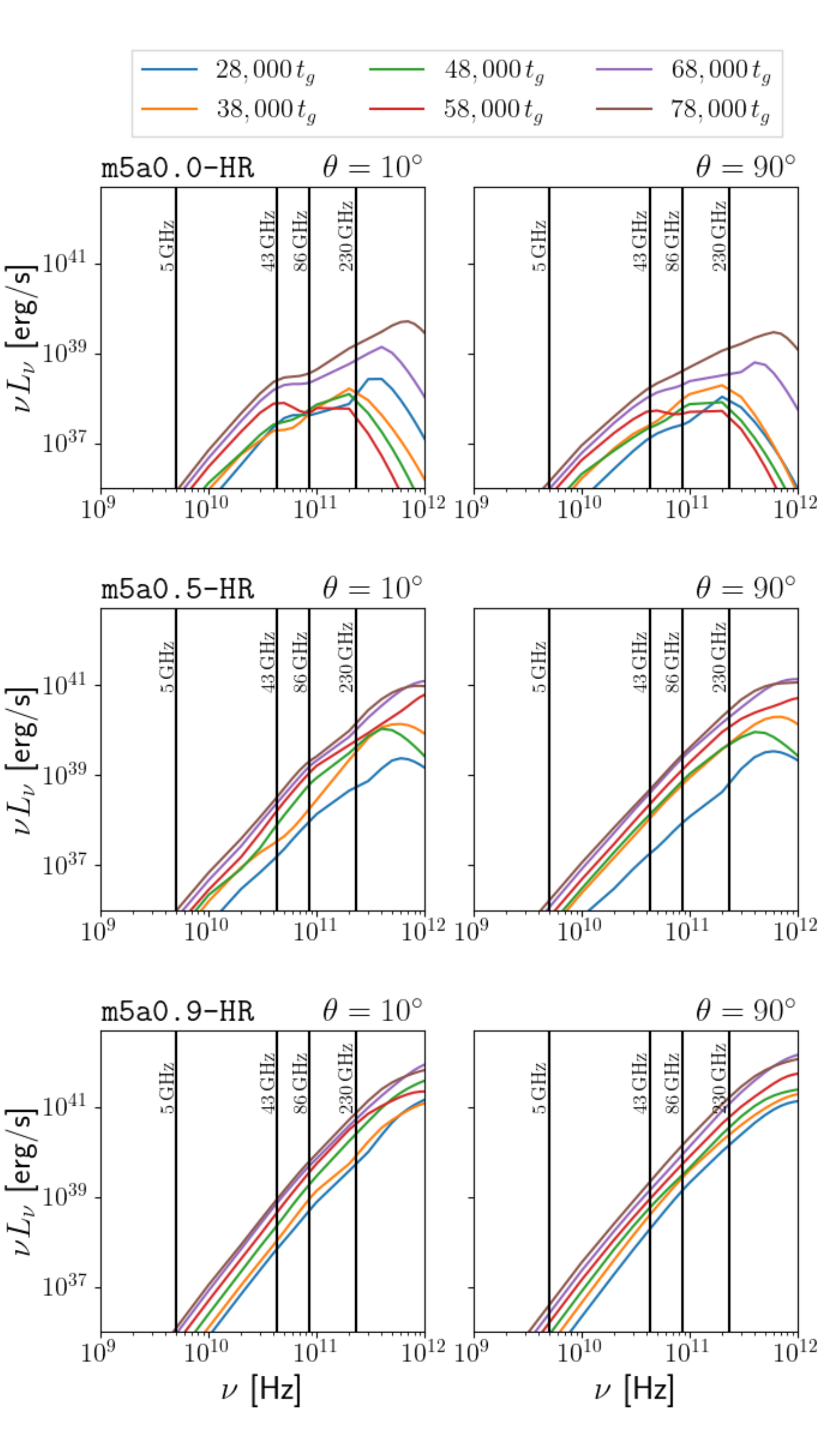}
    \caption{Here we show spectra for \texttt{m5a0.0-HR} (top),  \texttt{m5a0.5-HR} (middle), and \texttt{m5a0.9-HR} (bottom) computed for viewing angles of $10^\circ$ (left) and  $90^\circ$ (right) at snapshots ranging from $t=28,000-78,000\, t_g$. In each case, the jet is generally brightening across all frequencies; however, the jet becomes more energetic and the peak frequency becomes larger as the spin increases.}
    \label{fig:tdep_spec_m5}
\end{figure}

\begin{figure}
    \centering{}
	\includegraphics[width=\columnwidth]{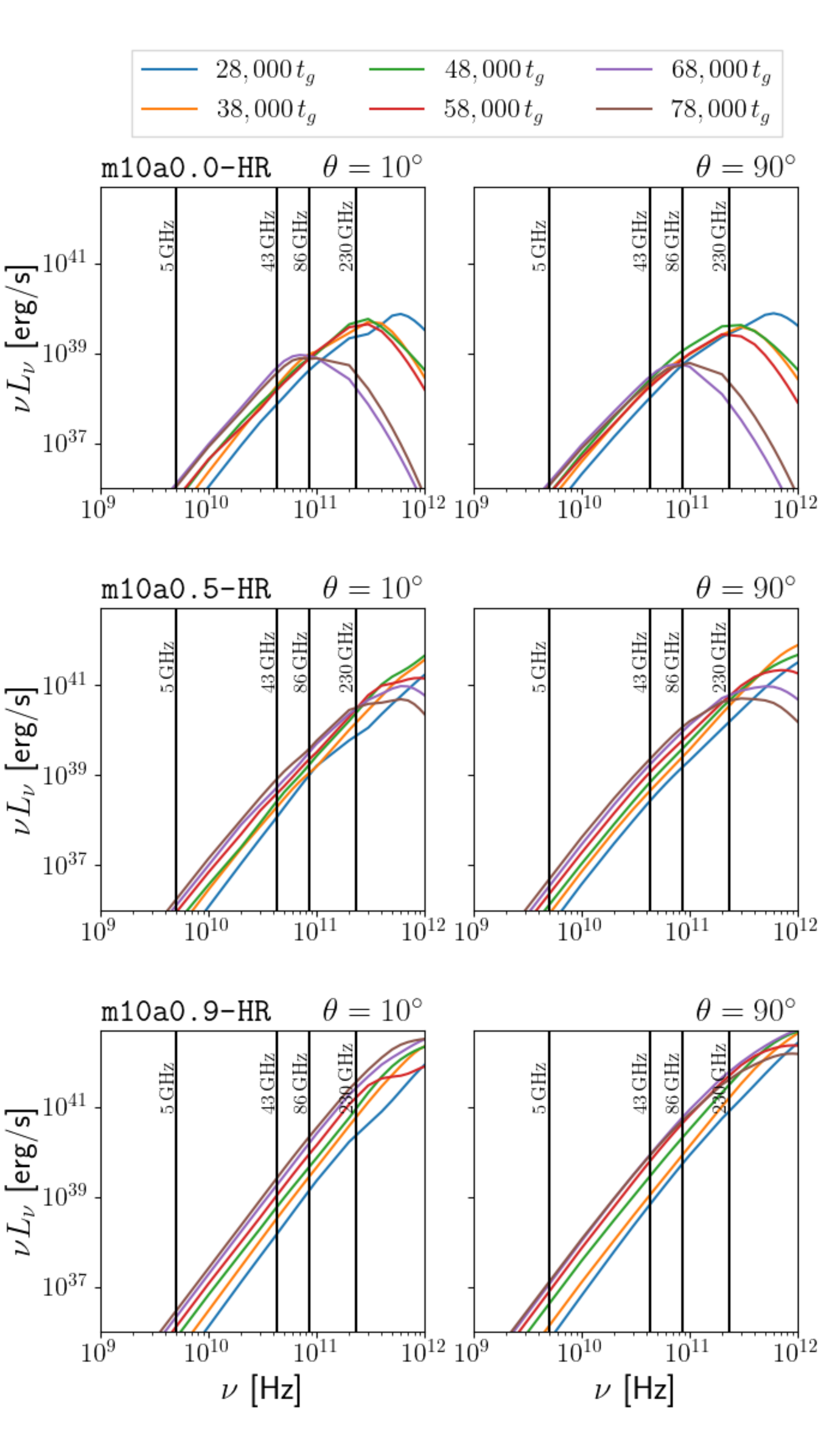}
    \caption{The same as Figure \ref{fig:tdep_spec_m5} but for models \texttt{m10a0.0-HR} (top), \texttt{m10a0.5-HR} (middle), and \texttt{m10a0.9-HR} (bottom). In all cases, the spectrum is generally shifting to the left and the peak frequency also decreases as the jet expands and some of the gas cools. However, \texttt{m10a0.0-HR} begins to dim substantially and the 230 GHz luminosity drops from $\sim 5\times 10^{39}\, \rm{erg\ s^{-1}}$ at $t\leq 58,000\, t_g$ to $\sim 10^{38}\, \rm{erg\ s^{-1}}$ after $t=68,000\,t_g$. Model \texttt{m10a0.9-HR} on the other hand is still brightening at 230 GHz by the final snapshot.}
    \label{fig:tdep_spec_m10}
\end{figure}

\begin{figure*}
    \centering{}
	\includegraphics[width=\textwidth]{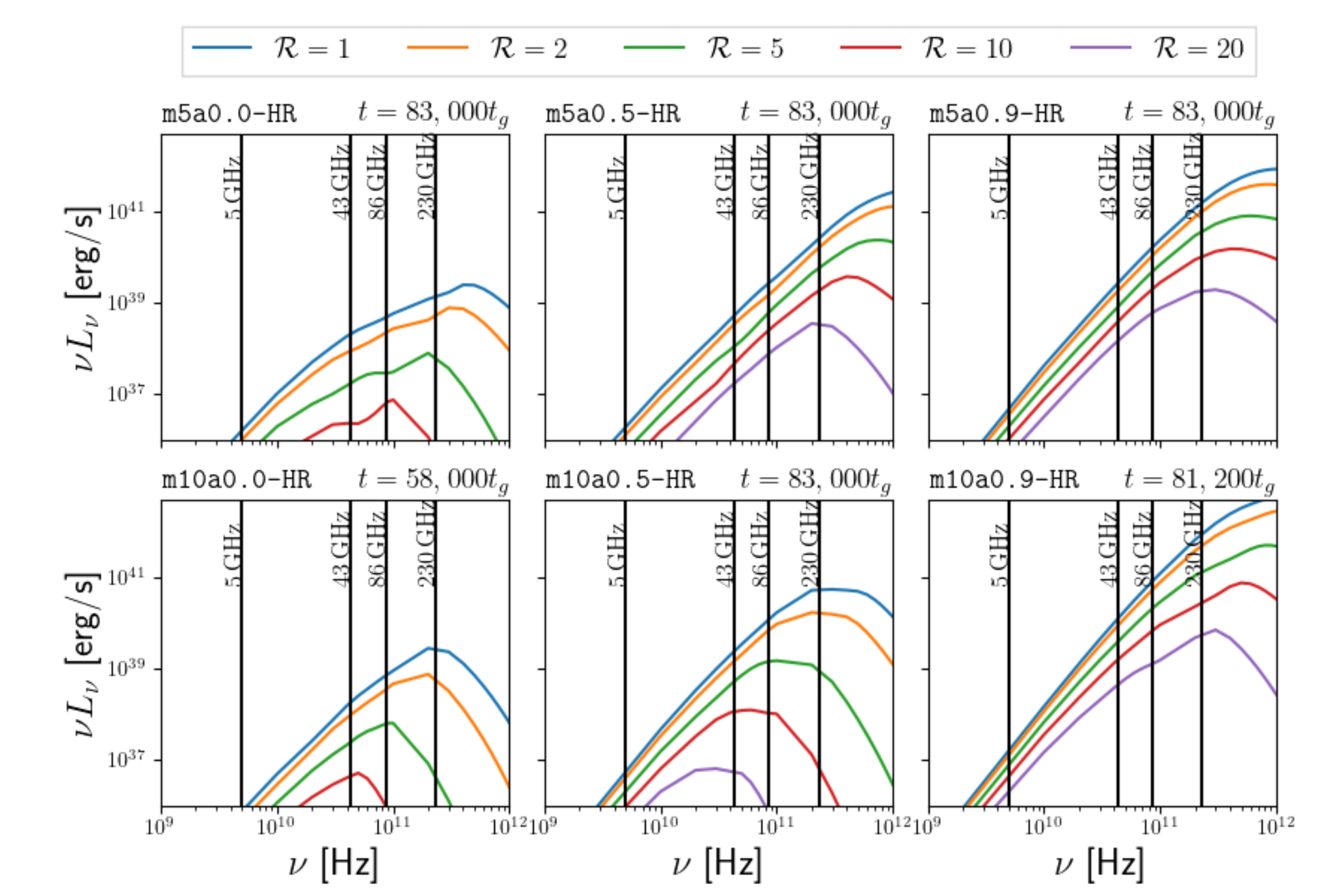}
    \caption{Model spectra of each model produced by varying $T_i/T_e = \mathcal{R}$. Here we use a camera angle of $\theta=90^\circ$ (viewing the disk/jet edge on). Increasing $\mathcal{R}$ has the effect of decreasing both the peak frequency and luminosity in each model. Even a modest temperature ratio of $\mathcal{R}=5$ can diminish the jet luminosity by 1-2 orders of magnitude. A temperature ratio of $\mathcal{R}\geq5$ may also substantially diminish the 230 GHz flux for colder jets which has important consequences for detecting and resolving potential sources.}
    \label{fig:ipole_spectra}
\end{figure*}

\subsection{Time Evolution and Viewing Angle Dependence of Emission for $T_e=T_i$}
\label{sec:tdepspec}

We present viewing angle dependent spectra for the $M_{\rm{BH}}=5\times10^{6}\, M_\odot$ models computed over $\nu=10^9 - 10^{12}$ Hz for snapshots ranging from $t=28,000-78,000\, t_g$ in Figure \ref{fig:tdep_spec_m5}. Here we assume a single-temperature gas with $T_e=T_i$.

Focusing on the spectrum over time for each BH spin, we find that the jet is generally brightening across all frequencies. The emission becomes significantly brighter as the spin increases, reflecting the substantial increase in jet power as the spin increases (see Table \ref{tab:tab1}). We observe weak beaming comparing the $10^\circ$ and $90^\circ$ spectra, as indicated by the shift to the left of each spectrum as the viewing angle increases. 

An interesting feature of \texttt{m5a0.0-HR} that is not apparent in the other models is delayed jet brightening. In fact, the jet is initially becoming less luminous and the peak frequency is decreasing until $t=58,000\, t_g$, where the peak frequency is only $\sim40$ GHz. At this stage, the jet emission is dominated by material near the head of the jet and not very much emission is present within the jet itself. As material at the jet head expands and cools, the brightness declines and the spectrum shifts to lower frequencies overall. After this point, faster moving gas which was accelerated earlier in the evolution catches up to material downstream which has slowed and drives internal shocks within the jet between $z\sim 5000-20,000\, r_g$. These shocks cause a significant amount of heating, driving the gas temperature where shocks occur to increase from $\sim 10^9$ K to $\sim 10^{10}$ K. The internal shocks lead to high frequency emission which is substantially brighter than that from the jet head alone. For instance, by $t=78,000\, t_g$ the jet has brightened to a luminosity of nearly $5\times 10^{39}\, \rm{erg\ s^{-1}}$ at the peak frequency versus the initial $\sim10^{38}\, \rm{erg\ s^{-1}}$ at $t=28,000\, t_g$. This demonstrates clearly how significant the contribution of emission due to internal shocks is in the context of these models.

For the $M_{\rm{BH}}=10^{7}\, M_\odot$ models (Figure \ref{fig:tdep_spec_m10}), we again observe only weak beaming and the jet becomes brighter as the BH spin increases, but there are several notable differences. Firstly, the jet in \texttt{m10a0.0-HR} and \texttt{m10a0.5-HR} is actually dimming over time. For instance, at $t=28,000\, t_g$, the jet in \texttt{m10a0.0-HR} has a peak luminosity of $\sim5\times10^{39}\,\rm{erg\ s^{-1}}$ and a peak frequency near 700 GHz, but by $t=78,000\, t_g$ the jet has become less bright with a peak luminosity of $\sim5\times10^{38}\,\rm{erg\ s^{-1}}$ and a peak frequency near 100 GHz. The jet in \texttt{m10a0.9-HR} on the other hand is still brightening by the time we stop the simulation.

The difference between the $M_{\rm{BH}}=5\times10^{6}\, M_\odot$ and $M_{\rm{BH}}=10^{7}\, M_\odot$ models is in part because the jets have propagated a shorter physical distance due to the shorter time in physical units. Although each model was run $81,200-83,000\, t_g$ this only corresponds to $\sim 24$ days for $M_{\rm{BH}}=5\times10^{6}\, M_\odot$ and $\sim 48$ days for $M_{\rm{BH}}=10^7\, M_\odot$. Consequently, the jets for $5\times10^{6}\, M_\odot$ have propagated roughly half the distance of the jets for $M_{\rm{BH}}=10^{7}\, M_\odot$. The jets for the $M_{\rm{BH}}=5\times10^{6}\, M_\odot$ models may also show dimming given a longer run time.

It is interesting to note the evolution of \texttt{m10a0.0-HR}, which is only bright at 230 GHz until $t = 58,000\, t_g$. This suggests that high frequency radio/submillimeter emission (i.e. $\nu > 100$ GHz) from TDE jets may only last for several weeks depending on the BH spin. However, we have not taken into account the effects of varying the external medium nor the possible misalignment between the BH spin and accretion disk in this work. These two factors will have a non-negligible effect on the jet power and the jet evolution. Future studies should explore these factors to better understand TDE jets.

\subsection{Effects of Two-Temperature Plasma} \label{sec:twotempspectra}

Internal shocks in the jet are expected to produce non-thermal electrons, which we do not model in this analysis. In addition, these shocks are expected to lead to differential heating of electrons and ions. The plasma will retain memory of this because electrons and ions and cannot efficiently thermalize in low density plasmas. This effect has been treated in studies of hot accretion flows using a simple two-temperature prescription. For instance, it is common to define an electron temperature which depends on $\beta_g$ to differentiate the electron temperature in the jet and disk separately \citep{2016A&A...586A..38M}. Since we only model the jet emission, we adopt a simple approach and define the electron temperature via:
\begin{equation}
    \dfrac{T_i}{T_e} = \mathcal{R}.
\end{equation}
This simplified prescription smooths over the microphysics, which depend on magnetic reconnection and shock properties; however, it does provide some handle on how the electron and ion populations must differ within a particular model to produce a specific emission property, albeit in a parameterized fashion. A recent study of electron heating in AGN jets \citep{2019Galax...7...14O,2020MNRAS.493.5761O} has demonstrated that $T_i/T_e$ can become as large as $10-100$ depending on the heating physics. In these simulations, shocks led to ion heating while weak Coulomb coupling prevented the ion and electron populations from equilibrating. Our simulation results indicate that internal shocks due to variable ejections take place. Furthermore, the time between Coulomb collisions is $>10^5$ years for typical densities and temperatures in the jet, thus it is reasonable to model the electrons in the jet with $T_i/T_e >1$.

We present full spectra at selected times for each model in Figure \ref{fig:ipole_spectra}. In all models, we find that increasing $\mathcal{R}$ decreases the peak frequency and overall luminosity. In model \texttt{m10a0.5-HR} at $t=83,000\, t_g$ for instance, as $\mathcal{R}$ is varied from $1$ to $20$, the peak frequency shifts from $\sim200$ GHz to $\sim 20$ GHz and the luminosity declines by nearly four orders of magnitude from $\sim10^{41} \, \rm{erg \ s^{-1}}$ to $\sim10^{37} \, \rm{erg \ s^{-1}}$. As models \texttt{m5a0.0-HR}, \texttt{m10a0.0-HR} and \texttt{m10a0.5-HR} illustrate, the electron temperature is extremely important for the high frequency emission as each of these models show greatly diminished emission at 230 GHz for $\mathcal{R}>2-5$. For models \texttt{m5a0.5-HR}, \texttt{m5a0.0-HR} and \texttt{m10a0.5-HR}, the jet luminosity at $\nu \gtrsim 230$ GHz is not as strongly diminished as $\mathcal{R}$ increases and $\mathcal{R}>20$ is required to drop the peak frequency below $230$ GHz. We explore the effects of $\mathcal{R}$ on the detectability of these models in the following section.

Studies of electron and ion heating suggest that the electron temperature might depend on $\beta_g$ (e.g. \citealt{2010MNRAS.409L.104H,2017ApJ...850...29R}) which would introduce spatial variation in $\mathcal{R}$. This would significantly change the behaviour in the spectra if regions that emit the most also have large values of $\mathcal{R}$. Without a heating prescription implemented during the simulation, however, we do not think it is justified to use a model which varies with $\beta_g$. That being said, a more accurate prescription may also need to account for the radiation pressure and use a two-temperature model which scales based on $\beta_t$. We leave a precise analysis of the spatial electron temperature ratio to a future analysis. 

\begin{figure}
	\includegraphics[width=\columnwidth]{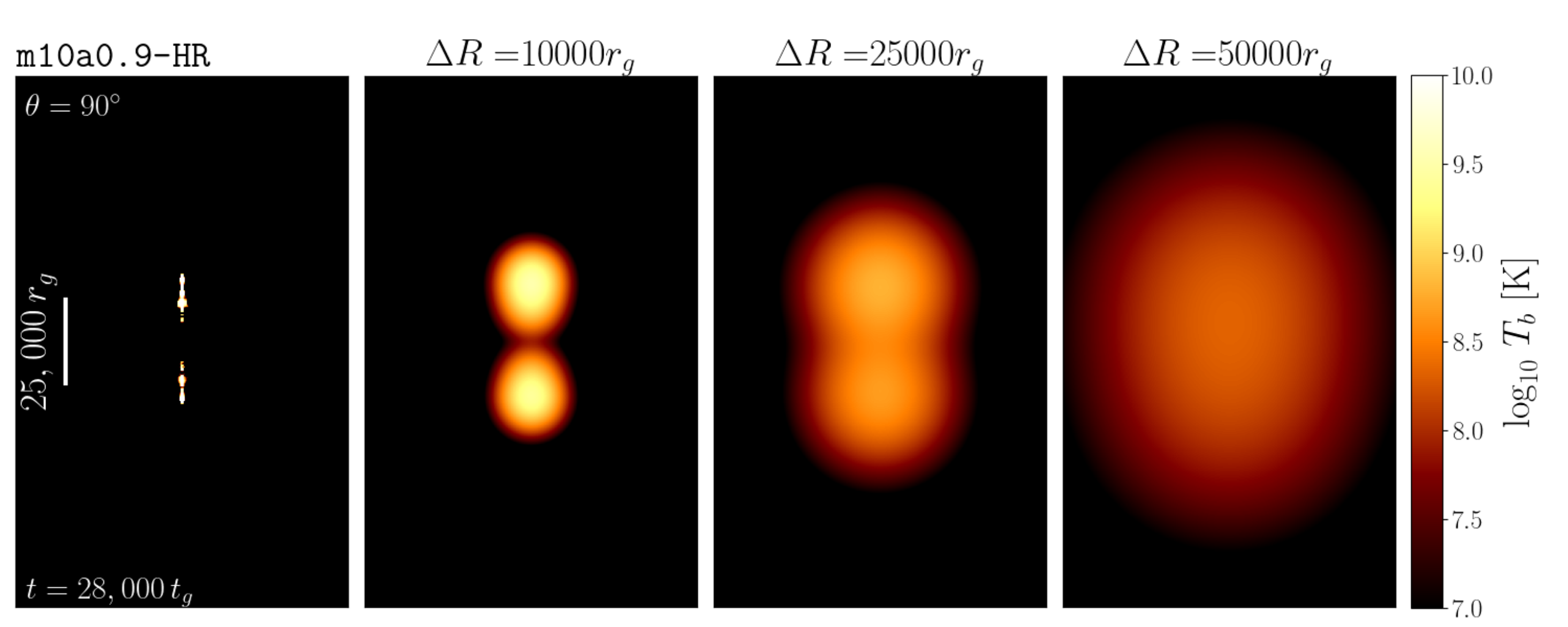}\\
	\includegraphics[width=\columnwidth]{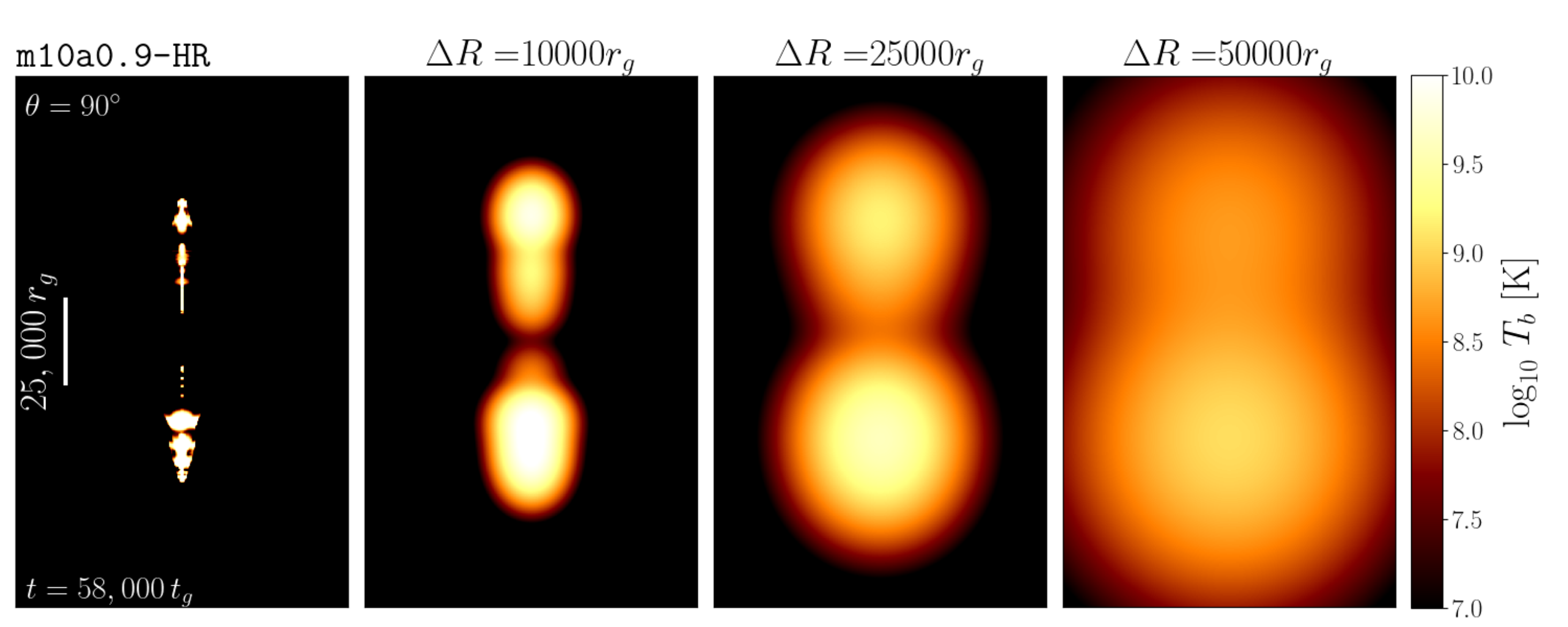}\\
	\includegraphics[width=\columnwidth]{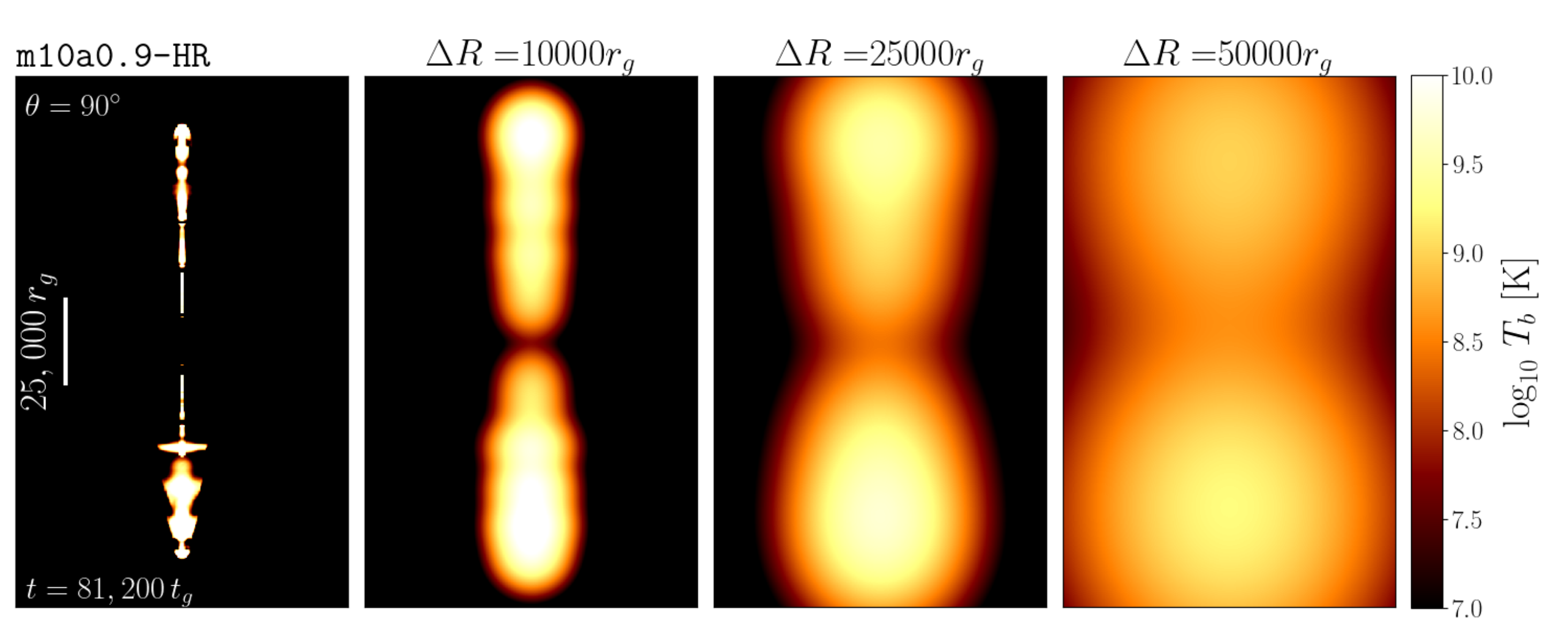}
    \caption{Here we present GRRT images of the thermal synchrotron emission (mapped via the brightness temperature $T_b$) at $\nu=230$ GHz with a viewing angle of $\theta=90^\circ$ for model \texttt{m10a0.9-HR}. Here we have simply assumed $T_e=T_i$. We plot a white line of length $25,000\, r_g$ in the left column for scale. We show snapshots at $28,000\, t_g$ (top row), $58,000\, t_g$ (middle row), and $81,200\, t_g$ (bottom row). We show the infinite resolution images in the 1st column from the left. In addition, we apply Gaussian smoothing with a FWHM of $\Delta R/r_g =$ 10000 (2nd column from the left), 25000 (3rd column from the left), and 50000 (right column) to illustrate the effects of blurring due to distance and resolution. The jet head is the brightest feature and in the infinitely resolved case appears to show bow shock features. For well resolved images, bright `bubbles' along the jet axis can be seen, similar to what we presented in the emissivity maps in Figure \ref{fig:emissivity}. The jet head features are best resolved when $\Delta R \lesssim 25,000\, r_g$. Observations with similar resolutions scales may allow for direct measurements of the ejecta velocity.}
    \label{fig:230GHz_m10a0.9_tdep}
\end{figure}

\begin{figure}
	\includegraphics[width=\columnwidth]{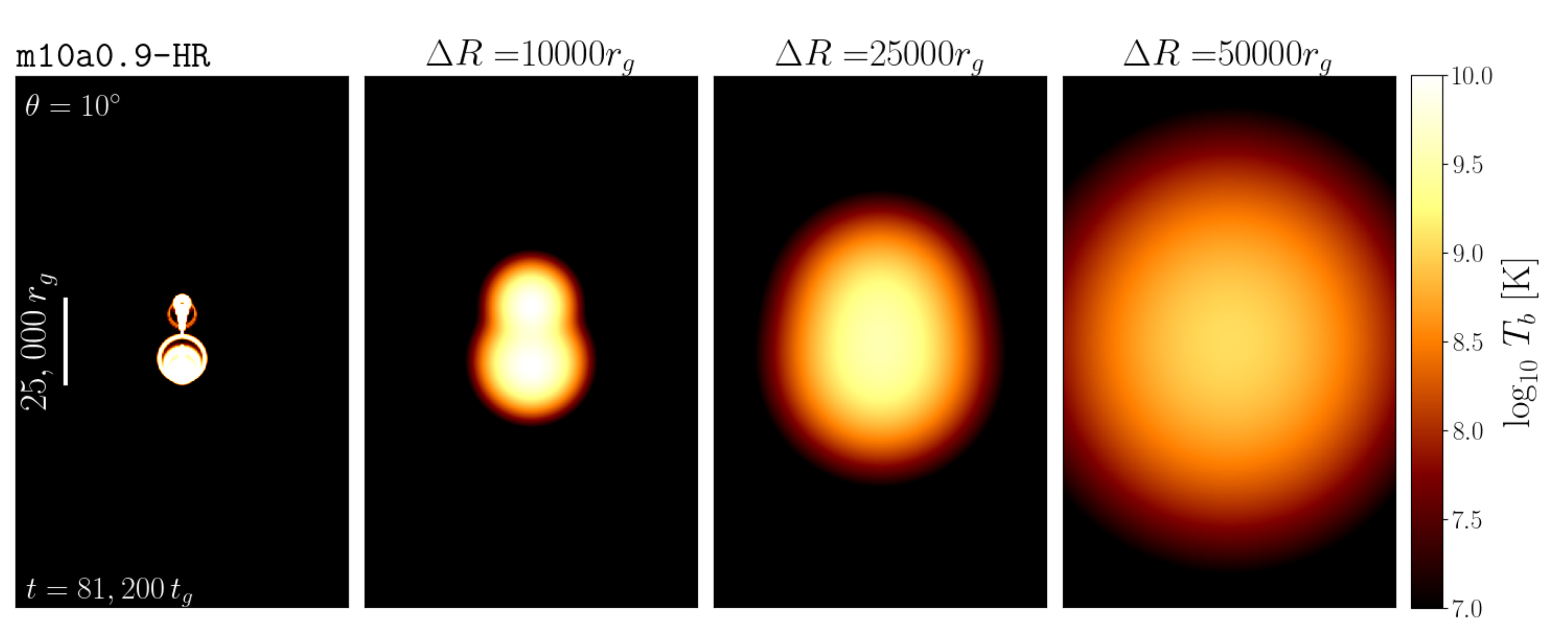}\\
	\includegraphics[width=\columnwidth]{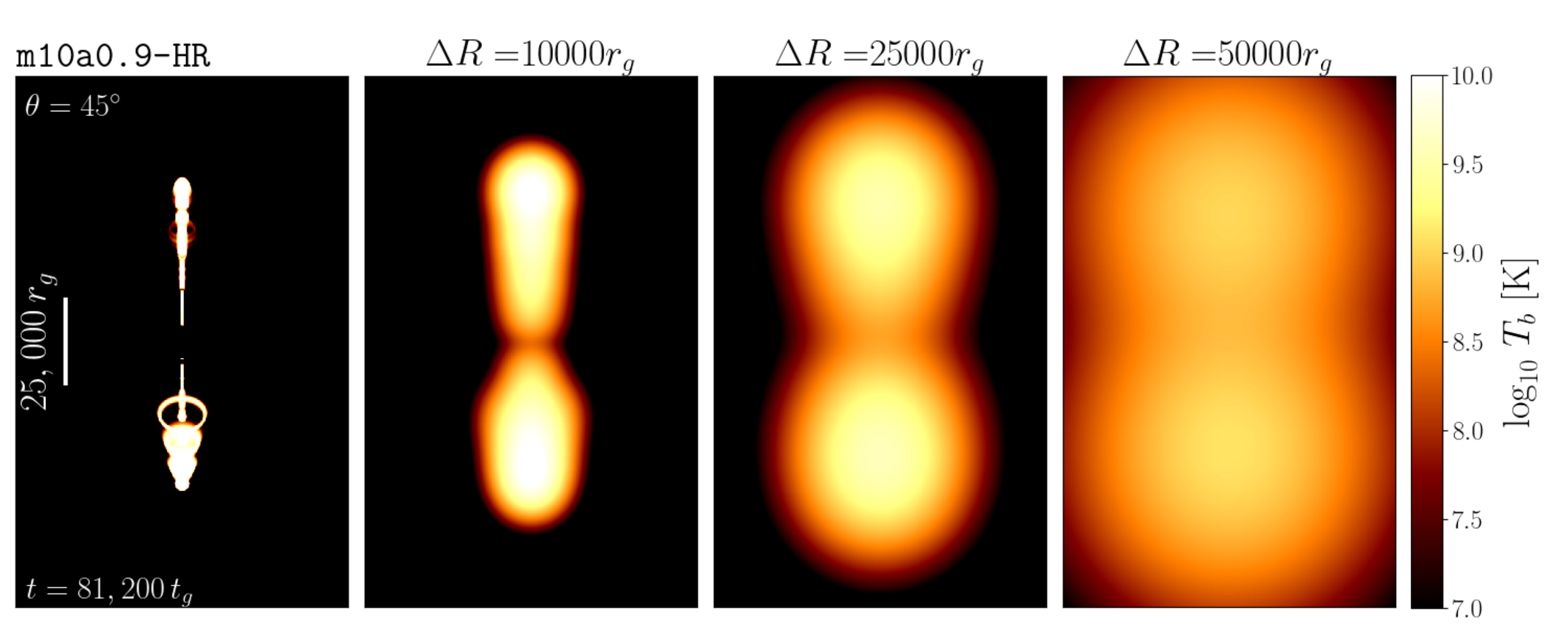}\\
	\includegraphics[width=\columnwidth]{images_pdf/7s09hi_HR_0812_th90.pdf}
    \caption{Similar to Figure \ref{fig:230GHz_m10a0.9_tdep}, but here each row shows \texttt{m10a0.9-HR} at $t=81,200\,t_g$ but with a viewing angle of $\theta=10^\circ$ (top row), $\theta=45^\circ$ (middle row), and $\theta=90^\circ$ (bottom row). The images at $\theta=45^\circ$ illustrate that internal shock features cannot be distinguished even for fairly well resolved sources as the viewing angle decreases from edge-on. For jets viewed nearly down the axis (top row) only the jet heads are seen and the jet appears as two blurred lobes. For poorly resolved sources, a nearly face-on jet begins to look like a compact source.} 
    \label{fig:230GHz_m10a0.9}
\end{figure}

\begin{figure}
	\includegraphics[width=\columnwidth]{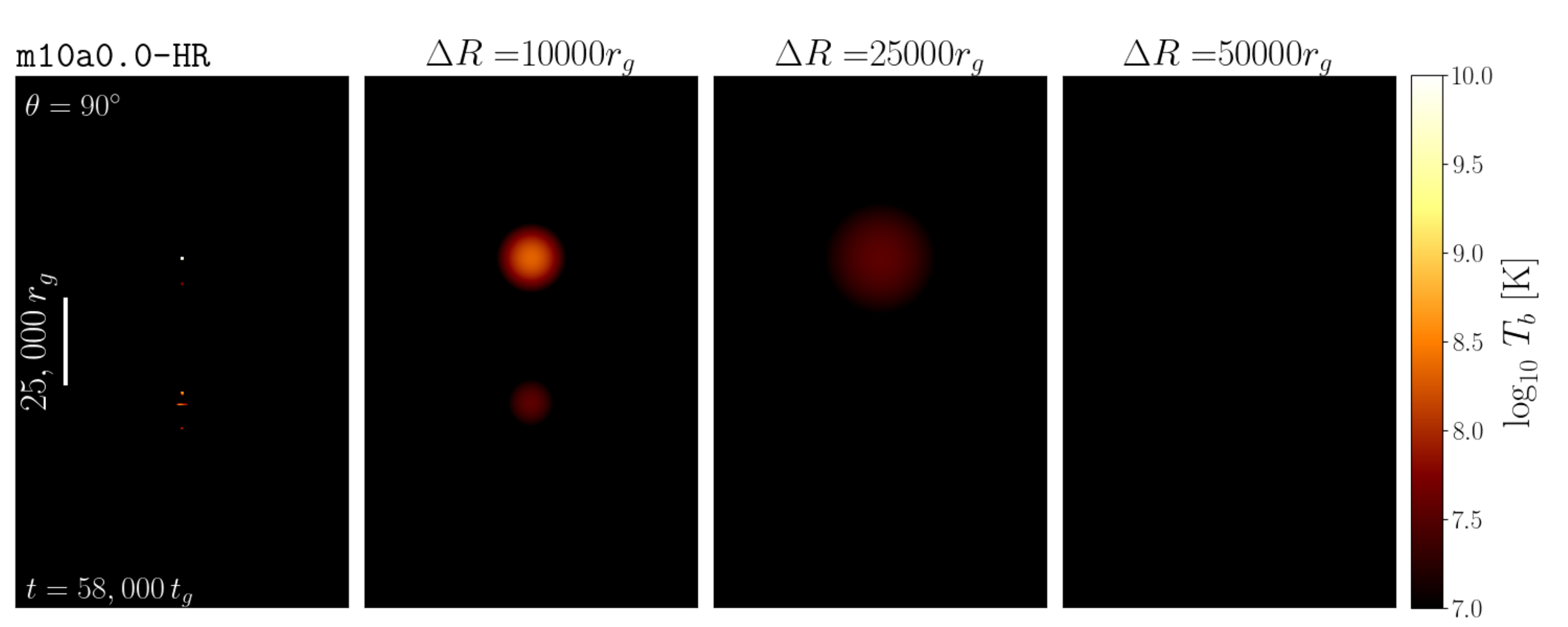}\\
	\includegraphics[width=\columnwidth]{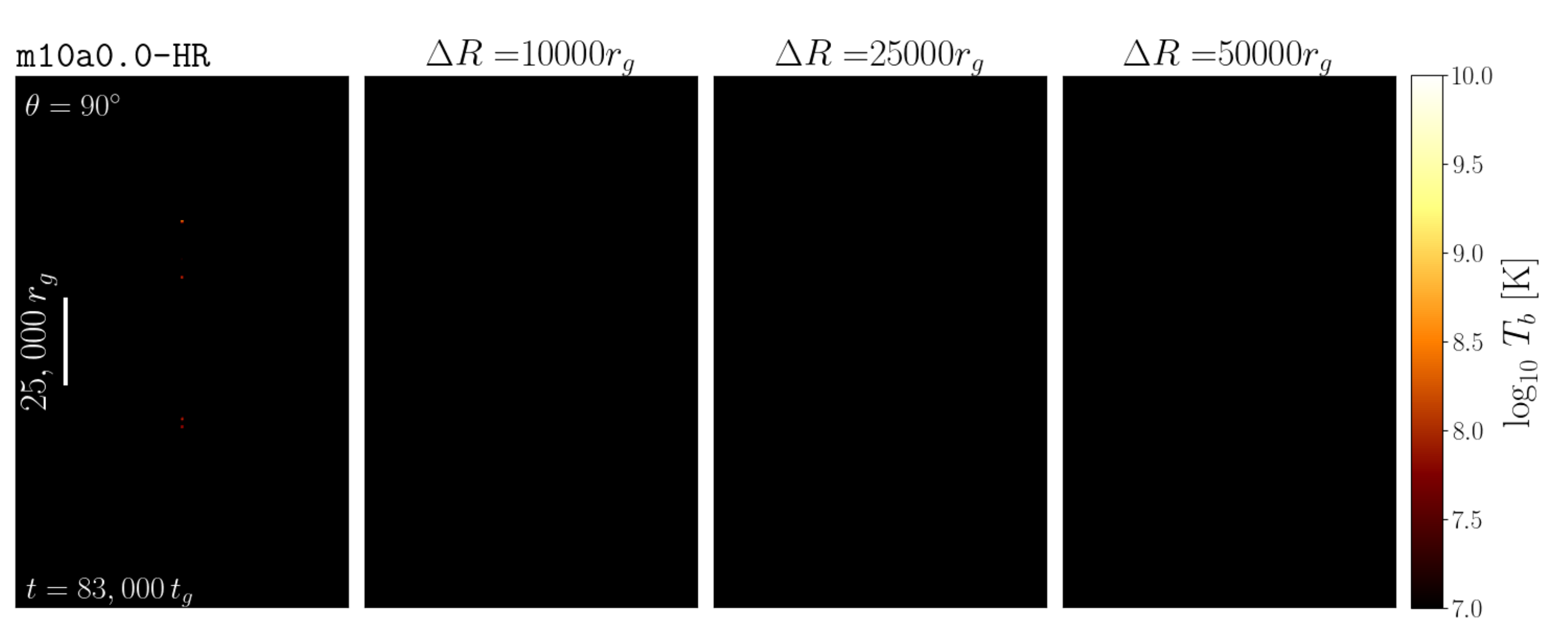}
    \caption{Similar to Figure \ref{fig:230GHz_m10a0.9}, but here we show \texttt{m10a0.0-HR} at $t=58,000\,t_g$ (top row) and $t=83,000\,t_g$ (bottom row) to illustrate the dimming of the jet. The jet in \texttt{m10a0.0-HR} did not show significant internal shock heating and as a result the jet head is responsible for most of the emission. As the jet head expands and cools, it eventually becomes dim, and possibly undetectable.}
    \label{fig:230GHz_m10a0.0}
\end{figure}

\begin{figure}
	\includegraphics[width=\columnwidth]{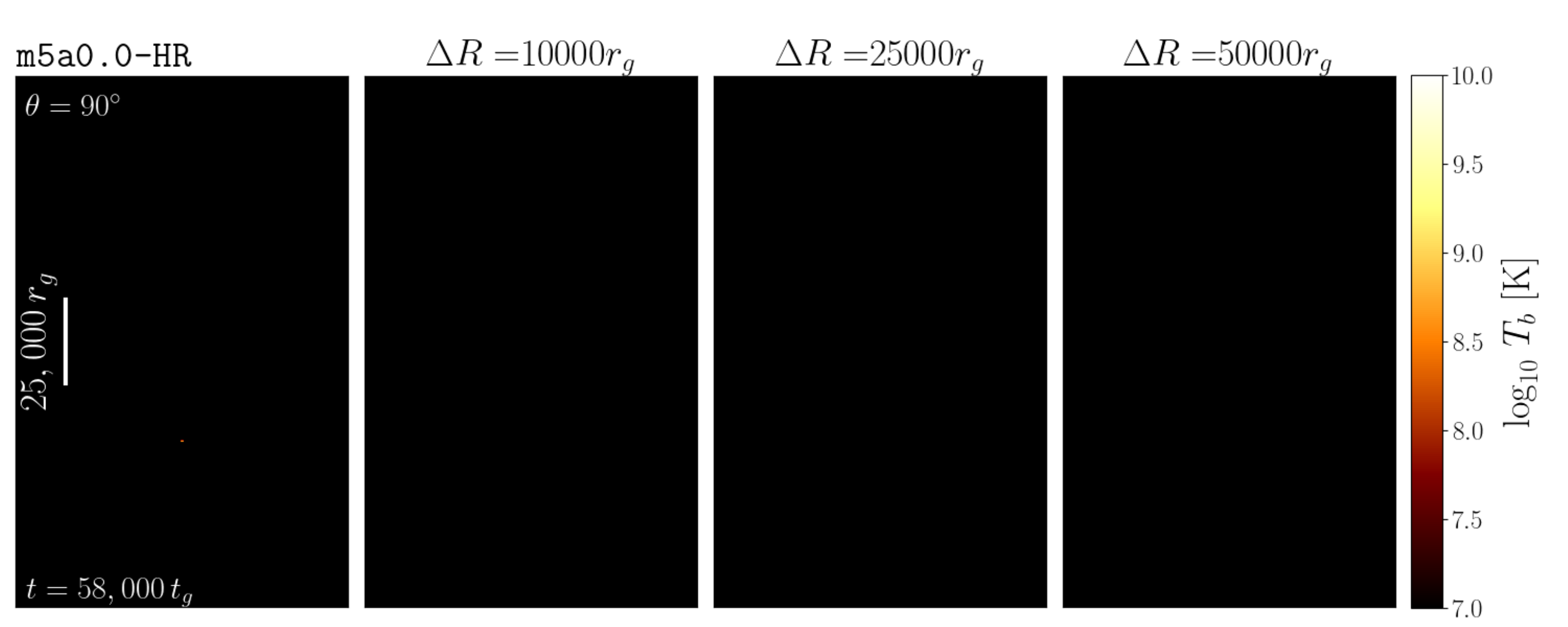}\\
	\includegraphics[width=\columnwidth]{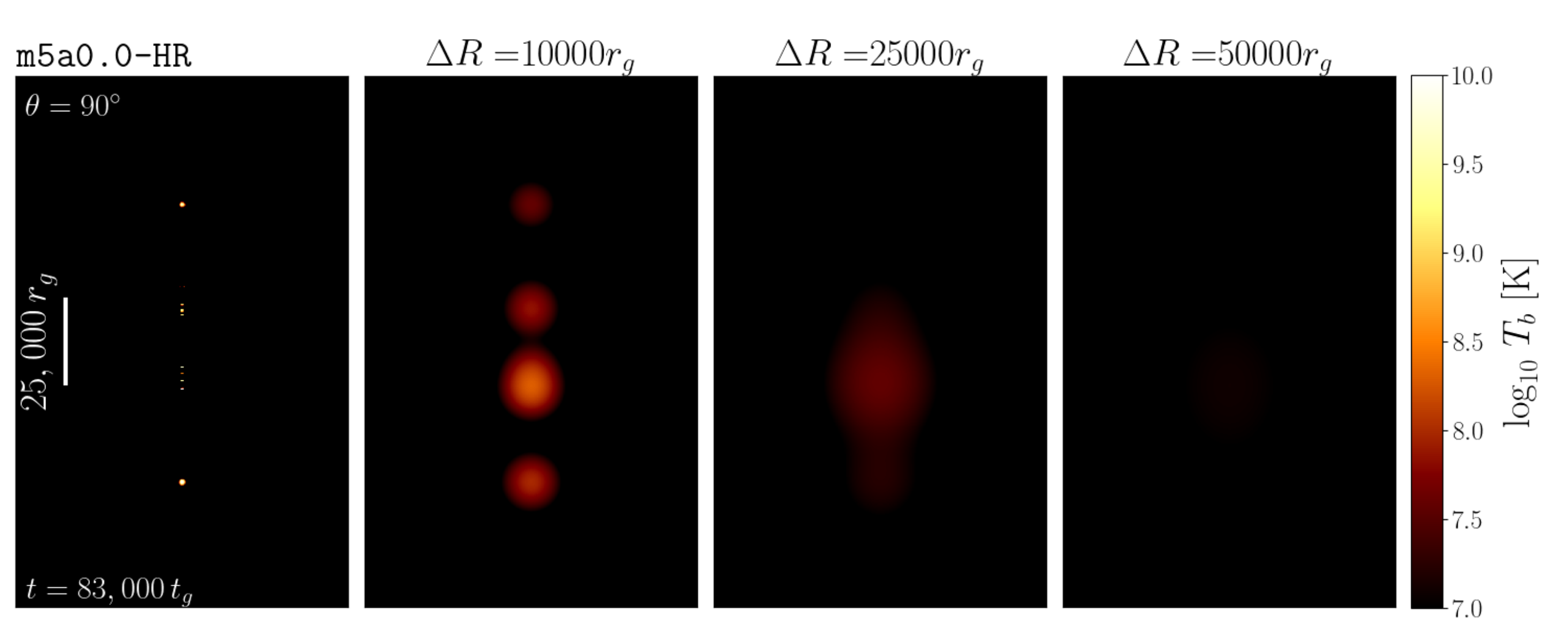}
    \caption{Similar to Figure \ref{fig:230GHz_m10a0.0}, but here we show \texttt{m5a0.0-HR} at $t=58,000\,t_g$ (top row) and $t=83,000\,t_g$ (bottom row) to illustrate the dimming of the jet. Similar to \texttt{m10a0.0-HR}, \texttt{m5a0.0-HR} also did not initially show significant internal shock heating but the initial jet was also quite weak and not significantly bright. By $t=68,000\, t_g$, the jet had brightened substantially due to internal shocks and shocks near the jet head where fast moving gas caught the slower moving jet head and energized it.}
    \label{fig:230GHz_m5a0.0}
\end{figure}

\subsection{230 GHz Images} \label{sec:detectability}

In this section, we analyze viewing angle dependent thermal synchrotron images at 230 GHz which were calculated assuming $T_e=T_i$ and employ the same methods as in Section \ref{sec:tdepspec}. In addition, we apply a Gaussian smoothing function with a FWHM of $\Delta R$ to each model to simulate the effects of the resolved angular scale ($\Delta \theta$) and distance to the source ($D$). The resolved scale in $r_g$ can be related to both quantities by:
\begin{equation}
    \dfrac{\Delta R}{r_g} \approx 2022 \left(\dfrac{M_{\rm{BH}}}{10^6 \, M_\odot}\right)^{-1} \left( \dfrac{D}{1 {\rm{\, Mpc}}}\right)\left(\dfrac{\Delta \theta}{20 {\rm{\mu as}}}\right) .
\end{equation}
For a given resolution scale, the image could represent a jet that is both distant and well resolved or nearby and poorly resolved. We blur each image with a Gaussian using $\Delta R/r_g = (10000, 25000, 50000)$. We first focus on \texttt{m10a0.9-HR} to describe key emission features as it is the brightest model with the largest physical scale.

In Figure \ref{fig:230GHz_m10a0.9_tdep}, we show the time evolution of the jet emission for a viewing angle of $90^\circ$ in the leftmost column. The jet in this model is extremely bright with $T_b > 10^{10}$ K. The jet emission is primarily from the jet head but there is also significant emission along the jet axis, where relativistic magnetized gas resides. Discrete features within the jet due to internal shocks which are spatially separated are also apparent in the emission features. In the next three columns, we apply a Gaussian smoothing function of size $\Delta R/r_g=10000, 25000,$ and $50000$, respectively. Even for the resolution scale $\Delta r/r_g = 25000$, the jet is incredibly bright with $T_b > 10^{9}$ K and the motion of the expanding jet head could be tracked if the source were resolved with $\Delta R \lesssim 25,000\, r_g$. Tracking the jet position may allow for the jet velocity to be constrained with a method independent of emission model assumptions.

In Figure \ref{fig:230GHz_m10a0.9}, we show the viewing angle dependence of the jet emission in \texttt{m10a0.9-HR} for a viewing angle of $10^\circ$ (top row), $45^\circ$ (middle row) and $90^\circ$ (bottom row). The distinguishing features of the jet (the jet head and discrete internal shock emission) cannot be distinguished at $45^\circ$ even with $\Delta R/r_g =10000$. At $10^\circ$, the jet appears as a double lobed structure due to the top and bottom jet, but this would be seen as a compact source unless the jet is well resolved. 

Additional 230 GHz images of the jets of each model near the brightest point in their evolution for a viewing angle of $90^\circ$ are shown in Figure \ref{fig:all_230GHz_images}. In general, the jets become brighter as the spin increases owing to the greater overall jet power. In addition, the jet head and internal shocks are also apparent for fairly well resolved jets viewed edge on for models with $M_{\rm{BH}}=5\times10^6 \, M_\odot$.

In Figure \ref{fig:230GHz_m10a0.0}, we illustrate the dimming in \texttt{m10a0.0-HR}. The emission in this model was dominated by the jet head with only weak internal shocks. As a result, the jet appears as simply two lobes separated spatially. As the jet head expands and cools, the high frequency emission declines rapidly as described in Section \ref{sec:tdepspec}. If detected as a radio/submillimeter source, \texttt{m10a0.0-HR} would appear as a bright radio/submillimeter source for several weeks before likely dropping below detection limits. 

On the other hand, the model \texttt{m5a0.0-HR} (Figure \ref{fig:230GHz_m5a0.0}) initially did not show internal shocks and was incredibly dim. Instead, the jet would have likely appeared to be dormant for several weeks before brightening if this model were observed immediately following the launch of the outflow. Although the timescale before the jet brightens ($\sim20$ days) is shorter than the delays seen in many radio quiet TDEs ($>30$ days, \citealt{2020SSRv..216...81A}) and we are imaging in a higher frequency than the 5-8.4 GHz most detections have been made at, this behaviour shows that delayed brightening can occur in super-Eddington accretion disk jets.

\begin{figure}
	\includegraphics[width=\columnwidth]{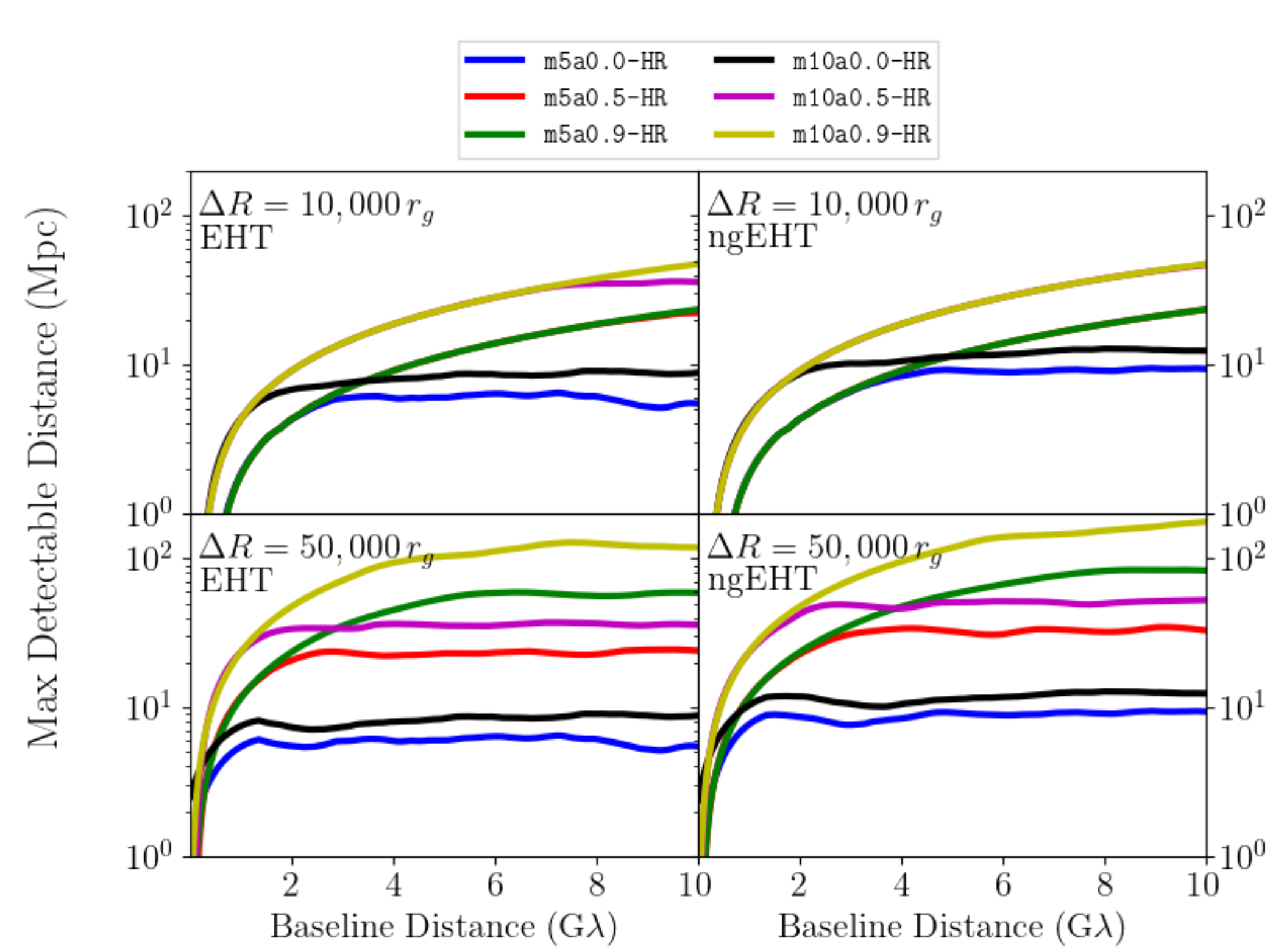}
    \caption{Here we show the maximum distance at which the jet in each model is detectable and resolved given a limiting baseline distance and a $5\sigma_{\rm{230\, GHz}}$ (with $\sigma_{\rm{230\, GHz}}=10$ mJy for the EHT and $\sigma_{\rm{230\, GHz}}=5$ mJy for the ngEHT) detection threshold. It is assumed that $T_e=T_i$ ($\mathcal{R}=1$). Focusing first on the results assuming observations with the EHT (left), the estimated maximum distance at which each model can be detected increases with BH spin due to the increase in jet power and brightness. The non-spinning models require $D\lesssim 6-9$ Mpc regardless of resolution scale and are thus unlikely targets. The spin $a_*=0.5$ models are substantially brighter and may be suitable targets at up to $D\lesssim35$ Mpc. In the brightest model (\texttt{m10a0.9-HR}), resolving the internal shocks ($\Delta R = 10,000\, r_g$) requires that the TDE occur within $D\lesssim45$ Mpc while resolving only the jet head ($\Delta R = 50,000\, r_g$) only requires $D\lesssim110$ Mpc. The increased detector sensitivity expected in the ngEHT (right) compared to the EHT (left) increases the maximum distance at which the jets can be detected and resolved by a factor of $\sim1.4-1.8$. The greatest improvement is for large baseline distances and $a_*=0.9$. For instance, the maximum distance increases from $\sim110$ Mpc to $\sim 180$ Mpc at a baseline distance of $10\, \rm{G\lambda}$.}
    \label{fig:detectability}
\end{figure}

\begin{figure}
	\includegraphics[width=\columnwidth]{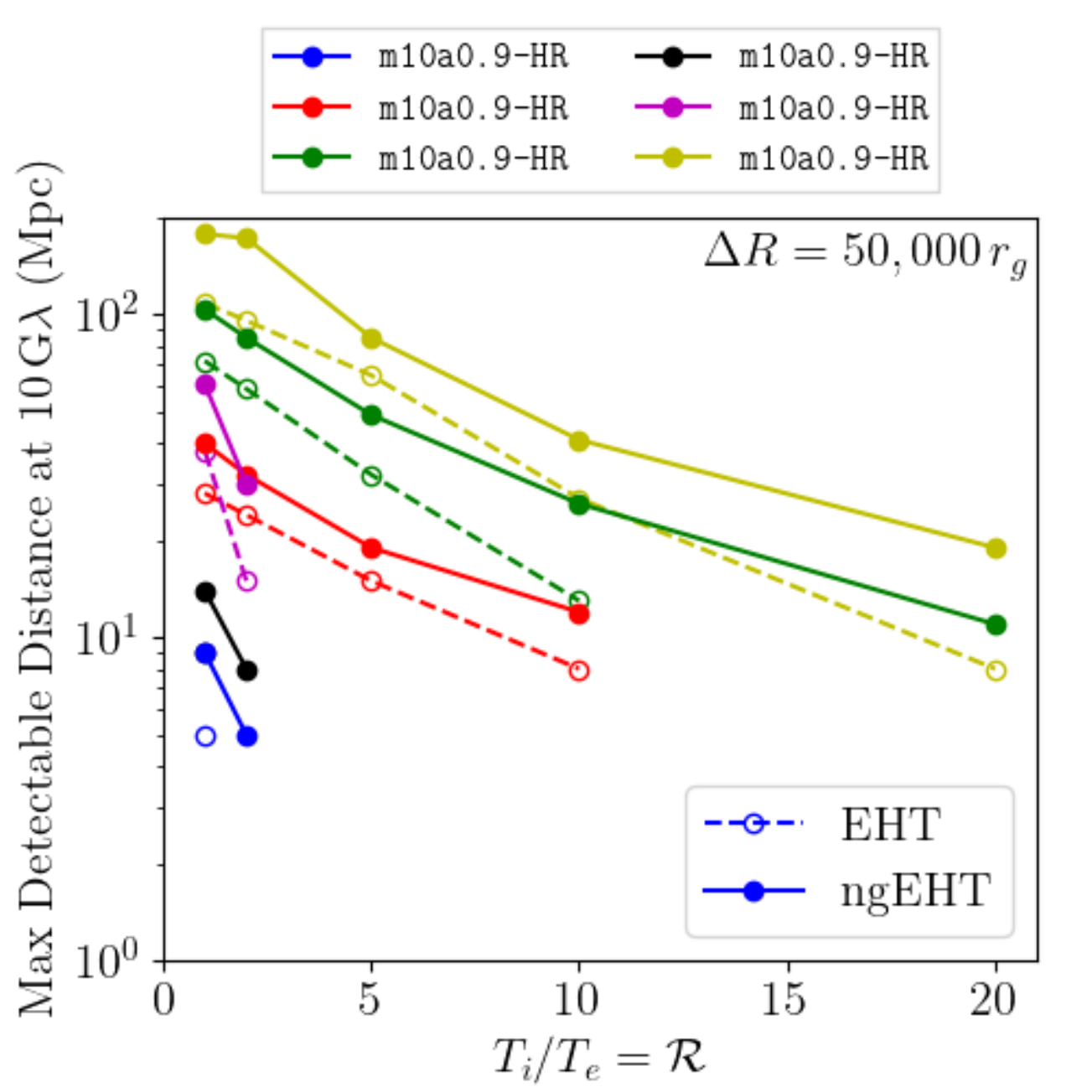}
    \caption{Here we show the maximum distance at which the jet in each model is detectable and resolved as $\mathcal{R}$ increases for $\Delta R = 50,000\, r_g$ at a limiting baseline distance of $10\ \rm{G\lambda}$. We compare EHT ($\sigma_{\rm{230\, GHz}}=10$ mJy, dashed lines with open circles) and ngEHT ($\sigma_{\rm{230\, GHz}}=5$ mJy, solid lines with filled circles) limits for the correlated flux density and a $5\sigma_{\rm{230\, GHz}}$ detection threshold. The curve is cut off at $\mathcal{R}$ where the jet becomes unresolved.}
    \label{fig:detectabilityRmodel}
\end{figure}


To estimate the viability of detection and resolution of each model, we estimate the visibility amplitude as a function of baseline distance assuming a detector limit of $\sigma_{\rm{230\, GHz}} = 10$ mJy for the EHT \citep{2019ApJ...875L...3E} and $\sigma_{\rm{230\, GHz}} = 5$ mJy for the ngEHT \citep{2019BAAS...51g.256D} and assuming that Earth baselines at 230 GHz will not exceed 10 G$\lambda$. Visibility amplitudes were calculated using the \texttt{eht-imaging} library \citep{2018zndo...1173414C}. We consider a jet model detectable and resolvable if (i) the total image (zero fringe spacing baseline) flux is above 5$\sigma_{\rm{230\, GHz}}$, (ii) the difference between the total image flux and any other, longer baseline (up to 10 G$\lambda$) flux is above 5$\sigma_{\rm{230\, GHz}}$, and (iii) the nominal resolution of the baseline ($1/u$ for baseline length $u$) is less than $\Delta R$. This definition assumes that the baseline length is the maximum baseline separation available for the observations. Given this assumption, it additionally requires that the baseline can resolve the jet at the desired length scale. We compute the maximum distance for viewing angles of $10^\circ$, $45^\circ$, and $90^\circ$ and then average over viewing angle to obtain a representative maximum distance. We also smooth the maximum distance profiles using a moving average with a window of $\sim 2.7\ {\rm{G}}\lambda$ in order to reduce scatter introduced by the variable behaviour in the visibility amplitude.

We present the estimated maximum distance for imaging a resolved jet near the brightest period in each simulation assuming $\mathcal{R}=1$ with $5\sigma_{\rm{230\, GHz}}$ significance at length scales $\Delta R = 10,000 \, r_g$ and $\Delta R = 50,000 \, r_g$ in Figure \ref{fig:detectability}. As expected, we find that the maximum distance at which the jet can be detected and resolved increases with BH spin due to the increased jet power. Assuming only the jet head is resolved, the brightest jets may be detected with the EHT (ngEHT) at up to $D \sim 110$ (180) Mpc depending on the baseline coverage available. The non-spinning models on the other hand are only detectable within $D\lesssim 6-9$ ($8-13$) Mpc. Assuming a middling BH spin of $a_*=0.5$, detection may be possible at $D\lesssim35$ (60) Mpc.  Resolving finer structure requires that the TDE occurs nearby. For instance, \texttt{m10a0.9-HR} is only detectable and resolved within $D\lesssim45$ Mpc if $\Delta R=10,000\, r_g$. The same maximum distance is found for the ngEHT assuming a resolution length scale of $\Delta R=10,000\, r_g$. A similar effect on the maximum distance is seen in the other models as the resolved length scale decreases.

We quantify the effects of a two-temperature plasma by determining the maximum detectable and resolved distance for each model as $\mathcal{R}$ is varied assuming a baseline separation of $10\ \rm{G\lambda}$ and a length scale $\Delta R=50,000\, r_g$ in Figure \ref{fig:detectabilityRmodel}. Due to the decline in overall luminosity and the shift in peak frequency, the maximum distance can decrease by up to an order of magnitude. For instance, the maximum distance for \texttt{m10a0.9-HR}, assuming EHT detector limits, declines from $\sim110$ Mpc for $\mathcal{R}=1$ to $\sim8$ Mpc for $\mathcal{R}=20$. In models \texttt{m5a0.0-HR}, \texttt{m10a0.0-HR}, and \texttt{m10a0.5-HR}, even a small increase in $\mathcal{R}$ shifts the peak frequency to less than 230 GHz which leads to unresolved 230 GHz jet emission for values of $\mathcal{R}>1-2$ (assuming a lower threshold on the distance to the TDE of 1 Mpc). A similar result is found assuming the expected detector sensitivity of ngEHT, but the maximum distance increases by a factor of $\sim1.4-1.8$. In some cases, the maximum temperature ratio $\mathcal{R}$ that the jet can have and still be detectable increases (i.e. model \texttt{m5a0.9-HR}).

Our results suggest that the number of sources that can be detected and resolved will depend on the spin distribution of SMBHs across cosmic distance due to the increase in jet power with BH spin. Current simulations and observations which have estimated the spins of BHs at redshift $z\approx0$ indicate that SMBHs in the mass range of $\sim 10^{6.5}-10^{7.5}\, M_\odot$ will tend towards $a_* \sim 1$ \citep{2013CQGra..30x4004R,2013ApJ...762...68D,2014MNRAS.440.1590D,2019MNRAS.490.4133B}. This is encouraging since, with the EHT (ngEHT), at least $\sim45$ (200) TDEs are expected per year within $D\lesssim 110$ (180) Mpc assuming $\mathcal{R}=1$. This suggests that the increased detector sensitivity expected in the ngEHT could increase the number of possible TDE targets by a factor of nearly 4 compared to the EHT. On the other hand, if $\mathcal{R}$ significantly exceeds unity, less than one resolvable source may occur per year, even with the improved sensitivity of the ngEHT. In addition, if accretion occurs in the MAD accretion state for a significant fraction of a BH's lifetime, this could skew the BH spin distribution towards lower spin values due to spin down \citep{2022MNRAS.511.3795N} which would certainly reduce the number of detectable TDE jet sources at larger distances based on our models.

Before we conclude, we must point out several caveats in this work. Our simulation results demonstrate that in super-Eddington accretion disks, the jet magnetization is an important factor since it can produce a relativistic component driven by Poynting acceleration. Simulations of TDE accretion disk formation have not yet demonstrated the presence of significant magnetic flux in the inner accretion flow nor the onset of jet launching so the highly magnetized, bright jets produced in the $a_*\geq 0.5$ simulations are somewhat of an idealization given the assumed initial torus configuration. The mass accretion rates that our models achieve is not precisely at the peak accretion rate. This will change the jet power and potentially the radio/submillimeter brightness since $L_{\rm{net}}\propto \dot{M}$ \citep{2015MNRAS.453.3213S}. In other words, the spectra we provide here may under/over estimate the jet brightness by a factor of a few. This may increase/decrease the maximum resolvable distance in some models, but the overall finding that some nearby TDEs may produce jets which are detectable and resolvable is robust.

The choice to cut out the $\sigma>1$ region of the jet before ray tracing is conservative and well motivated, but the effects on the ray traced images has not yet been well studied. Given that much of the emission in the brightest jets originates from near the poles, our spectra provide a lower limit on the luminosity of the jet. In addition, as we have emphasized throughout this section, jets with internal shocks are likely to produce a non-thermal electron population and may also consist of a two-temperature plasma. Neither of these were employed during the simulations, and they could modify the jet properties since radiation fuels the jet.

\section{Discussion}\label{sec:discussion}

\begin{figure}
	\includegraphics[width=\columnwidth]{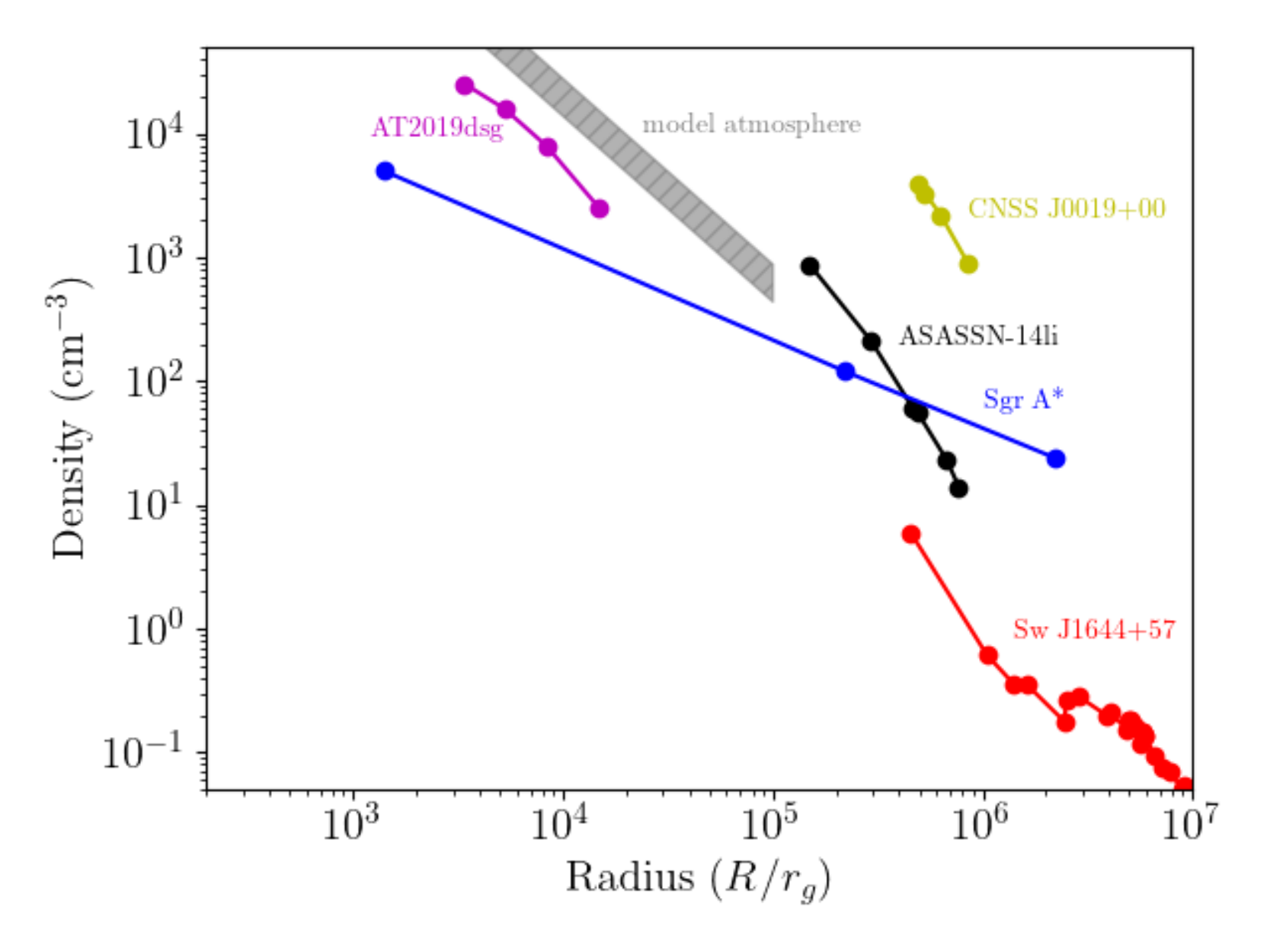}
    \caption{Here we compare our model atmospheres (gray hatched region) with CNM profiles of ASASSN-14li \citep{2016ApJ...819L..25A}, CNSS J0019+00 \citep{2020ApJ...903..116A}, \textit{Swift} J1644+57 (Sw J1644+57, \citealt{2018ApJ...854...86E}), AT2019dsg \citep{2021NatAs...5..510S}, and Sagittarius A* (Sgr A*, \citealt{2003ApJ...591..891B,2019ApJ...871..126G}). The scaling in our models is such that the density of the CNM at large radii is similar to that inferred in other TDEs and Sgr A*.}
    \label{fig:CNM}
\end{figure}

\subsection{Initial Atmosphere vs. Inferred Profiles in TDEs} \label{sec:atmosphere}

As noted in Section \ref{sec:initialconditions}, we initialized each simulation with a low density atmosphere where $\rho \approx (0.5-1\times10^{-14}\,\rm{g \, cm^{-3}})(r/r_0)^{-3/2}$, where $r_0=r_H$. The $r^{-3/2}$ profile that we have chosen is shallower than the inferred profile of roughly $r^{-5/2}$ for ASASSN-14li \citep{2016ApJ...819L..25A}. We note however that estimates of the density profile of ASASSN-14li were only obtained up to $\gtrsim 10^{5}r_g$ (or $\gtrsim 10^4\, r_g$ if the outflow geometry is conical). Assuming that such a profile does indeed continue to the BH, this implies that the atmosphere in ASASSN-14li may reach a density near the horizon of $\sim 10^{-11} \, \rm{g\ cm^{-3}}$, nearly three orders of magnitude denser than we have implemented. On the other hand, the profile of the TDE AT2019dsg \citep{2021NatAs...5..510S} appears to be turning over to a shallower profile of nearly $r^{-1}$ at $r\lesssim 10^{3}\, r_g$, which is similar to the inferred properties of the CNM in Sagittarius A$^*$ \citep{2003ApJ...591..891B,2019ApJ...871..126G}. Extrapolating the profile of AT2019dsg down to the horizon implies a maximum density of $\sim 10^{-17} \, \rm{g\ cm^{-3}}$, which is orders of magnitude less than what we have implemented.

Despite these differences in the behaviour at small radii, our model atmospheres are similar to that of previous radio TDEs over $r\sim 10^4-10^5\, r_g$ (see Figure \ref{fig:CNM}), the region over which we present our spectra and images. The density profiles of TDEs have only been estimated for a small handful of radio TDEs. As of yet, the behaviour of the density profile down to $r\sim 1000 r_g$ has only been measured in AT2019dsg while the other known TDEs have profile measurements down to $r\gtrsim 10^4 r_g$. Additionally, the overall scale of the gas density (i.e. the value of $\rho_{\rm{atm,max}}$ as defined in Section \ref{sec:initialconditions}) appears to vary between systems. For instance, the CNM in the radio TDE CNSS J0019+00 at $r\approx 10^6\, r_g$ has the same estimated density as ASASSN-14li at $r\approx 10^5\, r_g$. Simulations with more dense atmospheres should be explored, as well as more or less steep gas density profiles, as the jet may experience strong deceleration and/or collimation as it encounters the CNM (e.g. see \citealt{2017MNRAS.469.4957B}).

\subsection{Comparison with Radio TDEs}

The density weighted outflow velocity of each model is $\sim0.2-0.35c$, which is mildly relativistic but substantially faster than the $\lesssim 0.1 c$ found in most radio quiet TDEs so the outflow properties cannot explain the inferred velocities in these transients. Of note however is the production of $\gamma > 2$ gas in the jet for the $a_*\geq0.5$ models. \citet{2015MNRAS.453.3213S} found that the isotropic equivalent luminosity for highly super-Eddington ($\dot{M}>10^3\dot{M}_{\rm{Edd}}$) accretion flows can exceed $10^{47}\, \rm{erg \, s^{-1}}$, which is sufficient to explain the $\sim10^{48}\ \rm{erg \, s^{-1}}$ X-ray equivalent luminosity seen in \textit{Swift} J1644+57 \citep{2011Sci...333..203B,2011Natur.476..421B}. However, they performed simulations with $a_*=0$ models and suggested that a $\gamma\gtrsim 2$ component may be untenable with radiative acceleration. Our simulation results suggest that for mass accretion rates $\sim11-25\dot{M}_{\rm{Edd}}$ (compared to the $\sim45-4800\dot{M}_{\rm{Edd}}$ in \citealt{2015MNRAS.453.3213S}), the Poynting flux may result in a relativistic component. Not only does the jet power increase substantially as the BH spin increases in our simulations, but a relativistic jet component with $\gamma>2$ is produced. This is due to increased magnetic energy flux in the jet, which leads to Poynting acceleration of gas where $\sigma > 1$.

Assuming $T_e=T_i$, the thermal synchrotron spectra presented in this work suggest that super-Eddington accretion disks can produce highly energetic jets which primarily emit at $\nu > 100$ GHz and are bright with $L\sim 10^{41} \, \rm{erg \ s^{-1}}$ for $>48$ days. This is in staunch contrast with the radio properties of non-jetted TDEs. Non-jetted TDEs all appear to have a peak frequency of the order $\nu_{\rm{peak}}\sim 10$ GHz and have a luminosity of $L_{\rm{radio}}\approx 10^{37} - 10^{39} \, \rm{erg \ s^{-1}}$ at the time of detection.

The jet emission models which are brightest at 230 GHz, and thus the most favorable for detection in terms of ngEHT observations, are $\mathcal{R}=1$. However, these models are in conflict with TDE observations in terms of the peak frequency. The thermal synchrotron models which peak at smaller frequencies, and are thus more similar to known radio TDEs in terms of the peak frequency at least, have $\mathcal{R}>10$. Nevertheless, none of our models are able to remain bright enough ($L_{\rm{radio}}\gtrsim 10^{37}\ \rm{erg \, s^{-1}}$) and simultaneously peak at $\lesssim10$ GHz.

Another possibility is that jets in these models will continue evolving towards lower frequencies and luminosity and we are simply observing a higher energy stage of the jet. For example, the low frequency luminosity in each of our models at $\nu \approx 5-8.4$ GHz (where most radio TDEs have been observed) is dimmer than non-jetted TDEs. However, our models showed evolution towards lower frequencies, and in the case of \texttt{m10a0.0-HR} the luminosity at the peak frequency is also decreasing over time. This suggests that our models may evolve towards spectra more similar to radio quiet TDEs given enough time.

This is an observationally interesting possibility since the majority of follow-up observations of TDEs in the radio have been at $5-8.4$ GHz. Our models have a luminosity of $\lesssim 10^{36} \, \rm{erg \ s^{-1}}$ at 5-8.4 GHz, placing them below upper limits in the majority of TDEs where no radio emission was detected (mostly $>10^{37}\, \rm{erg \ s^{-1}}$, see Figure 1 in \citealt{2020SSRv..216...81A}). If there is indeed a radio/submillimeter component at $\nu > 100$ GHz that is bright for the first few weeks of non-jetted TDEs, low frequency searches are likely to miss this emission. Our results suggest earlier follow-up observations at high frequencies may be necessary to capture the full radio/submillimeter activity of some TDEs.

\section{Conclusions} \label{sec:conclusions}

We have analyzed outflows from SANE super-Eddington accretion disks across mass and spin parameter space with mass accretion rates $\dot{M} \gtrsim 11 \dot{M}_{\rm{Edd}}$. We confirm that dissipation takes place at the jet head as well as along the jet due to variable ejections of high velocity gas, which propagate along the jet axis and shock with slower moving gas within the jet. The jet power and maximum velocity of gas near the pole increases with BH spin and models with $a_*\geq 0.5$ launch $\gamma>2$ gas that is low density. However, we find that the density weighted outflow velocity is roughly $\sim0.2-0.35c$ for each model.

Through GRRT post-processing with \texttt{ipole}, we produce time dependent thermal synchrotron spectra assuming a single temperature plasma. The spectra peak at $>100$ GHz in each model and the overall radio/submillimeter luminosity increases with BH spin. In general, the jet head brightens as it expands into the CNM. However, models \texttt{m10a0.0-HR} and \texttt{m10a0.5-HR} appear to be dimming as they evolve. In addition, model \texttt{m5a0.0-HR} shows a delayed jet brightening with its jet remaining initially dim with $L\sim 10^{38}\, \rm{erg\ s^{-1}}$ and increasing to  $L\sim 5\times 10^{39}\, \rm{erg\ s^{-1}}$ at $t>58,000\, t_g$.

We also test a simple electron temperature model. We find that increasing $T_i/T_e = \mathcal{R}$ has the effect of reducing the peak frequency and the overall luminosity. For instance, we found that $\mathcal{R}=20$ reduced the peak frequency from $\sim200$ GHz to $\sim20$ GHz and the luminosity from $\sim10^{41} \, \rm{erg \ s^{-1}}$ to $\sim10^{37} \, \rm{erg \ s^{-1}}$ in model \texttt{m10a0.5-HR}. We assumed the temperature ratio was independent of $\beta_g = p_{\rm{gas}}/p_{\rm{mag}}$, but there may be spatial variation in $\mathcal{R}$. In addition, a more accurate prescription may require some accounting for the radiation pressure as well since our simulations are in GRRMHD.

The 230 GHz images show that the brightest feature in each jet is the jet head, which shocks on the CNM. Internal shocks driven by variable ejection events also produce bright `bubbles' of emission at smaller radii in the jet. For sources resolved on scales $\Delta R\lesssim 10000\, r_g$, we predict that the jet head and the internal shocks can be distinguished so long as the jet is not viewed at steep viewing angles. At poorer resolution ($\Delta R\lesssim 50000 r_g$), the jet head is the most distinguishable feature.

We tested the viability of detecting and resolving the jets in each model at 230 GHz assuming flux limits appropriate to the EHT (ngEHT). The spin $a_*=0.9$ models are bright enough to resolve the jet head within $\lesssim110$ (180) Mpc if $\mathcal{R}=1$. At this distance, $\sim45$ (200) TDEs are expected per year which suggests that several TDEs could be potential targets for radio/submillimeter follow up during future observing missions. Our calculations suggest that at most a factor of 4 increase in the number of TDE targets is possible if the detector sensitivity is improved to $5$ mJy in the ngEHT. If the electrons are significantly cooler than the ions, the maximum distance where resolved jets may be detected is significantly reduced. For example, we find a maximum resolved distance of $\sim 8$ (18) Mpc for $\mathcal{R}=20$ which would reduce the number of potential targets per year to less than one even for the ngEHT. Our simulations may apply to the super-Eddington phase of a TDE, roughly $\sim 2t_{\rm{fb}}$ ($\approx1-$several months). We suggest that high frequency radio/submillimeter follow-up of nearby TDEs during this early period be conducted to search for radio/submillimeter jets.

We must stress that much is unknown about jet launching in TDEs, thus while the number of TDEs increases with distance, only a fraction of these may ever appear in the radio/submillimeter. In this work, we have merely provided an analysis of a single mechanism by which radio/submillimeter emission can be produced, that is jets from SANE, super-Eddington accretion disks which undergo conical expansion, and determine the viability of detection of such jets with the EHT/ngEHT. As of this writing, the majority of known radio TDEs appear to have produced their radio emission through sub-relativistic gas shocking with the CNM. However, if the accretion flow is indeed super-Eddington, some fraction of future radio TDEs may launch jets similar to those described in this work. Our model spectra were below detection limits at 5-8.4 GHz, where most radio follow-ups to date have been conducted, which suggests that such jets may have escaped detection previously. Radio/submillimeter follow-up of TDEs at $>100$ GHz may reveal a higher energy component associated with a bipolar jet produced by a super-Eddington accretion disk.

\section*{Acknowledgements}

We thank Ben Prather and George Wong for their assistance with the \texttt{ipole} code in this work. We thank Angelo Ricarte, Dominic Pesce, and Lindy Blackburn for helpful comments and discussions. We also thank the anonymous EHT publication committee referee who provided useful feedback. This work was made possible thanks to the NSBP/SAO EHT Scholars program and was supported in part by NSF grant AST-1816420, and made use of computational support from NSF via XSEDE resources (grant TG-AST080026N). This work was carried out at the Black Hole Initiative at Harvard University, which is supported by grants from the John Templeton Foundation and the Gordon and Betty Moore Foundation.

\section*{Data Availability}
The data underlying this article will be shared on reasonable request to the corresponding author.



\bibliographystyle{mnras}
\bibliography{main}




\appendix

\section{Additional Figures}

\begin{figure*}
    \centering{}
	\includegraphics[width=\textwidth]{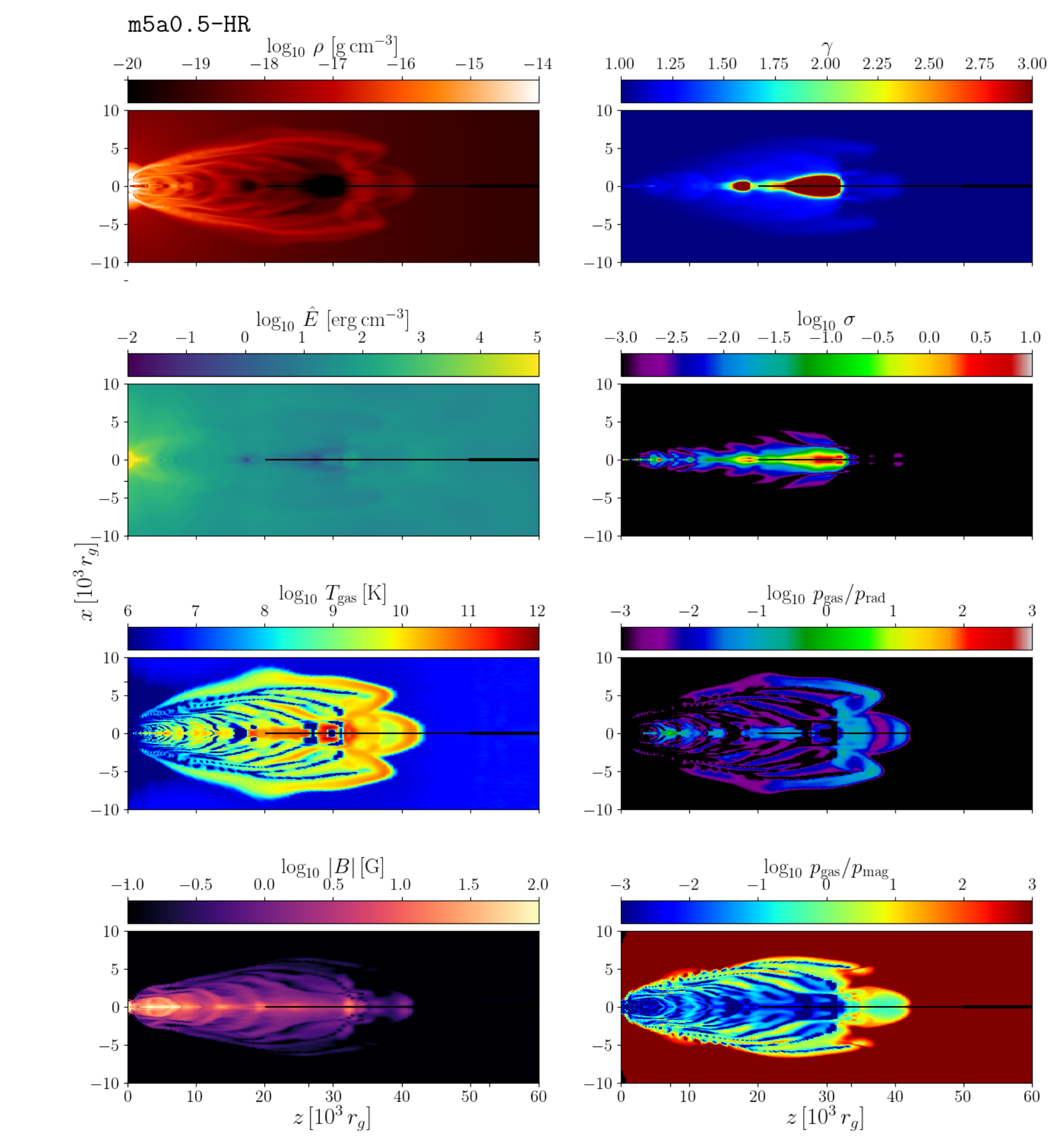}
    \caption{The same as Figure \ref{fig:m5a00HR} but for \texttt{m5a0.5-HR}.}
    \label{fig:m5a05HR}
\end{figure*}

Here we show additional figures for comparison with fiducial models described in the text. We show the jet properties of \texttt{m5a0.5-HR} in Figure \ref{fig:m5a05HR}. Compared to \texttt{m5a0.0-HR}, the jet has travelled nearly the same distance ($\sim 40,000\,r_g$), but the jet core is much faster and the gas temperature in the core is hotter. In addition, the magnetic field strength along the jet axis is larger. 

\begin{figure*}
    \centering{}
	\includegraphics[width=\textwidth]{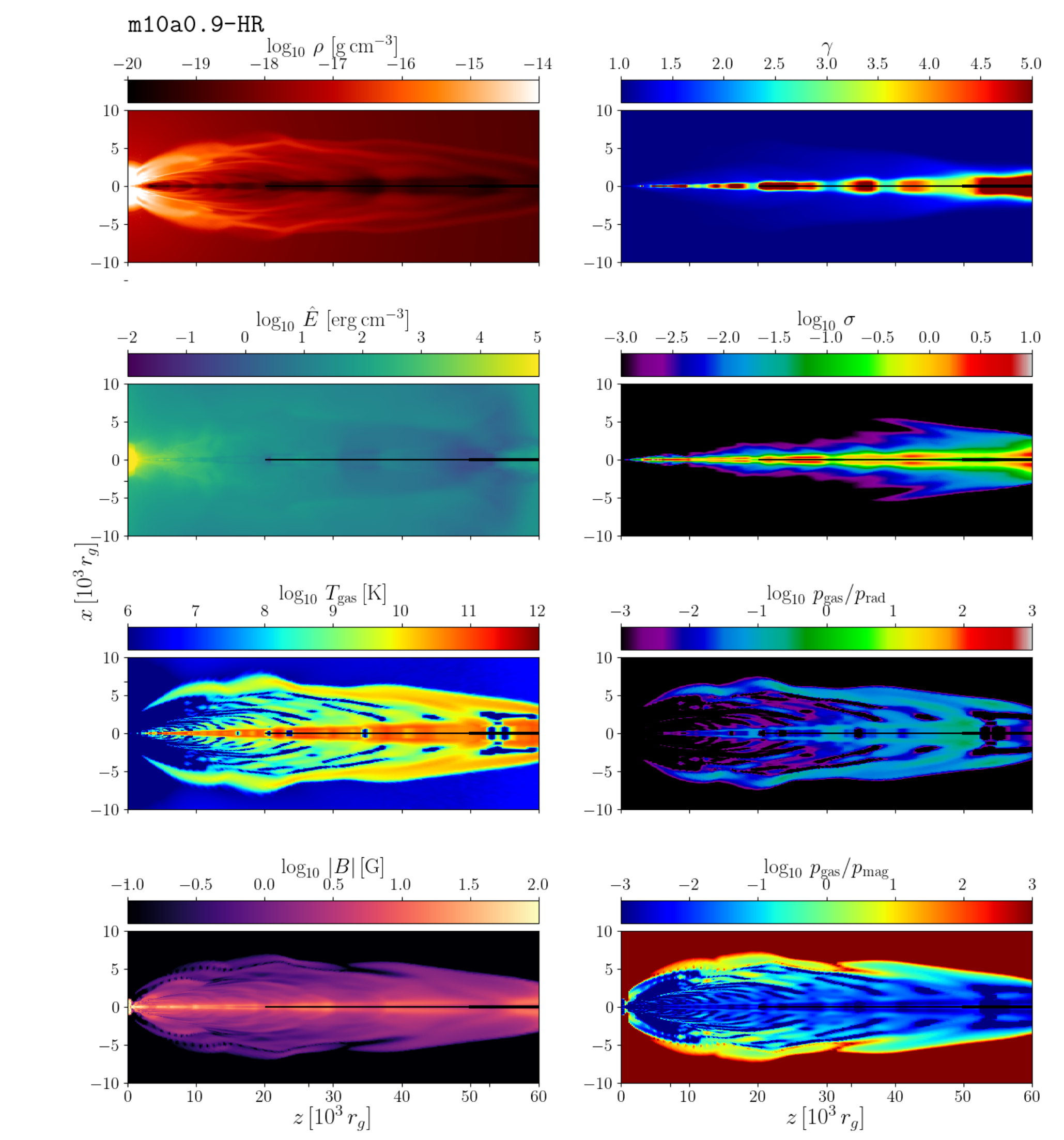}
    \caption{The same as Figure \ref{fig:m5a00HR} but for \texttt{m10a0.9-HR}.}
    \label{fig:m10a09HR}
\end{figure*}

We show the jet properties of \texttt{m10a0.9-HR} in Figure \ref{fig:m10a09HR}. Compared to \texttt{m5a0.9-HR}, the jet features are generally similar. This is in well agreement with the general result that we find throughout this work that the BH mass does not have a noticeable effect on the jet. The most important factor is instead the BH spin.

\begin{figure*}
	\includegraphics[width=\columnwidth]{images_pdf/6.7s00hi_HR_0830_th90.pdf}
	\includegraphics[width=\columnwidth]{images_pdf/7s00hi_HR_0580_th90.pdf}\\
	\includegraphics[width=\columnwidth]{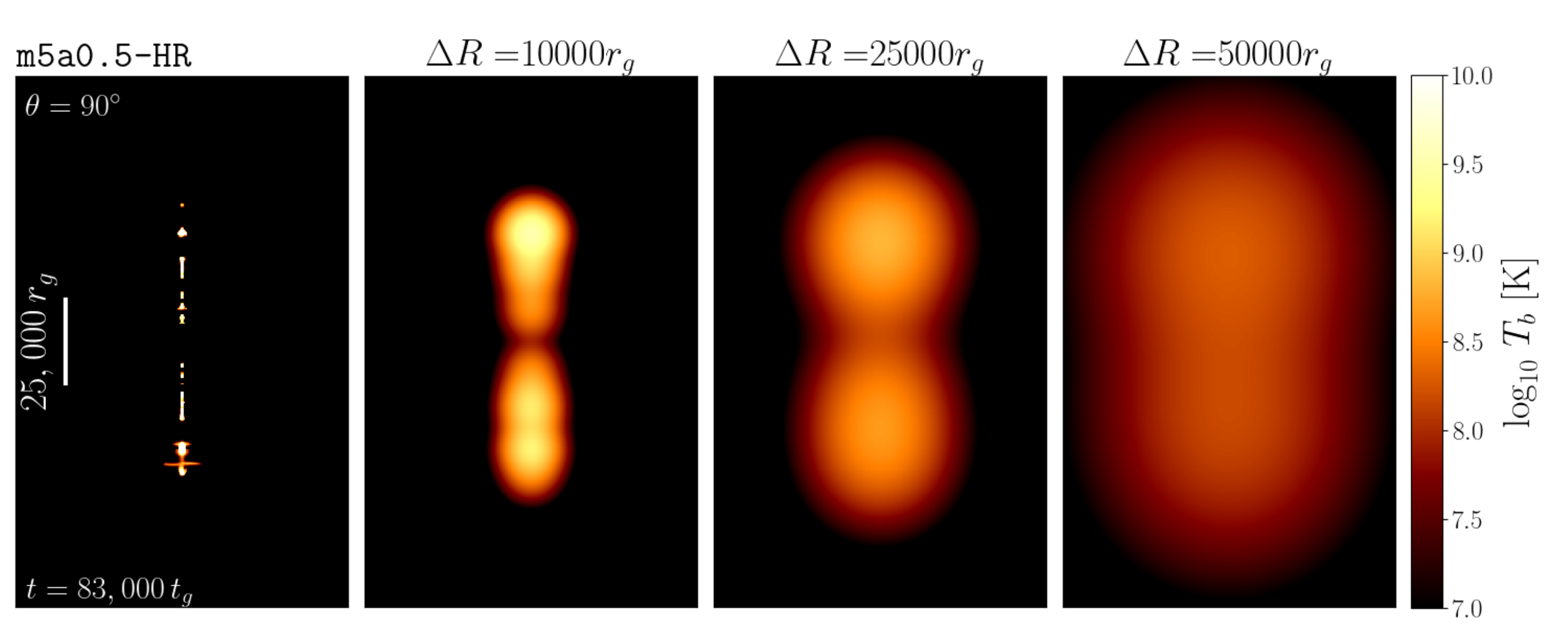}
	\includegraphics[width=\columnwidth]{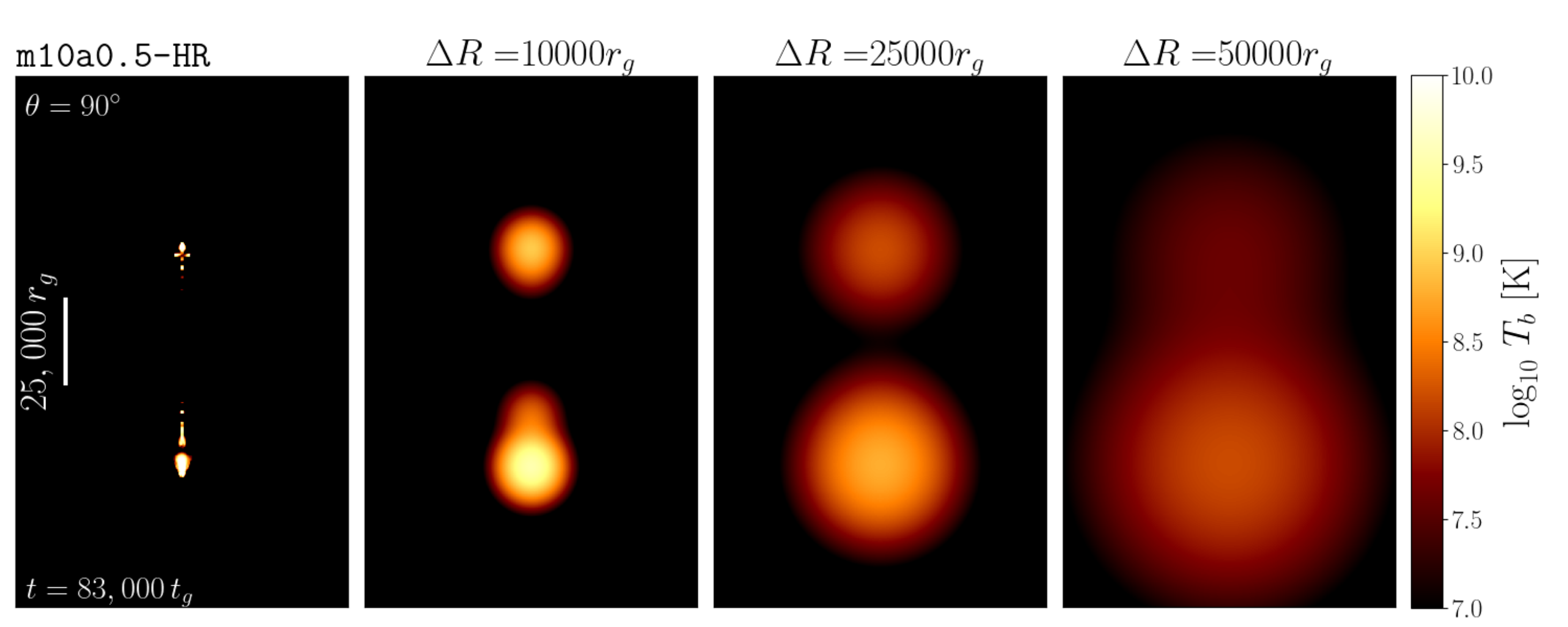}\\
	\includegraphics[width=\columnwidth]{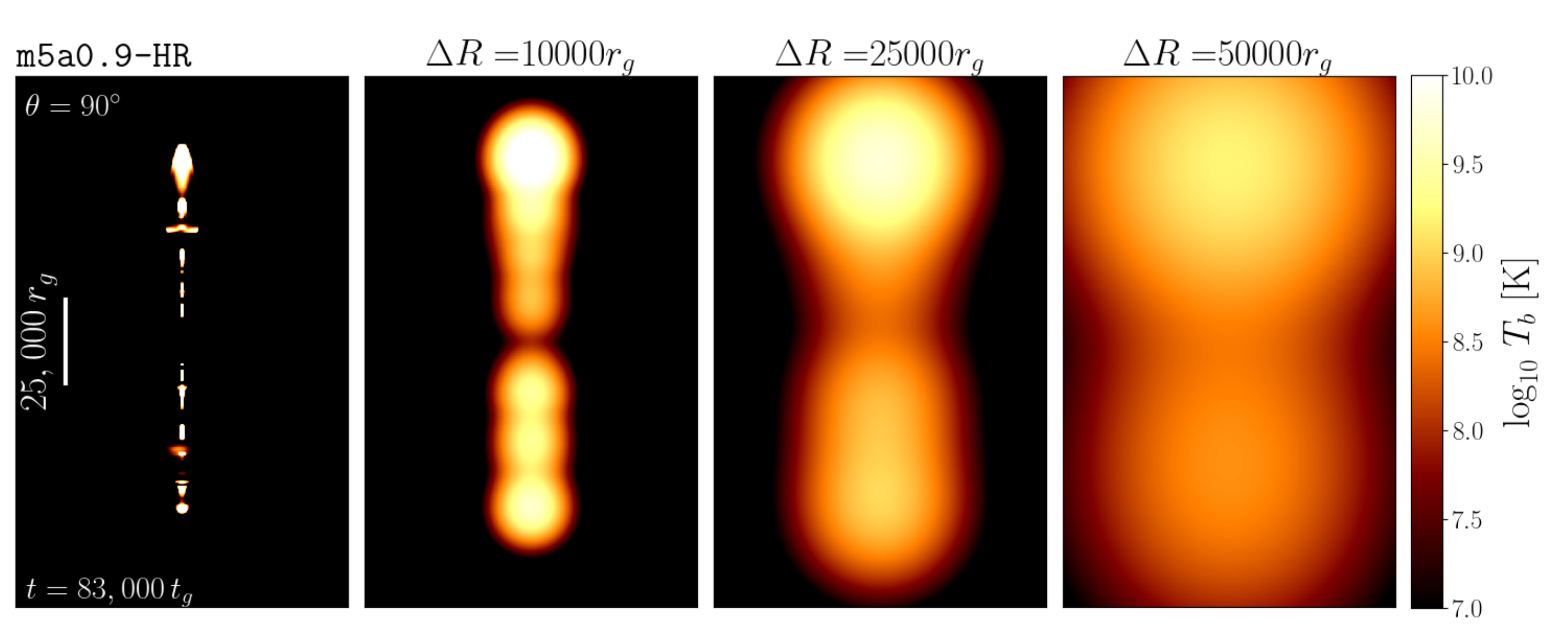}
	\includegraphics[width=\columnwidth]{images_pdf/7s09hi_HR_0812_th90.pdf}
    \caption{Here we compare the jets for each model near their brightest point viewed at $90^\circ$.}
    \label{fig:all_230GHz_images}
\end{figure*}

In Figure \ref{fig:all_230GHz_images}, we show the 230 GHz jet emission viewed at $90^\circ$ near its brightest point in each simulation.

\label{lastpage}
\end{document}